\documentclass[11pt]{article}

\usepackage{verbatim,amsmath,amssymb}
\usepackage{epsfig,float,color}
\usepackage{geometry}
\usepackage{setspace}
\usepackage[export]{adjustbox}
\usepackage{subfig}
\usepackage{natbib}
\usepackage{placeins}
\usepackage{amsthm,mathrsfs,amsfonts,dsfont} 
\usepackage{hyperref}
\usepackage{tikz}
\usetikzlibrary{calc,math}
\usepackage{nicefrac}
\usepackage{pgfplots}
\pgfplotsset{compat=newest}
\usepackage{framed}

\definecolor{darkblue}{rgb}{0,0,1}
\definecolor{dgreen}{rgb}{0,0.5,0}
\hypersetup{colorlinks=true, breaklinks=true, linkcolor=darkblue, menucolor=darkblue,  citecolor=darkblue, urlcolor=darkblue}

\newcommand{\lr}[1]{\left(#1\right)}
\newcommand{\Psiel}{\Psi_\mathrm{el}}
\newcommand{\Psimem}{\Psi_\mathrm{mem}}

\newcommand{\Psibend}{\Psi_\mathrm{bend}}
\newcommand{\Psielp}{\Psi_\mathrm{el}^{+}}
\newcommand{\Psielm}{\Psi_\mathrm{el}^{-}}

\newcommand{\ab}{{\alpha\beta}}

\newcommand{\gd}{{\gamma\delta}}

\newcommand{\abgd}{{\alpha\beta\gamma\delta}}

\DeclareRobustCommand{\rchi}{{\mathpalette\irchi\relax}}
\newcommand{\irchi}[2]{\raisebox{\depth}{$#1\chi$}}

\newcommand{\mrT}{\mathrm{T}}

\newcommand {\mrS}{\mathrm{S}}

\newcommand{\nablao}{\nabla_{\!\mrS}}

\newcommand {\eqb}[1]{\begin{equation}\begin{array}{#1}}
\newcommand {\eqe}{\end{array}\end{equation}}

\newcommand {\esb}[1]{\begin{equation*}\begin{array}{#1}}
\newcommand {\ese}{\end{array}\end{equation*}}
\newcommand {\ds}{\displaystyle}

\newcommand {\pa}[2]{\frac{\partial{#1}}{\partial{#2}}}

\newcommand {\paqq}[3]{\frac{\partial^2{#1}}{\partial{#2}\,\partial{#3}}}
\newcommand {\back}{\! \! \!}
\newcommand {\is}{\back &=& \back}
\newcommand {\dis}{\back &:=& \back}

\newcommand {\norm}[1]{\|#1\|}

\newcommand {\dif}{\mathrm{d}}

\newcommand {\II}{{I\kern-.3em I}}
\newcommand {\III}{{I\kern-.3em I\kern-.3em I}}

\newcommand {\mrc}{\mathrm{c}}

\newcommand {\mrf}{\mathrm{f}}

\newcommand {\mrm}{\mathrm{m}}
\newcommand {\mrn}{\mathrm{n}}

\newcommand {\mrx}{\mathrm{x}}

\newcommand {\mf}{\mathbf{f}}

\newcommand {\mk}{\mathbf{k}}

\newcommand {\mm}{\mathbf{m}}

\newcommand {\mx}{\mathbf{x}}

\newcommand {\ba}{\boldsymbol{a}}

\newcommand {\bff}{\boldsymbol{f}}

\newcommand {\bn}{\boldsymbol{n}}

\newcommand {\bv}{\boldsymbol{v}}

\newcommand {\bx}{\boldsymbol{x}}

\newcommand {\bnu}{\mbox{\boldmath$\nu$}}
\newcommand {\bmu}{\mbox{\boldmath$\mu$}}

\newcommand {\bphi}{\mbox{\boldmath$\phi$}}

\newcommand {\mK}{\mathbf{K}}

\newcommand {\mM}{\mathbf{M}}
\newcommand {\mN}{\mathbf{N}}

\newcommand {\mX}{\mathbf{X}}

\newcommand {\bA}{\boldsymbol{A}}
\newcommand {\bB}{\boldsymbol{B}}

\newcommand {\bE}{\boldsymbol{E}}
\newcommand {\bF}{\boldsymbol{F}}

\newcommand {\bK}{\boldsymbol{K}}

\newcommand {\bM}{\boldsymbol{M}}
\newcommand {\bN}{\boldsymbol{N}}

\newcommand {\bT}{\boldsymbol{T}}

\newcommand {\bX}{\boldsymbol{X}}

\newcommand {\sig}{\sigma}

\newcommand {\bsig}{\mbox{\boldmath$\sigma$}}

\newcommand {\bone}{\mathbf{1}}

\newcommand {\IR}{{\rm\kern.24em
   \vrule width.02em height1.53ex depth-.05ex
   \kern-.3em R}}
\newcommand {\ic}{{\rm\kern.20em
   \vrule width.02em height1.0ex depth-.05ex
   \kern-.22em c}}
\newcommand {\ia}{{\rm\kern.20em
   \vrule width.02em height1.05ex depth-.0ex
   \kern-.25em a}}
\newcommand {\IC}{{\rm\kern.24em
   \vrule width.02em height1.4ex depth-.05ex
   \kern-.26em C}}
\newcommand {\ID}{{\rm\kern.34em
   \vrule width.02em height1.5ex depth-.05ex
   \kern-.36em D}}
\newcommand {\IS}{{\rm\kern.24em
   \vrule width.02em height1.6ex depth.05ex
   \kern-.26em S}}
\newcommand {\IT}{{\rm\kern.50em
   \vrule width.02em height1.55ex depth-.05ex
   \kern-.52em T}}

\newcommand {\IE}{{\rm\kern.24em
   \vrule width.02em height1.55ex depth-.05ex
   \kern-.33em E}}
\newcommand {\IEa}{{\rm\kern.24em
   \vrule width.02em height1.55ex depth-.05ex
   \kern-.33em E}^{1}_{ijkl}}
\newcommand {\IEb}{{\rm\kern.24em
   \vrule width.02em height1.55ex depth-.05ex
   \kern-.33em E}^{2}_{ijkl}}

\newcommand {\sG}{\mathcal{G}}
\newcommand {\sH}{\mathcal{H}}

\newcommand {\sS}{\mathcal{S}}

\newcommand {\sU}{\mathcal{U}}
\newcommand {\sV}{\mathcal{V}}

\newcommand {\Ass}[2]{\kern 0.9ex \vrule width0.45em height0.2ex depth0ex \kern -2.1ex \bigwedge_{#1}^{#2}}
\newcommand {\ASS}[2]{\kern 1.45ex \vrule width0.5em height0.2ex depth0ex \kern -2.65ex \bigwedge_{#1}^{#2}}

\begin{document}

\begin{center}
\Large{\bf{An adaptive space-time phase field formulation for\\dynamic fracture of brittle shells based on LR NURBS}}\\

\end{center}

\begin{center}
\large{Karsten Paul$^\ast$, Christopher Zimmermann$^\ast$, Kranthi K. Mandadapu$^{\dagger\S}$,\\[1mm] Thomas J.R. Hughes$^\ddagger$, Chad M. Landis$^\ddagger$, Roger A. Sauer$^\ast$\footnote{corresponding author, email: sauer@aices.rwth-aachen.de}}
\vspace{4mm}

\small{\textit{$^\ast$Aachen Institute for Advanced Study in Computational Engineering Science (AICES), \\ RWTH Aachen University, Templergraben 55, 52062 Aachen, Germany}}

\small{\textit{$^\dagger$Department of Chemical and Biomolecular Engineering,\\ University of California at Berkeley,
110A Gilman Hall, Berkeley, CA 94720-1460, USA}}

\small{\textit{$^\S$Chemical Sciences Division, Lawrence Berkeley National Laboratory, CA 94720, USA}}

\small{\textit{$^\ddagger$The Oden Institute for Computational Engineering and Sciences,\\ The University of Texas at Austin, 201 E. 24th Street, POB 4.102,\\ 1 University Station (C0200), Austin, TX 78712-1229, USA}}

\vspace{4mm}

Published\footnote{This pdf is the personal version of an article whose final publication is available at \href{https://link.springer.com/article/10.1007/s00466-019-01807-y}{link.springer.com}.} 
in \textit{Comput. Mech.}, 
\href{https://link.springer.com/article/10.1007/s00466-019-01807-y}{DOI: 10.1007/s00466-019-01807-y} \\
Submitted on 28.~June 2019, Revised on 20.~September 2019, Accepted on 21.~November 2019 

\end{center}

\vspace{3mm}

\rule{\linewidth}{.15mm}
{\bf Abstract}\\
We present an adaptive space-time phase field formulation for dynamic fracture of brittle shells. Their deformation is characterized by the Kirchhoff-Love thin shell theory using a curvilinear surface description. All kinematical objects are defined on the shell's mid-plane. The evolution equation for the phase field is determined by the minimization of an energy functional based on Griffith's theory of brittle fracture. Membrane and bending contributions to the fracture process are modeled separately and a thickness integration is established for the latter.
The coupled system consists of two nonlinear fourth-order PDEs and all quantities are defined on an evolving two-dimensional manifold. Since the weak form requires $C^1$-continuity, isogeometric shape functions are used. The mesh is adaptively refined based on the phase field using Locally Refinable (LR) NURBS. Time is discretized based on a generalized-$\alpha$ method using adaptive time-stepping, and the discretized coupled system is solved with a monolithic Newton-Raphson scheme. The interaction between surface deformation and crack evolution is demonstrated by several numerical examples showing dynamic crack propagation and branching.


\vspace{-4mm}
\rule{\linewidth}{.15mm}

\section{Introduction} \label{sec:intro}
The need for shortening development cycles of engineering components requires efficient computational methods. 
The robustness requirements for these components are increasing so that the prediction of structural defects and failure plays a major role in current development processes. It is therefore important to have efficient and reliable computational methods for predicting fracture. Several computational methods have been introduced to model crack growth. The most important ones in the framework of finite elements are described subsequently.

\textit{Sharp interface models} introduce discontinuities within the body in order to model cracks. In the \textit{extended finite element method} by \cite{moes1999}, the basis functions are enriched by discontinuities to model the displacement jump across cracks. In contrast to this, a crack can be introduced by a modification of the finite element mesh as in the \textit{virtual crack closure technique} \citep{krueger2004}. Similar to the \textit{extended finite element method}, \cite{remmers2003} also enrich the basis in the \textit{cohesive segments method}.
Several of these sharp interface models have been used to model dynamic fracture and fragmentation. \cite{ortiz1999} introduce cohesive elements in a large deformation framework to track evolving cracks in a dynamic framework. Fragmentation stemming from high loading rates is investigated by \cite{molinari2007} within the small strain regime, based on the cohesive element approach. In \cite{papoulia2017}, a cohesive model based on a non-differentiable energy functional is outlined. They add a momentum term to the latter to enable the use of implicit time-stepping. The latter has been further advanced by \cite{vavasis2020}.
In \cite{hirmand2018}, a discontinuous Galerkin-formulation is used to model dynamic fracture. They employ Newmark's time integration scheme and use a trust region minimization approach to solve the smooth non-convex problems that occur in their formulation.
 \cite{geelen2018} combine a phase field formulation with an extended finite element method by using a diffuse crack tip and a sharp traction-free crack behind it. In \cite{radovitzky2011}, a combination of a discontinuous Galerkin-formulation and a cohesive zone model is presented. This combination ensures stability and robustness prior to the onset of fracture and shows good scalability with respect to computation time. \cite{geelen2019} consider cohesive fracture and investigate a novel degradation function and different approaches to enforce an irreversible fracture process. Explicit and implicit time integration schemes are compared in a dynamic cohesive fracture framework in \cite{hirmand2019}. Their formulation leads to a flexible framework that is easy to implement into existing standard finite element frameworks.
In general, the location of the crack has to be known in sharp interface models. Thus, it has to be numerically tracked, which tends to be a complex task, especially in three dimensions.

Thus, \textit{diffuse interface models} have gained popularity for modeling brittle fracture. In the \textit{phase field method} no discontinuities are introduced within the body. Instead, the crack is smoothed out and described by a small transition zone that ranges between undamaged and fully fractured material. Phase field methods describe the evolving cracks by an additional partial differential equation (PDE) such that there is no need for tracking the interface. For complex crack patterns including nucleation, branching, and merging, phase field formulations have been shown to be very effective.

Based on the thermodynamic considerations of brittle fracture by \cite{griffith1921}, a variational formulation of brittle fracture has been introduced by \cite{francfort1998}. Their formulation includes the minimization of a global energy functional to model the quasi-static fracture process. A corresponding phase field implementation within the finite element method has been presented by \cite{bourdin2000}. The robustness and accuracy of the variational formulation in two and three dimensions using phase field methods have been demonstrated by e.g. \cite{miehe2010a} and \cite{miehe2010b}. Successful extensions to dynamic problems have been presented by \cite{larsen2010a}, \cite{larsen2010b}, \cite{bourdin2011}, \cite{borden2012}, \cite{hofacker2012} and \cite{schlueter2014}.
In contrast to the variational formulation of brittle fracture, \cite{karma2001} and \cite{kuhn2010} use a phase transition framework based on the Ginzburg-Landau equation. The latter is more often used in the physics community. Its derivation is based on general phase separation processes and small adjustments are required for fracture, for instance to avoid crack healing. In these models, the onset of brittle fracture is not seen as instantaneous, but obeying its own gradient-based dynamics.
A stabilization for quasi-static simulations using a monolithic solution approach for the coupled system is proposed by \cite{gerasimov2016}. \cite{heister2015} convexify their energy functional to obtain a positive definite Hessian matrix for monolithic coupling. \cite{gerasimov2018} apply a non-intrusive global/local approach in a phase field framework for brittle fracture, in which at first the structural analysis of the whole domain is performed and, afterwards, local regions where fracture is predicted are re-analyzed. These steps are then repeated until convergence is obtained. \cite{ambati2015} summarize several phase field formulations for brittle fracture. In the work of \cite{kuhn2015}, the influence of different degradation functions on the solution is investigated. Similar investigations are made by \cite{sargado2018} who also study parametric degradation functions. Possibilities to enforce irreversibility of the fracture process are presented in detail in the work of \cite{gerasimov2019}, especially focusing on the penalty method. The authors also derive a lower bound for the penalty parameter for a quasi-static second-order phase field model  for brittle fracture.

The majority of the published phase field methods for fracture use a second-order phase field formulation.
The high order differential operators of the phase field PDE stemming from the crack density functional of \cite{borden2014}, which is used in this work, and the equation of motion of the shell framework require a spatial finite element discretization that is at least $C^1$-continuous. Isogeometric Analysis (IGA), proposed by \cite{hughes2005}, allows for user-defined smoothness of the solution within the finite element framework. Within IGA, the smoothness is most commonly achieved through the use of B-Spline- and NURBS-based shape functions. Since phase field methods require a highly resolved finite element mesh in the transition zone, local refinement methods are commonly used in the context of phase field methods for fracture. The introduction of hierarchical B-splines by \cite{forsey1988} has offered the possibility of local refinement within an IGA framework. The extension to the local refinement of NURBS is for instance given by \cite{sederberg2003} by introducing T-Splines. Another approach that allows local refinement is Locally Refinable (LR) splines. LR B-splines were first introduced by \cite{dokken13} and further advanced by \cite{johannessen2014}. Their extension to LR NURBS is provided by \cite{zimmermann17}. A combination of LR and T-splines is given by \cite{chen2018a} by the introduction of LR T-splines. In constrast to LR splines, LR T-splines take a T-mesh as input instead of a tensor-product mesh. Isogeometric collocation methods \citep{gomez2014,reali2015} for phase field models of fracture are also introduced, for instance by \cite{schillinger2015}.

\cite{hesch2016b} employ a hierarchical refinement scheme within a higher order phase field model. Similarly, \cite{hesch2016a} couple a model for frictional contact to a higher order phase field model using hierarchical NURBS. 
\cite{kaestner2016} investigate phase field models by comparing adaptive refinement based on locally refined hierarchical B-splines with uniformly refined discretizations.
\cite{borden2012} propose an adaptive refinement strategy using T-splines and use the phase field value itself to identify the need for local refinement.
Mesh adaptivity schemes, in which a predictor-corrector scheme is used, are employed by \cite{zhou2018} for modeling fracture in rocks and by \cite{badnava2018} to model mechanically and thermo-mechanically induced cracks. In these approaches, the system is solved and then checked for the need of mesh refinement. A similar approach is employed by \cite{heister2015}.
In the work by \cite{nagaraja2018}, a multi-level hp-refinement technique is established using the finite cell method \citep{parvizian2007} to model brittle fracture in two dimensions.
\cite{chen2018b} employ LR T-splines for discrete fracture analysis. They insert mesh lines to obtain discontinuous basis functions that are able to represent sharp cracks.

Many papers concerning the computational modeling of shells within an isogeometric framework have been published, for instance by \cite{benson2013}, \cite{echter2013}, \cite{kiendl2015} and \cite{duong2017}.
Since for shells the bending stress varies across the thickness, a suitable split of the energy within the fracture model has to be established.
In the work by \cite{ulmer2012}, brittle fracture in thin plates and shells is modeled. They combine a plate and a standard membrane to model the shell but only split the membrane and not the bending part of the elastic energy. Thus, the whole bending energy contributes to crack evolution and is degraded in regions of damage. \cite{amiri2014} do not employ an energy split, which limits their model to shells under pure tension. In the work by \cite{ambati2016b}, the shell and the phase field are also discretized over the thickness. \cite{areias2016} utilize two phase fields, one for the top and the other one for the bottom face of the shell. This framework is also used by \cite{reinoso2017} for a $6$-parameter shell model. Their formulation results in a non-constant phase field throughout the thickness. In contrast to this, \cite{kiendl2016} use a constant phase field over the thickness but use thickness integration to split the whole energy into a tensile part, which contributes to crack growth, and a compressive part, which does not. 

\cite{zimmermann2019} model Cahn-Hillard phase field equations on deforming surfaces based on the shell formulation of \cite{duong2017}. Even though a different physical process is modeled, the resulting coupled finite element formulation is similar to the one proposed here.

In this paper we establish a dynamic brittle fracture framework within the nonlinear IGA thin shell formulation of \cite{duong2017}, in which shells with arbitrarily large curvature or doubly curved shells can be modeled. Its hyperelastic material model allows for large deformations and is given as a sum of membrane and bending contributions. The proposed higher order phase field model of \cite{borden2014} is adopted because of its higher rate of convergence and it is formulated on the shell's mid-plane. Motivated by the work of \cite{kiendl2016}, bending effects on the fracture process are modeled based on thickness integration. Adaptive spatial refinement is based on LR NURBS \citep{zimmermann17} and temporal discretization is based on the generalized-$\alpha$ scheme \citep{chung93}. The time steps are adjusted based on the number of Newton-Raphson iterations required during the last time step.
In summary, the proposed formulation contains the following features:
\begin{itemize}
	\item It couples a higher-order phase field model for
	fracture with a nonlinear shell formulation.
	\item It is formulated in curvilinear coordinates, and applicable to general shell configurations.
	\item The coupled system is solved within a monolithic, fully implicit solution approach.
	\item It uses adaptive local refinement in space and time. 
	\item The spatial discretization is based on LR NURBS.
	\item An energy split is used in which the membrane and bending energies are split separately.
\end{itemize}
The subsequent sections are structured as follows: 
Sec.~\ref{Sec:def_s} summarizes the surface description and kinematics. 
The balance laws and the equation of motion are derived in Sec.~\ref{Sec:thin_s}. 
Sec.~\ref{sec:frac} introduces the energy minimization problem and the material model employed. 
Extensions to degradation, irreversibility and an energy split are also presented. Based on the Euler-Lagrange equation, the Helmholtz free energy is minimized, which leads to the governing equation for the phase field's evolution. 
The discretization of the coupled problem is described in Sec. \ref{sec:discr}. 
Numerical examples are presented in Sec.~\ref{sec:num_ex} to illustrate crack propagation on curved surfaces.
Conclusions are drawn in Sec.~\ref{sec:concl}.

\section{Deforming surfaces} 
\label{Sec:def_s}
This section summarizes the thin shell formulation in the framework of curvilinear coordinates and Kichhoff-Love kinematics. A more detailed presentation can be found in \citet{sauer2018}.

\subsection{Surface description} \label{s:surf_descr}

A curved surface $\sS$ in 3D space can be characterized by the parametric description at any time $t$ by the function
\eqb{l}
\bx = \bx(\xi^\alpha,t)\,,\quad \alpha=1,2\,,
\label{e:bx}\eqe
where $\xi^\alpha$ denote the curvilinear coordinates associated with a material point $\bx\in\sS$. $\xi^\alpha$ are convected along with the material deformation of the surface and hence, they are also called \textit{convected coordinates}.
The co-variant tangent vectors at $\bx$  are given by
\eqb{l}
\ba_\alpha := \ds\pa{\bx}{\xi^\alpha}\,.
\label{e:ba}\eqe
From these follow the surface metric
\eqb{l}
a_{\alpha\beta} := \ba_\alpha\cdot\ba_\beta\,,
\eqe
the surface normal
\eqb{l}
\bn := \ds\frac{\ba_1\times\ba_2}{\norm{\ba_1\times\ba_2}}\,,
\label{e:bn}
\eqe
and the contra-variant tangent vectors 
\eqb{l}
\ba^\alpha = a^{\alpha\beta}\,\ba_\beta\,,
\eqe
where $[a^{\alpha\beta}] = [a_{\alpha\beta}]^{-1}$.
All Greek indices range from 1 to 2 and are summed when repeated.
Based on the second parametric derivative $\ba_{\alpha,\beta} := \partial\ba_\alpha/\partial\xi^\beta$, the curvature tensor components
\eqb{l}
b_{\alpha\beta} = \ba_{\alpha,\beta}\cdot\bn\,,
\label{e:b_ab}
\eqe
follow.
The set of initial surface points $\bX\in\sS_0$ follows from $\bX := \bx(\xi^\alpha,0)$.
In analogy to Eqs.~\eqref{e:ba}--\eqref{e:b_ab}, we define the surface quantities
$\bA_\alpha := \partial\bX/\partial\xi^\alpha$,
$A_{\alpha\beta} :=  \bA_\alpha\cdot\bA_\beta$,
$\bN := \bA_1\times\bA_2/\norm{\bA_1\times\bA_2}$,
$\bA^\alpha := A^{\alpha\beta}\bA_\beta$,
$[A^{\alpha\beta}] := [A_{\alpha\beta}]^{-1}$
and
$B_{\alpha\beta} := \bA_{\alpha,\beta}\cdot\bN$
at $t=0$ as a reference configuration, denoted $\sS_0$.
The surface gradient
\eqb{lllll}
\mathrm{grad}_\mrS\phi \is \nablao\phi \dis \phi_{;\alpha}\,\bA^\alpha,
\label{e:diff1}\eqe
and surface Laplacian
\eqb{lllll}
\Delta_\mrS\phi \dis \nablao\cdot\nablao\phi \is \phi_{;\alpha\beta}\,A^{\alpha\beta}\,,
\label{e:diff2}\eqe
can be defined based on the parametrization in Eq.~\eqref{e:bx}.
Here, $\phi$ denotes a general scalar function and the subscript `;' indicates the co-variant derivative. 
It is equal to the parametric derivative for general scalars, i.e.
$\phi_{;\alpha} = \phi_{,\alpha}:=\partial\phi/\partial\xi^\alpha$.
But, $\phi_{;\alpha\beta} \ne \phi_{,\alpha\beta}$ and instead
\eqb{lll}
\phi_{;\alpha\beta} \is \phi_{,\alpha\beta} - \hat\Gamma^\gamma_{\alpha\beta}\,\phi_{,\gamma}\,,
\label{e:phiab}\eqe
where $\hat\Gamma^\gamma_{\alpha\beta} = \bA_{\alpha,\beta}\cdot\bA^\gamma$ are the Christoffel symbols of the second kind on surface $\sS_0$. On $\sS$, these read $\Gamma^\gamma_{\alpha\beta} = \ba_{\alpha,\beta}\cdot\ba^\gamma$.

\subsection{Surface kinematics}

The relation between reference surface $\sS_0$ and current surface $\sS$ is described by the surface deformation gradient
\eqb{l}
	\bF=\ba_\alpha\otimes\bA^\alpha\,.
\eqe
The left surface Cauchy-Green tensor then follows as
\eqb{l}
\bB = A^{\alpha\beta}\,\ba_\alpha\otimes\ba_\beta\,,
\eqe
with its two invariants
\eqb{lll}
I_1 := A^{\alpha\beta}\,a_{\alpha\beta}\,\quad\text{and}\quad
J := \sqrt{\det[A^{\alpha\beta}]\det[a_{\alpha\beta}]}\,.
\label{e:invs}
\eqe
The latter characterizes the surface stretch between $\sS_0$ and $\sS$. The surface Green-Lagrange strain tensor and the symmetric relative curvature tensor are
\eqb{lll}
	\bE\is\ds\frac{1}{2}\lr{a_{\alpha\beta}-A_{\alpha\beta}}\bA^\alpha\otimes\bA^\beta,\\[3mm]
	\bK\is\ds\lr{b_{\alpha\beta}-B_{\alpha\beta}}\bA^\alpha\otimes\bA^\beta.
\eqe
The material time derivative is denoted by
\eqb{l}
\dot{(...)} := \ds\pa{...}{t}\Big|_{\xi^\alpha=\,\mathrm{fixed}}\,.
\eqe
This leads to the material velocity at $\bx$
\eqb{l}\label{e:vel}
\bv := \dot\bx\,,
\eqe
and the rates
\eqb{l}
\dot\ba_\alpha = \bv_{,\alpha} = \ds\pa{\bv}{\xi^\alpha}\,,\quad \text{and}\quad \dot a_{\alpha\beta} = \ba_\alpha\cdot\dot\ba_\beta + \dot\ba_\alpha\cdot\ba_\beta\,.
\eqe

\subsection{Surface variations}
The variation of various surface measures is required for the formulation of the weak form of the thin shell equation. 
Particularly important are the variations
\eqb{lll}
\delta a_{\alpha\beta} \is \ba_\alpha\cdot\delta\ba_\beta + \delta\ba_\alpha\cdot\ba_\beta\,, \\[2mm]
\delta b_{\alpha\beta} \is \big(\delta\ba_{\alpha,\beta}-\Gamma^\gamma_{\alpha\beta}\,\delta\ba_\gamma\big)\cdot\bn\,, \\[2mm]
\delta\bn \is -(\ba^\alpha\otimes\bn)\,\delta\ba_\alpha\,,
\eqe
where $\delta\ba_\alpha = \delta\bx_{,\alpha}$ and $\delta
\ba_{\alpha,\beta} = \delta\bx_{,\alpha\beta}$. Here, $\delta\bx$ denotes a kinematically admissible variation of the deformation. Additional variations of surface quantities are provided in \citet{sauer2017a}.

\section{Thin shell theory}\label{Sec:thin_s}
The governing equations for the shell are summarized in the following. Equilibrium is given in strong and weak form. 
Considering Kirchhoff-Love kinematics, the constitutive behavior of thin shells can be fully characterized by the quantities $a_{\alpha\beta}$ and $b_{\alpha\beta}$.

\subsection{Balance of linear and angular momentum}\label{s:mombal}
The equation of motion
\eqb{l}
\rho\,\dot{\bv} = \bT^{\alpha}_{;\alpha} + \bff\,,\quad\forall\,\bx\in\sS\,,
\label{e:sfm}\eqe
follows from the balance of linear momentum for surface $\sS$. $\bff = f^\alpha\,\ba_\alpha + p\,\bn$ denotes prescribed body forces and 
\eqb{l}
\bT^{\alpha} = N^{\alpha\beta}\,\ba_\beta  + S^\alpha\,\bn\,,
\label{e:Ta}\eqe 
are the stress vectors that include the in-plane membrane components $N^{\alpha\beta}$ and the out-of-plane shear components $S^\alpha$ \citep{naghdi1971,steigmann1999,sauer2017a}. 
These are related to the stress tensor
\begin{equation}
\bsig = N^{\alpha \beta} \ba_\alpha \otimes \ba_\beta + S^\alpha \ba_\alpha \otimes \bn\,,
\label{e:bsig}\end{equation}
through Cauchy's formula $\bT^\alpha = \bsig^\mrT \ba^\alpha$. Given the outward pointing normal $\bnu=\nu_\alpha\ba^\alpha$ at a cut through $\sS$, the traction $\bT = \bsig^\mrT \bnu=\bT^\alpha\nu_\alpha$ acting on this cut follows.
\\
Likewise, the moment vector on the cut reads $\bM  = \bmu^\mrT \bnu$ with the moment tensor
\begin{equation}
\bmu = -M^{\alpha \beta} \ba_\alpha \otimes \ba_\beta\,, 
\end{equation}
where  $M^{\alpha\beta}$ denotes its in-plane components \citep{sauer2017a,sahu2017}.
The balance of angular momentum yields
\eqb{lll} 
S^\alpha \is - M^{ \beta \alpha}_{;\beta}\,, \\[1mm]
\sigma^{\alpha \beta} \is \sigma^{\beta\alpha}\,,
\label{e:amom}\eqe
where $\sigma^{\alpha \beta} := N^{\alpha \beta} - b^\beta_{\gamma} M^{\gamma\alpha}$.
The stress components $\sig^{\alpha\beta}$ and $M^{\alpha\beta}$ follow from constitution, which is discussed in Sec. \ref{Sec:matmodel}.
\\
The component form of the equation of motion 
\eqb{lll}
\rho\,a^\alpha \is f^\alpha + N^{\lambda \alpha}_{; \lambda} - S^\lambda b^\alpha_\lambda\,,\\[1mm]
\rho\,a_\mrn \is p + N^{\alpha \beta} b_{\alpha \beta} + S^\alpha_{; \alpha}\,,
\label{e:sfm2}\eqe
is obtained by combining Eqs.~\eqref{e:sfm}, \eqref{e:bsig} and (\ref{e:amom}.1). Here, $a^\alpha := \dot\bv\cdot\ba^\alpha$, $a_\mrn := \dot\bv\cdot\bn$, $f^\alpha := \bff\cdot\ba^\alpha$ and $p := \bff\cdot\bn$.

\subsection{Weak form for deforming thin shells}

The weak form for Kirchhoff-Love shells is given by \citep{sauer2017a, sauer2017b}
\eqb{l}
G_\mathrm{kin} + G_\mathrm{int} - G_\mathrm{ext} = 0\,, \quad\forall\,\delta\bx\in\sU~,
\label{e:wfu}\eqe
with
\eqb{lll}
G_\mathrm{kin} 
\dis \ds\int_{\sS} \delta\bx\cdot\rho\,\dot\bv\,\dif a\,, \\[4mm]
G_\mathrm{int} 
\dis \ds\int_{\sS} \frac{1}{2}\,\delta a_{\alpha\beta} \, \sig^{\alpha\beta} \, \dif a  + \int_{\sS} \delta b_{\alpha\beta} \, M^{\alpha\beta} \, \dif a\,, \\[4mm]
G_\mathrm{ext} 
\dis \ds\int_{\sS}\delta\bx\cdot\bff\,\dif a + \int_{\partial_t\sS}\delta\bx\cdot\bT\,\dif s + \int_{\partial_m\sS}\delta\bn\cdot\bM\,\dif s\,.
\label{e:Giie}\eqe
Here,
\eqb{l}
\mathcal{U} = \left\lbrace \delta\bx\in \tilde{\mathcal{H}}^2\big(\sS(\bx,t)^3\big)|\,\delta\bx=0\,\text{on}\,\partial_x\sS\, ,\,\delta\bn=0\,\text{on}\,\partial_n\sS\right\rbrace\,,
\eqe
is the space of suitable surface variations, where $\tilde{\mathcal{H}}^2$ is the Sobolev space of Lebesgue square integrable functions and $\partial_x\sS$ and $\partial_n\sS$ are the Dirichlet boundaries for displacements and rotations.
The prescribed edge tractions $\bT = \bsig^\mrT\bnu$ and edge moments $\bM = \bmu^\mrT\bnu$ act on the boundaries $\partial_t\sS$ and $\partial_m\sS$ with the outward normal $\bnu=\nu_\alpha\ba^\alpha$.
We note that the torsional components of the moment $\bM$ are perceived as an effective shear traction in Kirchhoff-Love shells, e.g. see \cite{sauer2017a}.
If desired, $\dif a = J\,\dif A$ and $\rho\,\dif a = \rho_0\,\dif A$ can be used to map integrals to the reference surface $\sS_0$. 
The components $\sig^{\alpha\beta}$ and $M^{\alpha\beta}$ follow from the constitutive laws as outlined in Sec.~\ref{Sec:matmodel}.

\section{Fracture of deforming surfaces} \label{sec:frac}
The formulation for the modeling of brittle fracture is based on Griffith's theory \citep{griffith1921}, in which the energy release rate $E_\mathrm{G}$ of a body, which describes the dissipated energy during crack evolution, is related to the fracture toughness $\sG_\mrc\,[\mathrm{J}\,\mrm^{-1}]$. The latter is also referred to as the critical fracture energy density or critical energy release rate. The corresponding Kuhn-Tucker conditions read
\eqb{l}
E_\mathrm{G}-\sG_\mrc\leq0,\qquad\dot{c}\geq0,\qquad\lr{E_\mathrm{G}-\sG_\mrc}\dot{c}=0\,,
\eqe
with $\dot{c}$ denoting the crack propagation velocity.
Since crack nucleation and branching are not captured by this formulation, Griffith's theory has been reformulated as a global energy minimization problem \citep{francfort1998}. The corresponding energy functional is derived subsequently.

\subsection{Helmholtz free energy} \label{sec:helmholtz}
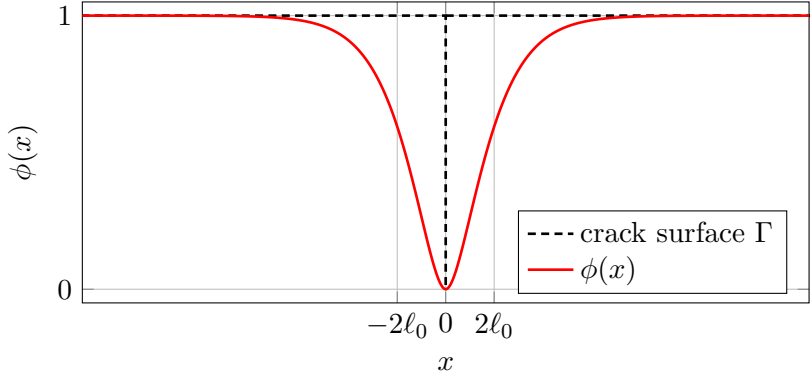
\begin{figure}[!ht]
	\centering
	\begin{tikzpicture}
		\begin{axis}[xlabel={$x$},ylabel={$\phi(x)$},width=0.7\textwidth,height=0.35\textwidth,xmin=-15,xmax=15,ymin=-0.05,ymax=1.05,xtick={-2,0,2},ytick={0,1},xticklabels={$-2\ell_0$,$0$,$2\ell_0$},grid=both,legend cell align={left},legend pos=south east]	
			\addplot[black,densely dashed,line width=1,forget plot] coordinates {(-20,1)  (0,1)  (0,0)};
			\addplot[black,densely dashed,line width=1] coordinates {(0,1) (20,1)};
			\addlegendentry{crack surface $\Gamma$};
			\addplot[red,line width=1,domain=-15:15,samples=501]{1-exp(-abs(x))*(1+abs(x))};
			\addlegendentry{$\phi(x)$};
		\end{axis}
	\end{tikzpicture}
	\caption{Phase field profile for the fourth-order theory of \cite{borden2014}. The crack surface $\Gamma=\{0\}$ is smoothed by the function $\phi(x)$ from Eq.~\eqref{e:phiprofile}.} \label{f:phiprofile}
\end{figure}

The total energy in the system is given by
\eqb{l}
\Pi := \Pi_\mathrm{int}+\Pi_\mathrm{kin}-\Pi_\mathrm{ext}\,,
\label{e:PiTot}
\eqe
where the three contributions denote the Helmholtz free energy $\Pi_\mathrm{int}$, the kinetic energy $\Pi_\mathrm{kin}$ and the external energy $\Pi_\mathrm{ext}$, respectively. Based on the formulation of energy minimization by \cite{francfort1998}, the Helmholtz free energy contains elastic and fracture energy contributions in the form
\eqb{l}
\Pi_\mathrm{int} = \ds\int_{\sS_0}\Psi\,\dif A = \ds\int_{\sS_0}\Bigl[g(\phi)\Psielp+\Psielm + \Psi_\mathrm{frac}\Bigr]\,\dif A\,,
\label{e:PiHF}
\eqe
where $\Psi$ denotes the Helmholtz free energy per reference area. 
Cracks resemble discontinuities in the deformation that are smeared out in the phase field formulation. Therefore, an indicator $\phi\in[0,1]$ is established that distinguishes between fully fractured, $\phi=0$, and undamaged, $\phi=1$, material. This field is referred to as the phase field or fracture field. Since it models the damage region, it is used to define the fracture energy appearing in Eq.~\eqref{e:PiHF}. The higher order phase field model by \citet{borden2014} is adopted here, which, expressed in variables of the present thin shell formulation, reads
\eqb{l}
\Psi_\mathrm{frac} = \ds\frac{\sG_\mrc}{4\ell_0}\Big[(\phi-1)^2+2\ell_0^2\,\nablao\phi\cdot\nablao\phi + \ell_0^4\,(\Delta_\mrS\phi)^2\Big]\,.
\label{e:Wfrac}\eqe
The length scale parameter $\ell_0\,[\mrm]$ controls the support width of the transition zone: $\mathrm{supp}(\phi)\sim\ell_0$. \cite{borden2014} have shown that the one-dimensional phase field approximation of the crack surface $\Gamma=\{0\}$ has the form
\eqb{l}
	\phi(x)=1-\exp\biggl(-\dfrac{|x|}{\ell_0}\biggr)\biggl(1+\dfrac{|x|}{\ell_0}\biggr)\,,
\label{e:phiprofile}\eqe
which is illustrated in Fig.~\ref{f:phiprofile}.

An additive energy split is required in which the elastic energy density is split into a part that contributes to crack evolution (`$+$') and a part that has no effect on crack growth (`$-$'): $\Psiel=\Psielp+\Psielm$. The two contributions are also referred to as the positive and negative part of the elastic energy density. The split is further motivated and derived in Sec. \ref{sec:energysplit}. 
According to Eq.~\eqref{e:PiHF}, the positive part of the elastic energy density $\Psielp$ is degraded through $g(\phi)$ along the damage regions. 
Here, it is assumed to take the form \citep{borden2016}
\eqb{l}
g(\phi) = (3-s)\phi^2 - (2-s)\phi^3\,,
\label{e:deg}
\eqe
where $s>0$ describes the slope of $g(\phi)$ at $\phi=1$. If $s=0$, a surface without initial damage
would fulfill the governing equation for crack evolution in Eq.~\eqref{e:strongphi} for any deformation implying that crack nucleation would not occur. 
Thus, $s$ is set to $10^{-4}$ \citep{borden2016} in all subsequent computations to allow crack nucleation in the absence of initial damage. Degradation functions with $g'(1)=0$ could be used but they require a perturbation in the first Newton-Raphson iteration to allow for crack nucleation in sound materials \citep{kuhn2015}.

\subsection{Hyperelastic material model} \label{Sec:matmodel}
The elastic energy density $\Psi_\mathrm{el}$ is taken as an additive composition of dilatational, deviatoric and bending energy densities in the form
\eqb{l}
\Psi_\mathrm{el} = \underbrace{\Psi_\mathrm{dil}(a_{\alpha\beta}) + \Psi_\mathrm{dev}(a_{\alpha\beta})}_{\Psi_\mathrm{mem}(a_{\alpha\beta})} + \Psi_\mathrm{bend}(b_{\alpha\beta})\,,
\label{e:Wel}\eqe
where the first two terms describe the membrane part of $\Psiel$. A Neo-Hookean surface material model \citep{sauer2017a} with
\eqb{l}\label{e:Wdil}
\Psi_\mathrm{dil} = \ds\frac{K}{4}\big(J^2-1-2\,\ln J\big)\,,
\eqe
and
\eqb{l}\label{e:Wdev}
\Psi_\mathrm{dev} = \ds\frac{G}{2}\big(I_1/J-2\big)\,,
\eqe
is used to model the isotropic in-plane constitutive response. $K$ refers to the 2D bulk modulus and $G$ to the 2D shear modulus. The bending response follows from the Koiter model \citep{ciarlet1993}
\eqb{l}
\Psi_\mathrm{bend} = \ds\frac{c}{2}\Bigl(b_{\alpha\beta}-B_{\alpha\beta}\Bigr)\Bigl(b^{\alpha\beta}_0-B^{\alpha\beta}\Bigr)\,,
\label{e:Psibend}
\eqe
with bending modulus $c$ and $b^{\alpha\beta}_0:=A^{\alpha\gamma}b_{\gamma\delta}A^{\beta\delta}$. Differentiating the Helmholtz free energy with respect to metric and curvature components, yields the stress and moment components
\eqb{lll}
\tau^{\alpha\beta} \is 2\ds\pa{\Psi}{a_{\alpha\beta}}\,, \\[4mm]
M_0^{\alpha\beta} \is \ds\pa{\Psi}{b_{\alpha\beta}}\,.
\label{e:tauM0}
\eqe
Here, these components are given with respect to the reference configuration but they can be mapped to the current configuration by dividing the expressions in Eq.~\eqref{e:tauM0} by the surface stretch $J$. The individual derivatives for the material model in Eqs.~\eqref{e:Wdil}, \eqref{e:Wdev} and \eqref{e:Psibend} read \citep{sauer2017a,zimmermann2019}
\eqb{lll}
\tau^\ab \is \tau^\ab_\mathrm{dil}+\tau^\ab_\mathrm{dev}\,,\\[4mm]
\tau_\mathrm{dil}^\ab \is \dfrac{K}{2}(J^2-1)\,a^\ab\,,\\[4mm]
\tau_\mathrm{dev}^\ab \is \dfrac{G}{2J}(2A^\ab-I_1\,a^\ab)\,,\\[4mm]
M_0^\ab \is c\,(b_0^\ab-B^\ab)\,.
\label{e:tauM0expl}\eqe

\subsubsection{Split of the elastic energy density} \label{sec:energysplit}
\begin{figure}[!ht]
\centering
\begin{tikzpicture}
	\centering
	\tikzmath{
		\wU = 1.5;
		\wL = 0.5;
		\w1 = (\wU+\wL)/2;
		\w2 = (\wU-\wL)/2;
		\h = 1; 
		\shift = 1.5;
		\del = 0.3;
		\arr = 0.5;
		\w = 0.7;
	};
	\coordinate (e1) at (1,0);
	\coordinate (e2) at(0,1);
	
	\coordinate (A) at (0,0);
	\coordinate (B) at ($ (A) + \h*(e2) $);
	\coordinate (C) at ($ (B) + \wU*(e1) $);
	\coordinate (E) at ($ (A) - \h*(e2) - \wL*(e1) $);
	\coordinate (D) at ($ (A) - \h*(e2) $);
	
	\draw (C) -- (E);
	\draw (B) -- (D);
	\draw[->,>=latex,line width = 1] (B) -- (C);
	\draw[->,>=latex,line width = 1] (D) -- (E);
	
	\node at ($ (B) - 1/3*\wU/(\wU+\wL)*2*\h*(e2) + 1/3*\wU*(e1) $) {\color{red}$\boldsymbol{+}$};
	\node at ($ (D) + 1/3*\wL/(\wU+\wL)*2*\h*(e2)-+ 1/3*\wL*(e1) $) {\color{blue}$\boldsymbol{-}$};
	\node[text height=1.5ex,text depth=.25ex,text width=11em,text centered] at ($ (B) + 1/2*\wU*(e1) + \del*(e2) $) {$\tilde{E}_{\alpha\beta}$};

	\node at ($ (A) + \wU*(e1) + 1/2*\shift*(e1) $) {$\boldsymbol{=}$};

	\coordinate (F) at ($ (C) + \shift*(e1) $);
	\coordinate (G) at ($ (F) + \w1*(e1) $);
	\coordinate (H) at ($ (G) - \h*2*(e2) $);
	\coordinate (I) at ($ (H) - \w1*(e1) $);
	
	\draw (F) -- (I);
	\draw (G) -- (H);
	\draw[->,>=latex,line width = 1] (F) -- (G);
	\draw[->,>=latex,line width = 1] (I) -- (H);
	
	\node at ($ (F) -\h*(e2) + 1/2*\w1*(e1) $) {\color{red}$\boldsymbol{+}$};
	\node[text height=1.5ex,text depth=.25ex,text width=11em,text centered] at ($ (F) + 1/2*\w1*(e1) + \del*(e2) $) {$E_{\alpha\beta}$};
	
	\node at ($ (G) - \h*(e2) + 1/2*\shift*(e1) $) {$\boldsymbol{+}$};
	
	\coordinate (L) at ($ (H) + \shift*(e1) $);
	\coordinate (M) at ($ (L) + \w2*(e1) $);
	\coordinate (J) at ($ (M) + 2*\h*(e2) $);
	\coordinate (K) at ($ (J) + \w2*(e1) $);
	
	\draw[black!45!green,->] ($ (J) - \h*(e2) + \del*\h*(e2) $) arc [radius=\del,start angle=90,end angle=50];
	\draw[black!45!green,->] ($ (J) - \h*(e2) - \del*\h*(e2) $) arc [radius=\del,start angle=-90,end angle=-130];
	\draw[color=black!45!green] ($ (J) + \del/2.3*(e1) + \del/2*(e2) $) -- ($ (J) - \h*(e2) + \del*1.5*(e2) + \del/8*(e1) $);
	
	\draw (K) -- (L);
	\draw (J) -- (M);
	\draw[->,>=latex,line width = 1] (J) -- (K);
	\draw[->,>=latex,line width = 1] (M) -- (L);
	
	\node at ($ (J) -1/3*\h*(e2) + 1/3*\w2*(e1) $) {\color{red}$\boldsymbol{+}$};
	\node at ($ (M) +1/3*\h*(e2) - 1/3*\w2*(e1) $) {\color{blue}$\boldsymbol{-}$};
	\node[text height=1.5ex,text depth=.25ex,text width=11em,text centered] at ($ (J) + 1/2*\w2*(e1) + \del*(e2) $) {$K_{\alpha\beta}$};
	
	\coordinate (CoSyUpper) at ($ (K) + \w*(e1) $);
	\coordinate (CoSyLower) at ($ (CoSyUpper) - 2*\h*(e2) $);
	\coordinate (CoSyMid) at ($ .5*(CoSyUpper) + .5*(CoSyLower) $);
	\coordinate (CoSyArrowEnd) at ($ (CoSyMid) + \arr*(CoSyUpper) - \arr*(CoSyMid) $);
	\draw[|-|] (CoSyLower) -- (CoSyMid);
	\draw[-|] (CoSyMid) -- (CoSyUpper);
	\draw[->,>=latex,line width=1] (CoSyMid) -- (CoSyArrowEnd);	
	
	\node at ($ (CoSyMid) + \del*(e1) + \del*(e2) $) {$\xi$};
	\node at ($ (CoSyUpper)+ 1.2*\del*(e1) $) {\small{$\frac{T}{2}$}};
	\node at ($ (CoSyLower)+ 1.2*\del*(e1) $) {\small{-$\frac{T}{2}$}};
	
\end{tikzpicture}
\caption{Strains over the shell's thickness \citep{kiendl2016}. In this example, the membrane part shows purely positive strains, whereas the strains of the bending part are skew-symmetric around the mid-plane. If the negative strains are not supposed to contribute to crack growth, the strain distribution over the thickness has to be taken into consideration. \cite{kiendl2016} introduce a thickness integration and split the total strain with a spectral decomposition. Their work motivates the necessity of our thickness integration (cf. Eq.~\eqref{e:psibendint}).} \label{fig:strains}
\end{figure}
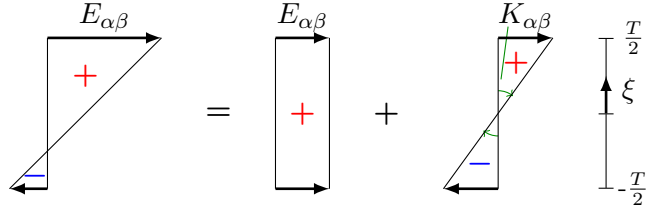

Crack evolution shows anisotropic behavior since cracks will not propagate for every state of stress. To avoid cracking in compression an energy split is required as follows
\eqb{l}
\Psiel=\Psielp+\Psielm\,,
\label{e:psisplit}
\eqe
where $\Psielm$ refers to the part of the elastic energy density that does not contribute to the fracture process. 
\cite{amor2009} make use of a split into deviatoric and dilational parts in which crack evolution is not permitted in volumetric compression but allowed in states of volumetric expansion and shear.
In the work of \cite{miehe2010a}, a spectral decomposition of the strain tensor is introduced in which only positive strains contribute to the fracture process. Likewise, \cite{kiendl2016} establish a spectral decomposition within a small deformation framework in plates and shells. They outline that it is not possible to consider a split into tension and compression as well as a split into membrane and bending contributions at the same time if such a spectral decomposition of the total strain is used.
In our formulation, the elastic energy density is already split into membrane and bending parts according to Eq.~\eqref{e:Wel} such that these terms can be decomposed separately
\eqb{l}
\Psiel^\pm=\Psimem^\pm+\Psibend^\pm\,.
\label{e:splitpsimb}\eqe
In the following, we show an example taken from \cite{kiendl2016} that they use to motivate the need for a thickness integration for the energy split. We use their example to motivate the proposed split of the bending energy density. The strain distribution over the shell's thickness is illustrated in Fig.~\ref{fig:strains}. The total strain $\tilde{\bE}=\bE-\xi\bK$ with components $\tilde{E}_\ab$ and thickness coordinate $\xi\in\left[-\frac{T}{2},\frac{T}{2}\right]$ can have both, positive and negative parts over the thickness $T$. It follows that there is a region of compression, which must not contribute to the fracture process. The membrane strains (due to the surface Green-Lagrange strain tensor $\bE$) are purely positive in this example, whereas the strains associated with the curvature part are asymmetrically distributed around the mid-plane of the shell. Since \cite{kiendl2016} are only interested in the tensile contributions, thickness effects for the elastic energy need to be considered to correctly distinguish between tensile and compressive contributions to the total strain. In contrast to this, the kinematical objects on the mid-plane include enough information for a suitable split of the membrane part. Subsequently, the individual splits of in-plane and out-of-plane parts are derived.

As already mentioned, a spectral decomposition of the strain tensor is not suitable in the present formulation since our elastic energy density is given as a sum of membrane and bending contributions. Instead, we follow the decomposition introduced by \cite{amor2009}, which has also been used by e.g. \cite{ambati2016a} and \cite{borden2016}. Corresponding to whether the surface stretch $J$ is greater than/equal to $1$ or smaller than $1$, the dilatational part will contribute to crack growth or not. The split of the membrane energy density required in Eq.~\eqref{e:splitpsimb} then yields
\eqb{l}
	\Psimem^{+}=\begin{cases}\Psi_\mathrm{dev}+\Psi_\mathrm{dil}\,, &J\geq1 \\ \Psi_\mathrm{dev}\,, & J<1 \end{cases}, \qquad
	\Psimem^{-}=\begin{cases}0\,, & J\geq1 \\ \Psi_\mathrm{dil}\,, & J<1 \end{cases}.
\label{e:psimemsplit}
\eqe
Thus, crack evolution is not permitted in states of volumetric compression ($J<1$) but allowed in states of pure shear ($J=1$) or volumetric expansion ($J>1$). For instance \cite{ambati2015} shown that this split works well for fracture prediction, but we note that a suitable split of the deviatoric energy density might be missing in Eq.~\eqref{e:psimemsplit}.\\
The thickness has to be taken into account in order to obtain a suitable split of the bending energy density in Eq.~\eqref{e:Psibend}. This is obtained from following relation \citep{duong2017}
\eqb{l}
	\Psibend = \ds\int_{-\frac{T}{2}}^{\frac{T}{2}}\tilde{\Psi}_\mathrm{bend}(\xi)\,\dif\xi\,,
\label{e:psibendintegrate}\eqe
where the corresponding three-dimensional constitutive model\footnote{This is a part of the Saint Venant-Kirchhoff model, see \cite{duong2017}.} is given by
\eqb{l}
\tilde{\Psi}_\mathrm{bend}\lr{\bK,\xi,T}=\xi^2\dfrac{12}{T^3}\dfrac{c}{2}\,\mathrm{tr}\!\lr{\bK^2}.
\label{e:psibendtilde}
\eqe
The split of $\Psibend$ is then modeled as
\eqb{l}
\Psibend^\pm=\ds\int_{-\frac{T}{2}}^{\frac{T}{2}}\tilde{\Psi}_\mathrm{bend}^\pm\!\lr{\xi}\dif\xi\,.
\label{e:psibendint}
\eqe
Still, Eq.~\eqref{e:psibendtilde} has to be additively decomposed according to $\tilde{\Psi}_\mathrm{bend}=\tilde{\Psi}_\mathrm{bend}^++\tilde{\Psi}_\mathrm{bend}^-$. Already in Eq.~\eqref{e:psimemsplit} the surface stretch at the mid-plane has been employed as an indicator for a possible contribution to the fracture process.
The surface stretch of other shell layers is obtained in analogy to Eq.~(\ref{e:invs}.2) as 
\eqb{lll}
\tilde{J}=\sqrt{\det[\tilde{A}^{\alpha\beta}]\det[\tilde{a}_{\alpha\beta}]}\,.
\eqe 
The metrics $\tilde{A}^{\alpha\beta}$ and $\tilde{a}_{\alpha\beta}$ follow from the tangent vectors $\tilde{\bA}_{\alpha}$ and $\tilde{\ba}_{\alpha}$ of the shell layer at points $\bx+\xi\,\bn$ and $\bX+\xi\,\bN$, respectively \citep{duong2017}. 
The split of $\tilde{\Psi}_\mathrm{bend}$ then follows as
\eqb{l}
	\tilde{\Psi}_\mathrm{bend}^{+}(\xi)=\begin{cases}\xi^2\dfrac{12}{T^3}\dfrac{c}{2}\,\mathrm{tr}\!\lr{\bK^2}\,, &\tilde{J}(\xi)\geq1 \\ 0\,, & \tilde{J}(\xi)<1 \end{cases}, \quad
	\tilde{\Psi}_\mathrm{bend}^{-}(\xi)=\begin{cases}0\,, & \tilde{J}(\xi)\geq1 \\ \xi^2\dfrac{12}{T^3}\dfrac{c}{2}\,\mathrm{tr}\!\lr{\bK^2}\,, & \tilde{J}(\xi)<1 \end{cases}.
\label{e:psibendsplit}
\eqe
This energy split corresponds to a combination of the split based on the surface stretch \citep{amor2009} and the split based on thickness integration \citep{kiendl2016}. The physical meaning of the split in Eq.~\eqref{e:psibendsplit} is the same as in \cite{kiendl2016}, see Fig.~\ref{fig:strains}. The surface stretch $\tilde{J}$ can be seen as an alternative to the indicator from a spectral decomposition that is able to model large deformations, similar as in \cite{amor2009}. We note that Eq.~\eqref{e:psibendtilde} is a simple bending model. More complicated bending energy models can also be used.\\
Based on Eq.~\eqref{e:psibendint}, the decomposition of the bending energy density follows from thickness integration of Eq.~\eqref{e:psibendsplit}. Thickness integration is performed numerically using Gaussian quadrature. We note that an analytical integration of Eq.~\eqref{e:psibendsplit} over the thickness is in general not possible due to the strong nonlinear dependence of the surface stretch $\tilde{J}$ on $\xi$. But there are two special cases for which Eq.~\eqref{e:psibendint} can be solved analytically, i.e.
\eqb{lllll}
\tilde{J}(\xi)\geq1\,,\:\:\forall\,\xi\in\Bigl[-\dfrac{T}{2},\dfrac{T}{2}\Bigr]: &\Psibend^+= \ds\frac{c}{2}\Bigl(b_{\alpha\beta}-B_{\alpha\beta}\Bigr)\Bigl(b^{\alpha\beta}_0-B^{\alpha\beta}\Bigr)\,,\quad \Psibend^-=0\,,
\label{e:spliteff1}\eqe
and
\eqb{lllll}
\tilde{J}(\xi)<1\,,\:\:\forall\,\xi\in\Bigl[-\dfrac{T}{2},\dfrac{T}{2}\Bigr]: &\Psibend^+=0\,,\quad \Psibend^-= \ds\frac{c}{2}\Bigl(b_{\alpha\beta}-B_{\alpha\beta}\Bigr)\Bigl(b^{\alpha\beta}_0-B^{\alpha\beta}\Bigr)\,.
\label{e:spliteff2}\eqe
These relations can then be used for an efficient FE implementation.\\
For loading-unloading scenarios, the non-physical interpenetration of the fracture surfaces has to be prohibited. The energy split presented above is able to avoid this interpenetration since, in cases of crack closure, the negative part of the membrane energy density in Eq.~(\ref{e:psimemsplit}.2) is non-vanishing. The resulting stresses then counteract the penetration of the crack faces, see \cite{amor2009}. Due to both, this membrane split, and the fact that the phase field is solely defined on the shell's mid-plane, the interpenetration of crack surfaces is avoided.

\subsubsection{Stresses and moments}
Based on the energy split from the previous section, the stress and moment components follow. In the reference configuration, the stress components read
\eqb{l}
\tau^\ab = g(\phi)\,\tau^\ab_++\tau^\ab_-\,,
\label{e:tauabsplt}
\eqe
with the individual contributions
\eqb{l}
	\tau^\ab_+=\begin{cases}\tau_\mathrm{dev}^\ab+\tau_\mathrm{dil}^\ab\,, & J\geq1 \\ \tau_\mathrm{dev}^\ab\,, & J<1 \end{cases}\, ,\qquad
	\tau^\ab_-=\begin{cases}0\,, & J\geq1 \\ \tau_\mathrm{dil}^\ab\,, & J<1 \end{cases}\,.
\label{e:frctaumem}\eqe
The individual contributions in Eq.~\eqref{e:frctaumem} are given in Eq.~(\ref{e:tauM0expl}.2)-(\ref{e:tauM0expl}.3). 
The moment components read
\eqb{l}
M_0^\ab=g(\phi)\,M_{0,+}^\ab+M_{0,-}^\ab\,,
\eqe
where the contributions are computed based on thickness integration via
\eqb{l}
	M_{0,\pm}^\ab=\ds\int_{-\frac{T}{2}}^{\frac{T}{2}}\tilde{M}_{0,\pm}^\ab(\xi)\,\dif\xi\,,
\label{e:M0absplt}
\eqe
with
\eqb{l}
\tilde{M}_{0,+}^\ab(\xi)=\begin{cases}\dfrac{\partial\tilde{\Psi}_\mathrm{bend}(\xi)}{\partial b_\ab}\,, & \tilde{J}(\xi)\geq1 \\ 0\,, & \tilde{J}(\xi)<1\end{cases}\, \qquad 
\tilde{M}_{0,-}^\ab(\xi)=\begin{cases}  0\,, & \tilde{J}(\xi)\geq1 \\ \dfrac{\partial\tilde{\Psi}_\mathrm{bend}(\xi)}{\partial b_\ab}\,, &\tilde{J}(\xi)<1\end{cases}\,.
\label{e:frcM0pm3d}\eqe
The required derivative in Eq.~\eqref{e:frcM0pm3d} is given by
\eqb{l}
\dfrac{\partial\tilde{\Psi}_\mathrm{bend}(\xi)}{\partial b_\ab}=\xi^2\dfrac{12}{T^3}c\,(b_0^\ab-B^\ab)\,,
\eqe
with $b_0^\ab=A^{\alpha\gamma}b_{\gamma\delta}A^{\beta\delta}$. We note that we have assumed that the order of integration $\int_{-T/2}^{T/2}(\cdot)\,\dif\xi$ and differentiation $\partial(\cdot)/\partial b_\ab$ can be exchanged.

\subsection{Irreversible fracture} \label{sec:irreversibility}
Crack evolution is an irreversible process since cracks cannot heal. Thus, the irreversibility condition $\Gamma(t+\Delta t)\supseteq\Gamma(t)\,,\:\forall\Delta t>0$ where $\Gamma$ is the crack surface needs to be enforced algorithmically. As described in \cite{gerasimov2019}, several methods exist to enforce this constraint within a phase field model for fracture. The constraint is rewritten in terms of the phase field as $\phi(\bx,t+\Delta t)\leq\phi(\bx,t)\,,\:\forall\Delta t>0$. In our work we make use of a history field

\eqb{l}
\sH\!\lr{\bx,t} := \ds\max\limits_{\tau\in[0,t]}\Psielp\lr{\bx,\tau} \,,
\label{e:his}
\eqe
which keeps track of the fracture contributing part of the elastic energy density \citep{miehe2010a}. $\Psielp$ in Eq.~\eqref{e:PiHF} is then replaced by the history field $\sH$. Complex initial crack patterns can also be realized by means of the history field \citep{borden2012}. The history field is often viewed as a \textit{driving force} for fracture \citep{miehe2010a}, but this viewpoint is questionable, see \cite{gerasimov2019}. Also, the replacement of $\Psielp$ by $\sH$ violates the variational nature of the formulation \citep{linse2017,gerasimov2019}. The new formulation with $\sH$ is thus, not equivalent to the one with the original energy functional. Despite the approximation of the irreversibility constraint, the new formulation leads to an easy implementation and an easy introduction of initial cracks. Initial cracks can also be inserted as discontinuities in the geometry. But this is more complicated in isogeometric discretizations, than in standard finite element discretizations, especially for complicated initial crack patterns.

\subsection{Euler-Lagrange equation and strong form}
Combining Eqs.~\eqref{e:PiTot}--\eqref{e:PiHF} and \eqref{e:his}, the total energy in the system follows as
\eqb{l}
\Pi := \ds\int_{\sS_0}\Bigl[g(\phi)\sH+\Psielm+\Psi_\mathrm{frac}(\phi)\Bigr]\,\dif A-\Pi_\mathrm{ext}+\Pi_\mathrm{kin}\,.
\label{e:tot}
\eqe
The kinetic energy $\Pi_\mathrm{kin}$ and the potential energy $\Pi_\mathrm{ext}$ do not depend on $\phi$. 
The elastic energy density occurring from volumetric compression $\Psielm$ does not contribute to crack propagation and is thus, not degraded in the domain of fracture. In contrast to this, $\sH$ is degraded by the degradation function $g\!\lr{\phi}$, but is not a function of $\phi$ itself. Only the energy density $\Psi_\mathrm{frac}$ depends on $\phi$, as seen in Eq.~\eqref{e:Wfrac}.
The minimization of the energy functional can be expressed by setting its variation to zero: $\delta\Pi=0$. The latter is solved by making use of the Euler-Lagrange equation, which then leads to the strong form for the phase field's evolution.
Given the Helmholtz free energy per reference area $\Psi=\Psi(\phi,\phi_{,\alpha},\phi_{;\alpha\beta})$, its variation reads
\eqb{l}
	\delta\Psi=\dfrac{\partial\Psi}{\partial\phi}\,\delta\phi + \dfrac{\partial\Psi}{\partial\phi_{,\alpha}}\,\delta(\phi_{,\alpha}) + \dfrac{\partial\Psi}{\partial\phi_{;\alpha\beta}}\,\delta(\phi_{;\alpha\beta})\,.
\label{e:psivariation}\eqe
Integration over the reference surface and applying integration by parts twice, yields
\eqb{l}
\ds\int_{\sS_0}\delta\Psi\dif A=\ds\int_{\sS_0}\left(\dfrac{\partial\Psi}{\partial\phi}-\left(\dfrac{\partial\Psi}{\partial\phi_{,\alpha}}\right)_{\back,\alpha}+\left(\dfrac{\partial\Psi}{\partial\phi_{;\alpha\beta}}\right)_{\back;\alpha\beta}\right)\delta\phi\,\dif A + \mathrm{boundary\:terms}\,.
\label{e:eleder}
\eqe
The boundary terms vanish by choosing appropriate boundary conditions. Boundary conditions for $\phi$ are given in Eq.~\eqref{e:bcphi}.
The energy minimization problem now reads $\delta\Psi=0$. Since Eq.~\eqref{e:eleder} holds true for all $\delta\phi$, the Euler-Lagrange equation follows from applying the fundamental lemma of variational calculus, yielding
\eqb{l}
\ds\pa{\Psi}{\phi}-\bigg(\pa{\Psi}{\phi_{,\alpha}}\bigg)_{\back,\alpha} + \bigg(\pa{\Psi}{\phi_{;\alpha\beta}}\bigg)_{\back;\alpha\beta} = 0\,.
\eqe
Inserting the Helmholtz free energy per reference area described in Sec. \ref{sec:frac} yields the strong form of the phase field fracture equation
\eqb{l}
\ds\frac{2\ell_0}{\sG_\mrc}g'(\phi)\,\sH + \phi-1 - 2\,\ell^2_0\, A^{\alpha\beta}\,\phi_{;\alpha\beta} + \ell_0^4\,A^{\gamma\delta}\bigl(A^{\alpha\beta}\,\phi_{;\alpha\beta}\bigr)_{\!;\gamma\delta} = 0\,,\quad\forall\,\phi\in\sS\,,
\label{e:strongphi}
\eqe
with $g'(\phi)=\partial g(\phi)/\partial\phi$.

\subsection{Weak form for the phase field fracture equation}
Integrating Eq.~\eqref{e:psivariation} over the domain $\sS_0$, the weak form for the phase field fracture equation becomes
\eqb{l}
\ds\int_{\sS_0}\delta\phi\,f(\phi)\,\dif A 
+ \int_{\sS_0}\nablao(\delta\phi)\cdot2\ell_0^2\,\nablao\phi\,\dif A 
+ \int_{\sS_0}\Delta_\mrS(\delta\phi)\,\ell_0^4\,\Delta_\mrS\phi\,\dif A = 0\,,\quad\forall\,\delta\phi\in\sV\,,
\label{e:wfFrac}\eqe
with
\eqb{l}
f(\phi) := \ds\frac{2\ell_0}{\sG_\mrc}g'(\phi)\sH+\phi - 1\,,
\eqe
and the space of suitable test functions  $\mathcal{V} = \left\lbrace \delta\phi\in \mathcal{H}^2\big(\sS(\phi,t)\big)\right\rbrace$. The boundary terms arising during the derivation of Eq.~\eqref{e:wfFrac} vanish due to the choice of the following boundary conditions
\eqb{rll}
	\Delta_\mrS\phi &= 0\,,\\[2mm]
	\nabla_\mrS\bigl(\ell_0^4\,\Delta_\mrS\phi-2\ell_0^2\,\phi\bigr) \cdot\bn&= 0\,,
\label{e:bcphi}\eqe
for all $\phi(\bx,t)$ with $\bx\in\partial\sS$.

\section{Discretization of the coupled problem} \label{sec:discr}
This section presents the monolithic discretization of the coupled system consisting of the thin shell equation, the phase field evolution equation, and their interaction. For the numerical examples presented in Sec.~\ref{sec:num_ex}, the shell surface is discretized by isogeometric finite elements \citep{hughes2005} since the high order operators of the coupled weak form require at least global $C^1$-continuity.
For the spatial discretization, LR NURBS \citep{zimmermann17} are employed to construct locally refined meshes in the domain of fracture.
For the temporal discretization, the generalized-$\alpha$ scheme of \cite{chung93} is used.

\subsection{Adaptive local surface refinement}
\subsubsection{LR NURBS} \label{sec:lr}
The fundamental work of \cite{dokken13} and their introduction of LR B-splines has been extended to LR NURBS by \cite{zimmermann17}. 
A knot vector $\Xi$ of size $n+p+1$ defines $n$ linearly independent basis functions of order $p$.
In the framework of LR NURBS, the global knot vector $\Xi = [\xi_1,...,\xi_{n+p+1}]$ is split into local knot vectors $\Xi_i = [\xi_i,...,\xi_{i+p+1}]$ ($i=1,\dots,n$) to represent local parameter domains. Each of these local knot vectors defines a single basis function.
By construction the basis function has minimal support on the local knot vector. 
Local refinement is performed by mesh line extensions in the parameter space. This includes insertion of new mesh lines, joining or elongation of existing ones or an increase of their multiplicity. The latter results in a decrease of continuity. 
Local refinement is based on knot insertion \citep{dokken13}, which is described for LR NURBS in the work of \cite{zimmermann17}.
LR NURBS inherit several mathematical properties from standard NURBS: The basis forms a partition of unity, it is non-negative and the geometry lies within the convex hull of the control points.

\subsubsection{Criteria for surface refinement} \label{sec:lrsurf}
\begin{figure}[!ht]
	\centering
	\begin{tikzpicture}
	\fill[color=cyan!35] (1,1) rectangle (4,4);
				\draw[line width=0.7] (0,0) -- (6,0);
				\draw[line width=0.7] (0,0.1) -- (6,0.1);
				\draw[line width=0.7] (0,0.2) -- (6,0.2);
				\draw[line width=0.7] (0,1) -- (6,1);
				\draw[line width=0.7] (0,2) -- (6,2);
				\draw[line width=0.7] (0,3) -- (6,3);
				\draw[line width=0.7] (0,4) -- (6,4);
				\draw[line width=0.7] (0,4.8) -- (6,4.8);
				\draw[line width=0.7] (0,4.9) -- (6,4.9);
				\draw[line width=0.7] (0,5) -- (6,5);
				\draw[line width=0.7] (0,0) -- (0,5);
				\draw[line width=0.7] (0.1,0) -- (0.1,5);
				\draw[line width=0.7] (0.2,0) -- (0.2,5);
				\draw[line width=0.7] (1,0) -- (1,5);
				\draw[line width=0.7] (2,0) -- (2,5);
				\draw[line width=0.7] (3,0) -- (3,5);
				\draw[line width=0.7] (4,0) -- (4,5);
				\draw[line width=0.7] (5,0) -- (5,5);
				\draw[line width=0.7] (5.8,0) -- (5.8,5);
				\draw[line width=0.7] (5.9,0) -- (5.9,5);
				\draw[line width=0.7] (6,0) -- (6,5);
				\node[anchor=east] at (0,0.1) {\footnotesize 000};
				\node[anchor=east] at (0,1) {\footnotesize 1};
				\node[anchor=east] at (0,2) {\footnotesize 2};
				\node[anchor=east] at (0,3) {\footnotesize 3};
				\node[anchor=east] at (0,4) {\footnotesize 4};
				\node[anchor=east] at (0,4.9) {\footnotesize 555};
				\node[anchor=north] at (0.1,0) {\footnotesize 000};
				\node[anchor=north] at (1,0) {\footnotesize 1};
				\node[anchor=north] at (2,0) {\footnotesize 2};
				\node[anchor=north] at (3,0) {\footnotesize 3};
				\node[anchor=north] at (4,0) {\footnotesize 4};
				\node[anchor=north] at (5,0) {\footnotesize 5};
				\node[anchor=north] at (5.9,0) {\footnotesize 666};
					\draw[line width=1.5,densely dashed,red] (1.5,1) -- (1.5,4);
					\draw[line width=1.5,densely dashed,red] (2.5,1) -- (2.5,4);
					\draw[line width=1.5,densely dashed,red] (3.5,1) -- (3.5,4);
					\draw[line width=1.5,densely dashed,red] (1,1.5) -- (4,1.5);
					\draw[line width=1.5,densely dashed,red] (1,2.5) -- (4,2.5);
					\draw[line width=1.5,densely dashed,red] (1,3.5) -- (4,3.5);
	\end{tikzpicture}
\caption{Refinement strategy \textit{Structured mesh}: The blue shaded area resembles the support domain of a basis function that is flagged for refinement. The dashed red lines are then inserted into the parameter domain. \citep{johannessen2014}}\label{f:strucmesh}
\end{figure}
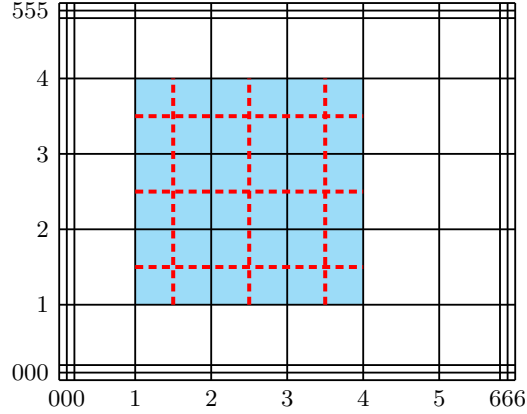

An accurate phase field approximation of the discontinuity across the crack is achieved by using a small length scale parameter $\ell_0$. This requires a highly resolved finite element mesh in the vicinity of the crack. The phase field $\phi$ is used as an indicator for refinement: As soon as a control point's phase field value is smaller or equal to $\phi_\mathrm{bound}$, all elements that lie in the support domain of the corresponding basis functions will be flagged for refinement.
If these elements are not yet refined up to a prescribed refinement depth, mesh line extensions are performed until the desired refinement depth is achieved. The latter can be computed based on the element areas. This refinement strategy is called \textit{Structured mesh} \citep{johannessen2014} and is illustrated in Fig.~\ref{f:strucmesh}. The blue shaded area in the parameter domain resembles the support domain of a basis function that is flagged for refinement. The dashed red lines are then inserted into the parameter domain. This is done recursively for all newly created basis functions up to the prescribed refinement depth. The refinement based on mesh line insertion and modification is described in Sec. \ref{sec:lr}. We have found $\phi_\mathrm{bound}=0.975$ to be a suitable choice for the threshold. We note that in the case of crack nucleation, the last time step needs to be recomputed to ensure crack initiation in a region of highly resolved mesh. But in case of crack propagation, the last time step does not need to be resolved. 
Since the threshold value $\phi_\mathrm{bound}$ is set very close to the undamaged state, where $\phi=1$, the region around the crack tip is always refined up to the highest prescribed refinement depth. 
The physically limited crack tip velocity and the chosen minimum time step size (see. Sec.~\ref{Sec:ATS}) prevents the crack from propagating into regions of coarser elements within one time step.

\subsection{Spatial discretization of primary fields} \label{s:discrfields}
Subsequently, the finite element approximations of the surface deformation and the phase field are described. It follows the work of \cite{sauer2014b}, \cite{sauer2017b}, \cite{duong2017} and \cite{zimmermann2019}. Let $n_e$ denote the number of spline basis functions on parametric element $\Omega_e$. They are numbered with global indices $i_1,\dots,i_{n_e}$. The surface representation follows from this as
\eqb{l}
	\bX^h = \mN\,\mX_e\,,\quad\text{and}\quad\bx^h = \mN\,\mx_e\,,
\label{e:xXh}
\eqe 
for the reference and current surface, respectively. The corresponding shape function array reads
\eqb{l}
\mN := [N_{i_1}\bone,\,N_{i_2}\bone,\,...,\,N_{i_{n_e}}\bone]\,.
\eqe
Here, the element-level vectors are denoted $\mX_e$ and $\mx_e$ and $\bone$ refers to the $(3\times3)$ identity matrix.
Likewise, the phase field is approximated via
\eqb{l}
	\phi^h = \bar\mN\,\bphi_e\,,
\label{e:phih}
\eqe
with element-level nodal values $\bphi_e$ and shape function array
\eqb{l}
	\bar\mN := [N_{i_1},\,N_{i_2},\,...,\,N_{i_{n_e}}]\,.
\eqe
The local vectors contain the nodal values with indices $i_1,\dots,i_{n_e}$. These can be extracted from the global ones  $\mX$, $\mx$ and $\bphi$ which contain all nodal values. In analogy to Eqs.~\eqref{e:xXh} and \eqref{e:phih}, the corresponding variations read
\eqb{l}
	\delta\bX^h = \mN\,\delta\mX_e\,,\quad\text{and}\quad\delta\bx^h = \mN\,\delta\mx_e\,,
\label{e:deltaxXh}
\eqe
and
\eqb{l}
\delta\phi^h = \bar\mN\,\delta\bphi_e\,.
\eqe 
Based on Eq.~\eqref{e:xXh}, the discretized tangent vectors follow as
\eqb{lll}
\bA_\alpha^h \is \mN_{\!,\alpha}\,\mX_e\,,\quad\text{and}\quad \ba_\alpha^h = \mN_{\!,\alpha}\,\mx_e\,,
\label{e:bah}\eqe
with $\mN_{\!,\alpha}:=\partial\mN/\partial\xi^\alpha$. 
From this, the discretized normals $\bn^h$ and $\bN^h$ follow according to Eq.~\eqref{e:bn}.\footnote{To avoid confusion, we write discrete arrays, such as the shape function array $\mN$, in roman font, whereas continuous tensors, such as the normal vector $\bN$, are written in italic font.}
The metric and curvature tensor components in the reference configuration are then given by
\eqb{lll}
A_{\alpha\beta}^h \is \mX_e^\mrT\,\mN^\mrT_{\!,\alpha}\,\mN_{\!,\beta}\,\mX_e\,, \quad\text{and}\quad
B_{\alpha\beta}^h = \bN^h\cdot\mN_{\!,\alpha\beta}\,\mX_e\,,
\eqe 
and similarly for the current surface
\eqb{lll}
a_{\alpha\beta}^h \is \mx_e^\mrT\,\mN^\mrT_{\!,\alpha}\,\mN_{\!,\beta}\,\mx_e\,,\quad\text{and}\quad b_{\alpha\beta}^h = \bn^h\cdot\mN_{\!,\alpha\beta}\,\mx_e\,.
\eqe 
From this, the contra-variant metrics $[A^{\alpha\beta}_h]=[A_{\alpha\beta}^h]^{-1}$ and $[a^{\alpha\beta}_h]=[a_{\alpha\beta}^h]^{-1}$ follow. In analogy, the discretized variations of the surface metric and curvature are given by
\eqb{lll}
\delta a_{\alpha\beta}^h \is \delta\mx_e^\mrT\big(\mN^\mrT_{\!,\alpha}\,\mN_{\!,\beta}+\mN^\mrT_{\!,\beta}\,\mN_{\!,\alpha}\big)\,\mx_e\,, \quad\mathrm{and}\quad
\delta b_{\alpha\beta}^h =\delta\mx_e^\mrT\,\mN^\mrT_{\!;\alpha\beta}\,\bn^h\,,
\eqe 
with
\eqb{l}
\mN_{\!;\alpha\beta} := \mN_{\!,\alpha\beta} - \Gamma^\gamma_{\alpha\beta}\,\mN_{\!,\gamma}\,,
\eqe
and discretized Christoffel symbols (cf. Sec.~\ref{s:surf_descr})
\eqb{l}
\Gamma^\gamma_{\alpha\beta} = \mx_e^\mrT\,\mN^\mrT_{,\alpha\beta}\,a^{\gamma\delta}_h\,\mN_{,\delta}\,\mx_e\,.
\eqe

Using Eqs.~\eqref{e:diff1},~\eqref{e:diff2} and~\eqref{e:phih}, the derivatives of the phase field follow as
\eqb{lll}
\phi^h_{;\alpha} \is \bar\mN_{\!,\alpha}\,\bphi_e\,, \\[1mm]
\nablao\phi^h \is \bA^\alpha_h\,\bar\mN_{\!,\alpha}\,\bphi_e\,, \\[1mm]
\nablao\delta\phi^h \is \bA^\alpha_h\,\bar\mN_{\!,\alpha}\,\delta\bphi_e\,, \\[1mm]
\Delta_\mrS\phi^h \is \Delta_\mrS\bar\mN\,\bphi_e\,, \\[1mm]
\Delta_\mrS\delta\phi^h \is \Delta_\mrS\bar\mN\,\delta\bphi_e\,,
\label{e:approx_diff}\eqe
with $\bA_h^\alpha=A^\ab_h\bA^h_\beta$ and $\bar\mN_{\!,\alpha}:=\partial\bar\mN/\partial\xi^\alpha$
and
\eqb{lll}
\Delta_\mrS\bar\mN := A^{\alpha\beta}_h\,\hat{\bar\mN}_{\!;\alpha\beta}\,,
\eqe
where
\eqb{lll}
\hat{\bar\mN}_{\!;\alpha\beta} = \bar\mN_{\!,\alpha\beta} - \hat\Gamma^\gamma_{\alpha\beta}\,\bar\mN_{\!,\gamma}\,.
\label{e:bN;ab}\eqe
Note that here, the discretized Christoffel symbols need to be taken from the reference surface (cf. Sec.~\ref{s:surf_descr}), i.e.
\eqb{l}
\hat\Gamma^\gamma_{\alpha\beta} = \mX_e^\mrT\,\mN^\mrT_{,\alpha\beta}\,A^{\gamma\delta}_h\,\mN_{,\delta}\,\mX_e\,.
\eqe

\subsection{Spatial discretization of the mechanical weak form}\label{sec:spaceDisc_mech}
Inserting the above approximations into Eq.~\eqref{e:wfu} yields the discretized mechanical weak form
\eqb{l}
\delta\mx^\mrT\,\big[\mf_\mathrm{kin} + \mf_\mathrm{int} - \mf_\mathrm{ext}\big] = 0~, \quad \forall~\delta\mx\in\sU^h\,,
\label{e:fx}
\eqe
with global force vectors $\mf_\mathrm{kin}$, $\mf_\mathrm{int}$ and $\mf_\mathrm{ext}$. These are assembled from their respective elemental contributions
\eqb{lll}
\mf^e_\mathrm{kin} \dis \mm_e\,\ddot\mx_e\,,\quad\mm_e := \ds\int_{\Omega^e}\rho\,\mN^\mrT\mN\,\dif a\,,  \\[4mm]
\mf^e_\mathrm{int} \dis \ds\int_{\Omega^e}\left(g(\phi^h)\,\sig^{\alpha\beta}_++\sig^{\alpha\beta}_-\right)\mN_{\!,\alpha}^\mrT\,\ba^h_\beta\,\dif a
+ \int_{\Omega^e}\left(g(\phi^h)\,M^{\alpha\beta}_++M^{\alpha\beta}_-\right)\mN^\mrT_{\!;\alpha\beta}\,\bn^h\,\dif a\,, \\[4mm] 
\mf^e_\mathrm{ext} \dis \ds\int_{\Omega^e}\mN^\mrT\,p(\phi)\,\bn^h\,\dif a + \ds\int_{\Omega^e}\mN^\mrT\,f^\alpha\,\ba^h_\alpha\,\dif a\,.
\label{e:finext}\eqe
The terms $\sigma^{\alpha\beta}_\pm$ and $M^{\alpha\beta}_\pm$ are given by the energy split outlined in Sec. \ref{sec:energysplit}.
In $\mf^e_\mathrm{ext}$ we have taken the boundary loads $\bT$ and $\bM$ acting on $\partial\sS$ as zero. 
The extension to boundary loads can be found in \citet{duong2017}. Apart from the dependence on $\mx_e$, the force $\mf^e_\mathrm{int}$ depends on $\bphi_e$ through the degradation of $\sigma^{\alpha\beta}_+$ and $M^{\alpha\beta}_+$ by $g(\phi^h)$.\\
From a physical point of view, the load-bearing capability vanishes in fully damaged regions where $\phi=0$. Thus, no pressure can act on the corresponding regions. We account for this by scaling the pressure linearly based on the phase field, i,.e.
\eqb{l}
	p(\phi) = \phi\,\bar{p}\,,
\label{e:pres}\eqe
with $\bar{p}$ denoting the pressure imposed on undamaged elements. Huge deformations and distorted elements at regions of full damage are prevented by means of the pressure function in Eq.~\eqref{e:pres}.
Putting everything together, the resulting equation system for the free nodes\footnote{The free nodes refer to the degrees of freedom, which are not given by boundary conditions.} reads
\eqb{l}
\mf(\mx,\bphi) = \mM\,\ddot\mx + \mf_\mathrm{int}(\mx,\bphi) - \mf_\mathrm{ext}(\mx,\bphi) = \mathbf{0}\,.
\label{e:ODEx}
\eqe
The global mass matrix $\mM$ is assembled from the elemental contributions $\mm_e$.

\subsection{Spatial discretization of the phase field}
Inserting the approximations from Sec.~\ref{s:discrfields} into the discretized weak form of Eq.~\eqref{e:wfFrac} yields
\eqb{l}
\delta\bphi^\mrT\,\big[\bar\mf_\mathrm{kin} + \bar\mf_\mathrm{int} - \bar\mf_\mathrm{ext}\big] = 0\,, \quad\forall\,\delta\bphi\in\sV^h\,,
\eqe 
where the global vectors $\bar\mf_\mathrm{kin}$, $\bar\mf_\mathrm{int}$ and $\bar\mf_\mathrm{ext}$ follow from the assembly of their corresponding elemental contributions
\eqb{lll}
\bar\mf^e_\mathrm{kin} \dis \mathbf{0}\,,  \\[2mm]
\bar\mf^e_\mathrm{int} \dis \bar\mk_0^e\,\bphi_e + \bar\mf^e_\mathrm{el}-\bar\mf^e_0\,,\quad
\bar\mk_0^e := \ds\int_{\Omega_0^e}\bigg[\bar\mN^\mrT\bar\mN + \bar\mN_{,\alpha}^\mrT\,2\ell_0^2\,A^{\alpha\beta}\,\bar\mN_{,\beta}+\Delta_\mrS\bar\mN^\mrT\,\ell_0^4\,\Delta_\mrS\bar\mN\bigg]\,\dif A\,, \\[4mm]
\bar\mf^e_\mathrm{el} \dis \ds\int_{\Omega_0^e}\bar\mN^\mrT\ds\frac{2\ell_0}{\sG_\mrc}g'(\phi)\sH\,\dif A\,, \\[4mm]
\bar\mf^e_0 \dis \ds\int_{\Omega_0^e}\bar\mN^\mrT\,\dif A\,, \\[4mm]
\bar\mf^e_\mathrm{ext} \dis \mathbf{0}\,.
\eqe
Apart from the dependence on $\bphi_e$, these expressions depend on $\mx_e$ through $\sH$. 
The resulting equations at the free nodes simplify to
\eqb{l}
\bar\mf(\mx,\bphi) =  \bar\mf_\mathrm{int}(\mx,\bphi)= \mathbf{0}\,.
\label{e:ODEphi}
\eqe

\subsection{Temporal discretization}\label{sec:timeDisc}
\subsubsection{Generalized-\texorpdfstring{$\alpha$}{alpha} method}
The fully implicit generalized-$\alpha$ method of \cite{chung93} is used as a monolithic time integration scheme. Given the quantities $(\mx_{n},\dot\mx_{n},\ddot\mx_{n},\bphi_{n})$ at time $t_n$, the new values $(\mx_{n+1},\dot\mx_{n+1},\ddot\mx_{n+1},\bphi_{n+1})$ at time $t_{n+1}$ need to be found. Additionally, equilibrium has to be fulfilled at intermediate states $(\mx_{n+\alpha_\mrf}, \dot\mx_{n+\alpha_\mrf}, \ddot\mx_{n+\alpha_\mrm}, \bphi_{n+1})$, i.e.
\eqb{lll}
\label{e:gen_a1}
\begin{bmatrix}
\mf\left(\mx_{n+\alpha_\mrf},\ddot{\mx}_{n+\alpha_\mrm},\bphi_{n+1} \right) \\[2mm]
\bar\mf\left(\mx_{n+\alpha_\mrf},\bphi_{n+1} \right) 
\end{bmatrix}
\is \mathbf{0}\,.
\eqe
The complete scheme has been described in the work of \cite{zimmermann2019}. Since there are no temporal derivatives of the phase field in our framework, the corresponding equations simplify as outlined in Appendix~\ref{sec:gena}. As shown for instance in \cite{heister2015} and \cite{gerasimov2016}, a monolithic coupling of the shell and phase field evolution equation leads to a non-convex optimization problem. While in their work, a stabilization scheme or a convexification of the energy functional is employed, we do not encounter any numerical instabilites in our implicit time integration scheme. This is a result of the spatial and temporal adaptivity approach. The first ensures a highly refined mesh around the crack tip, see Sec.~\ref{sec:lrsurf}, while the adaptive time stepping scheme (presented subsequently) provides sufficiently small time steps in case of crack propagation, see the numerical examples in Sec.~\ref{sec:num_ex}. The combination always ensured good convergence behavior, similar to the model presented by \cite{borden2012}.

\subsubsection{Adaptive time-stepping}\label{Sec:ATS}
The time step size should be chosen sufficiently small so that the crack does not propagate across too many elements in one time step. In contrast to this, large time steps can be used in cases of no crack propagation. This motivates the adaptive adjustment of the time step size. Since the phase field is not time-dependent, we cannot apply the adaptive time stepping scheme from \cite{zimmermann2019}. We therefore follow the subsequent approach:
The need for smaller or the possibility of larger time steps can be indicated by the required number of Newton-Raphson iterations $n_\mathrm{NR}$ during the last iteration, as for instance done by \cite{schlueter2014}. We adjust the new time step size at time step $n+1$ as
\eqb{l}
\Delta t_{n+1}=\begin{cases}1.5\,\Delta t_n\,, & n_\mathrm{NR}<4\\1.1\,\Delta t_n\,, & n_\mathrm{NR}=4\\0.5\,\Delta t_n\,, & n_\mathrm{NR}>4\\0.2\,\Delta t_n\,, & \mathrm{local~spatial~refinement}\end{cases}\,.
\label{e:adptvt}\eqe
The coefficients in Eq.~\eqref{e:adptvt} have been chosen based on the numerical examples presented in Sec.~\ref{sec:num_ex}. Note that the time step size is also reduced after each spatial refinement step to ensure good convergence behavior.
If not specified otherwise, a maximum time step size $\Delta t_\mathrm{max}=0.1\,T_0$ and the initial time step size $\Delta t_0=1.5\cdot10^{-5}\,T_0$ are used for the numerical results\footnote{$T_0$ refers to a reference time used to obtain a dimensionless formulation, see Sec.~\ref{s:numdim}}. In the following numerical examples, we have observed that time step sizes smaller than $10^{-8}-10^{-7}$ lead to ill-conditioned stiffness matrices. The maximum time step size has been mainly determined based on numerical investigations and set in a way, such that the cracks do not propagate over too many elements within one time step. We note that the latter can also be determined based on the stress wave propagation speeds or the natural frequencies of the system \citep{borden2012}, also see Sec.~\ref{s:numex_dynbrnch}.

\subsection{Stabilization of jump conditions}
\begin{figure}[!ht]
	\centering
	\begin{tikzpicture}
		\begin{axis}[xlabel={$\tilde{J}(\xi)$},ylabel={$\hat\rchi\bigl(\tilde{J}(\xi)\bigr)$},width=0.7\textwidth,height=0.4\textwidth,xmin=0,xmax=2,ymin=-0.05,ymax=1.05,xtick={0,1,2},ytick={0,0.5,1},grid=both,legend cell align={left},legend pos=south east]	
			\addplot[olive,line width=1,domain=0:2,samples=501]{1/(1+exp(-25*(x-1)))};
			\addlegendentry{$p_\rchi=25$};
			\addplot[red,line width=1,domain=0:2,samples=501]{1/(1+exp(-100*(x-1)))};
			\addlegendentry{$p_\rchi=100$};
			\addplot[blue,line width=1,domain=0:2,samples=501]{1/(1+exp(-250*(x-1)))};
			\addlegendentry{$p_\rchi=250$};
			\addplot[black,densely dashed,line width=1] coordinates {(0,0)  (1,0)  (1,1)  (2,1)};
			\addlegendentry{$p_\rchi\rightarrow\infty$};
		\end{axis}
	\end{tikzpicture}
	\caption{Smoothed jump function (cf. Eq.~\eqref{e:smoothchi}) used to stabilize the Newton-Raphson solution scheme.} \label{f:smoothedheavyside}
\end{figure}
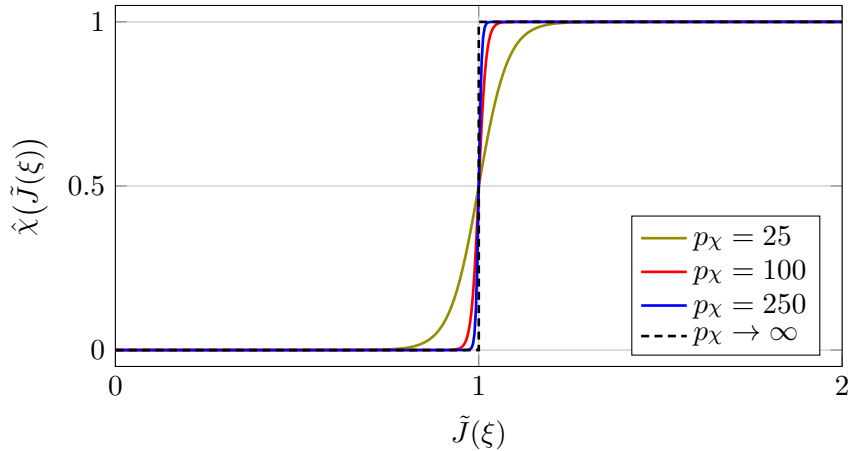

In Eqs.~\eqref{e:psibendsplit} and \eqref{e:M0absplt} and in the corresponding linearizations (cf. Appendix~\ref{s:lin}), integrals of the form
\eqb{l}
	\ds\int_{-\frac{T}{2}}^{\frac{T}{2}}\xi^2\,\rchi\bigl(\tilde{J}(\xi)\bigr)\,\dif\xi\,,\quad\mathrm{with}\quad\rchi\bigl(\tilde{J}(\xi)\big)=\begin{cases}1,& \tilde{J}(\xi)\geq1\\0,&\tilde{J}(\xi)<1\end{cases}\,,
\label{e:stabint}\eqe
have to be computed. In the numerical examples presented in Sec.~\ref{sec:num_ex}, we have observed that the jump function $\rchi\bigl(\tilde{J}(\xi)\bigr)$ leads to convergence problems in which the Newton-Raphson iteration may alternate between different states. This occurs when the surface stretch $\tilde{J}(\xi)$ has values close to one so that $\rchi\bigl(\tilde{J}(\xi)\bigr)$ may change its value after a Newton-Raphson update. We have tested two strategies to avoid these convergence problems: At first, an active set strategy can be employed. During a Newton-Raphson iteration the expressions in Eq.~\eqref{e:stabint} are kept constant and the coupled system is solved for these values. Afterwards, the expressions are recomputed and another Newton-Raphson iteration is performed. This active set iteration is performed until either there is no change in the active set (the integral expressions), a maximum number of active set iterations is reached or the solution alternates again between different states. Since this strategy introduces another iteration it can increase the computational effort significantly. We thus propose another approach in which we smooth the discontinuity in $\rchi\bigl(\tilde{J}(\xi)\bigr)$ by
\eqb{l}
	\hat{\rchi}\bigl(\tilde{J}(\xi)\bigr) := \dfrac{1}{1+e^{-p_\rchi\bigl(\tilde{J}(\xi)-1\bigr)}}\,.
\label{e:smoothchi} \eqe
This regularization is illustrated in Fig.~\ref{f:smoothedheavyside} for different values of the regularization parameter $p_\rchi\in(0,\infty)$.
The black dashed line shows the discontinuous function. As the parameter $p_\rchi$ increases, the smoothed function $\hat\rchi\bigl(\tilde{J}(\xi)\bigr)$ approximates the discontinuous function $\rchi\bigl(\tilde{J}(\xi)\bigr)$ more precisely. By means of this smoothed function, the Newton-Raphson iteration does not alternate between different states and, in contrast to the active set strategy depicted above, no additional iteration is necessary. We note that an increase in the regularization parameter $p_\rchi$ leads to a decrease in the average time steps computed by the adaptive time-stepping scheme in Sec.~\ref{Sec:ATS}.

\subsection{Dimensionless form} \label{s:numdim}
The preceding formulation is normalized by introduction of the reference length $L_0$, surface density $\rho_0$\footnote{Note that $\rho_0$ is the surface density and has units $[\mathrm{kg}/\mathrm{m}^2]$.} and time $T_0$. The corresponding dimensionless quantities are
\eqb{l}
	\bx^\star=\dfrac{\bx}{L_0}\,,\quad \rho^\star=\dfrac{\rho}{\rho_0}\,,\quad t^\star=\dfrac{t}{T_0}\,.
\eqe
The normalization quantities for the in-plane material parameters $K$ and $G$, the bending modulus $c$ and the critical energy density $\sG_\mrc$ then follow as
\eqb{lllll}
K^\star=\dfrac{K}{E_0}\,,\quad G^\star=\dfrac{G}{E_0}\,,\quad c^\star=\dfrac{c}{E_0\,L_0}\,,\quad \sG_\mrc^\star=\dfrac{\sG_\mrc}{E_0\,L_0}\,,
\eqe
where $E_0:=\rho_0\,L_0^2\,T_0^{-2}$ has units $[\mathrm{N}/\mathrm{m}]$. The surface stress $\sigma^\ab$, the surface moment $M^\ab$, the surface tension $\gamma$, the elastic energy density $\Psi$ and potential $\Pi$ are then given by
\eqb{llllll}
\sigma^\ab_\star=\dfrac{\sigma^\ab}{E_0}\,,\quad M^\ab_\star=\dfrac{M^\ab}{E_0\,L_0}\,,\quad \gamma^\star=\dfrac{\gamma}{E_0}\,,\quad \Psi^\star=\dfrac{\Psi}{E_0}\,,\quad \Pi^\star=\dfrac{\Pi}{E_0\,L_0^2}\,.
\eqe
The temporal and spatial derivatives are \citep{zimmermann2019}
\eqb{l}
	\dfrac{\partial\dots}{\partial t^\star}=T_0\,\dfrac{\dots}{\partial t}\,,\quad \nabla_\mrS^\star=L_0\,\nabla_\mrS\,,\quad\Delta_\mrS^\star=L_0^2\,\Delta_\mrS\,.
\eqe
In the following, the superscript $\star$ will be omitted for notational simplicity.

\section{Numerical examples}
\label{sec:num_ex}
This section shows several numerical examples of the proposed phase field formulation of brittle shells. The material parameters of the elastic energy density (cf.  Sec. \ref{Sec:matmodel}) are given via
\eqb{lll}
  K=\dfrac{E\,\nu}{\lr{1+\nu}\lr{1-2\nu}}\,,\qquad G=\dfrac{E}{2\lr{1+\nu}}\,,\qquad c=0.1\,E_0\,L_0\,,
\eqe
with stiffness $E$ and Poisson's ratio $\nu$. For all subsequently presented results, bi-quadratic LR NURBS are used and numerical integration on the bi-unit parent element is performed using Gaussian quadrature with $3\times3$ quadrature points. Numerical thickness integration is performed using four Gaussian quadrature points. For the visualization, the surface tension
\eqb{l}
	\gamma=\dfrac{1}{2}N_\alpha^\alpha\,,
\label{e:gam}\eqe
is plotted, where $N_\alpha^\alpha$ are the mixed components from the stress occurring in the equation of motion \eqref{e:sfm}. All crack patterns are illustrated as follows: Red color resembles the fractured state ($\phi=0$) and blue color indicates undamaged material ($\phi=1$). In between these states, a transition based on the colors yellow-green-cyan is used.

\textbf{Remark:} The examples in this section exhibit stress waves. The present formulation does not consider any damping such that stress waves do not dissipate but continue to propagate and reflect. An artificial damping, e.g. based on energy absorbing boundary elements, could be employed. Alternatively, physical viscosity can be introduced in the system, similar as is done by \cite{zimmermann2019}. The challenge for the latter is to correctly split the viscous terms in analogy to the elastic split outlined in Sec.~\ref{sec:energysplit}. Especially, the propagation of stress waves over elements of different size needs to be investigated further. Stress waves can be emitted from the crack, where the mesh is finest.
As they cross mesh interfaces (where elements of different sizes meet), it can happen that very fine waves are not represented on the coarse mesh. 
It can be expected that for high loading intensities, these mesh interfaces thus lead to unintentional and unphysical reflections of stress waves that may affect the fracture pattern. 
For a physically correct assessment realistic damping formulations are needed. 
The development of such formulations along with the investigation of stress waves is subject of future work.

\subsection{2D shear test}
\begin{figure}[!ht]
	\centering
	\includegraphics[scale=1]{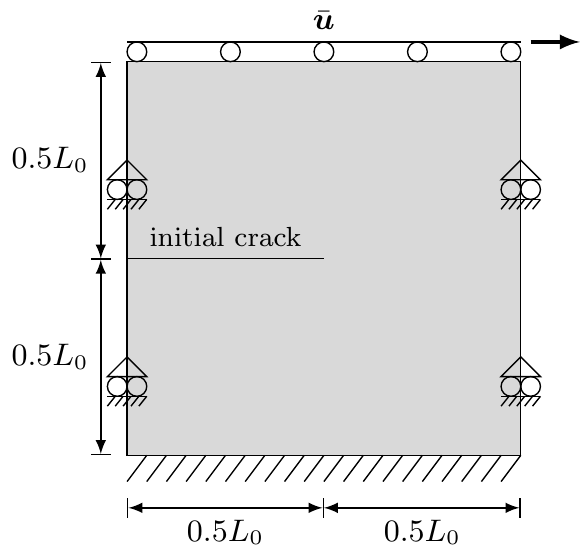}
	\caption{2D shear test: Specimen geometry, boundary and loading conditions.}
	\label{fig:sheargeom}
\end{figure}

\begin{figure}[!ht]
\captionsetup[subfigure]{labelformat=empty}
\centering
	\subfloat[$\bar{u}=0L_0$]{\includegraphics[trim= 770 390 680 320,, clip,height=0.3\textwidth]{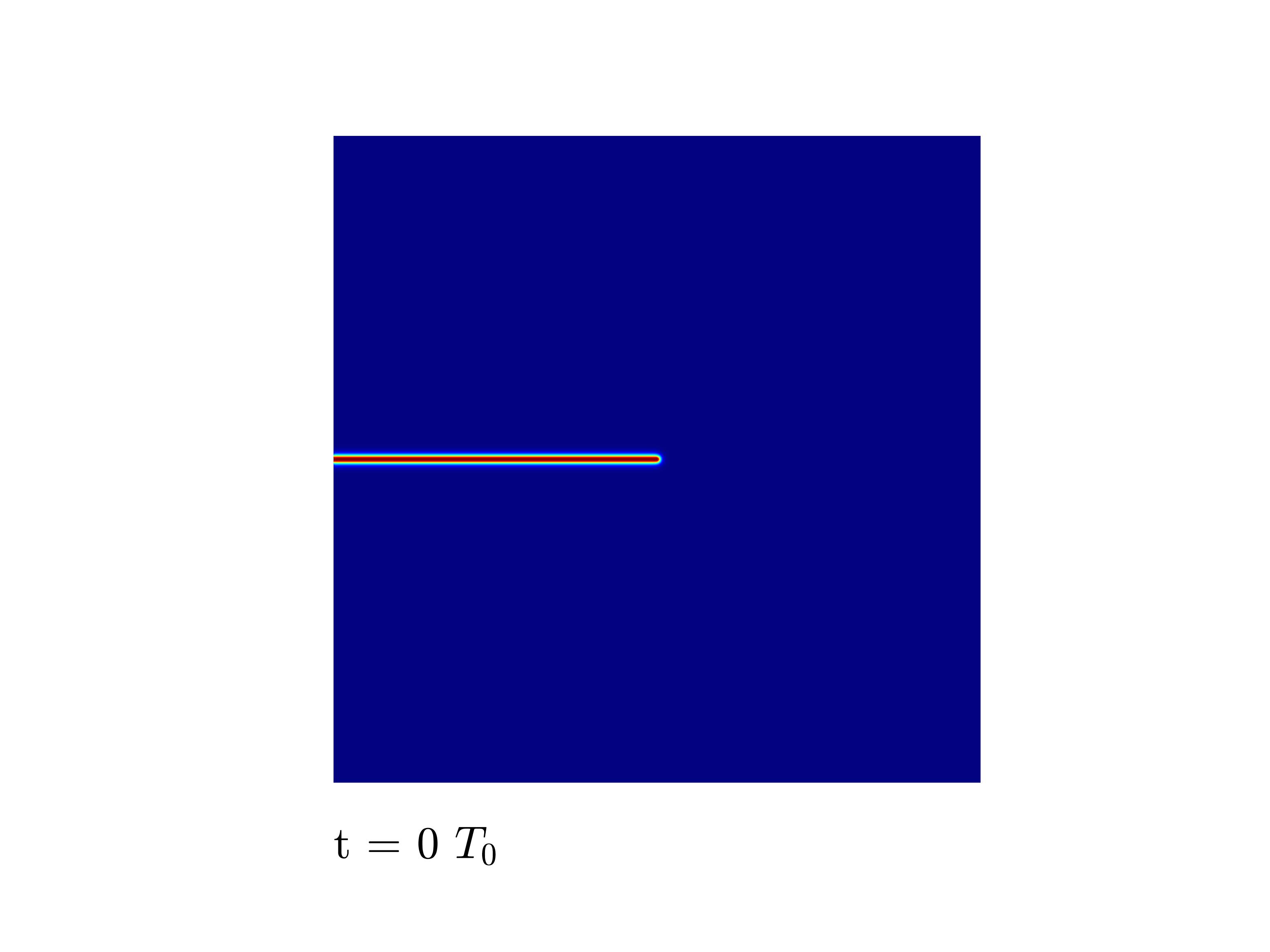}}
	\quad
	\subfloat[$\bar{u}=0.0094L_0$]{\includegraphics[trim= 770 390 680 320,, clip,height=0.3\textwidth]{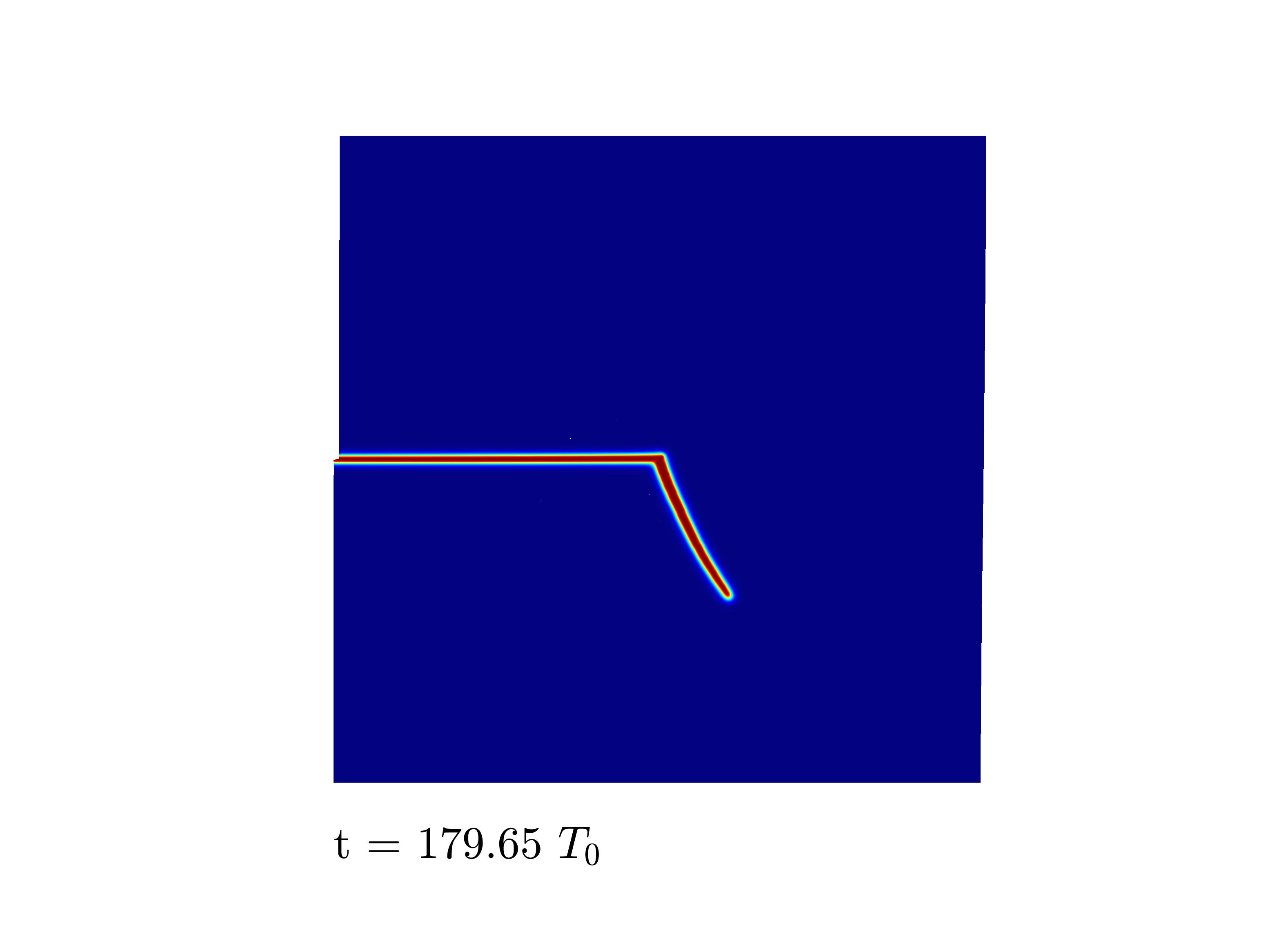}}
	\quad
	\subfloat[$\bar{u}=0.0128L_0$]{\includegraphics[trim= 770 390 680 320,, clip,height=0.3\textwidth]{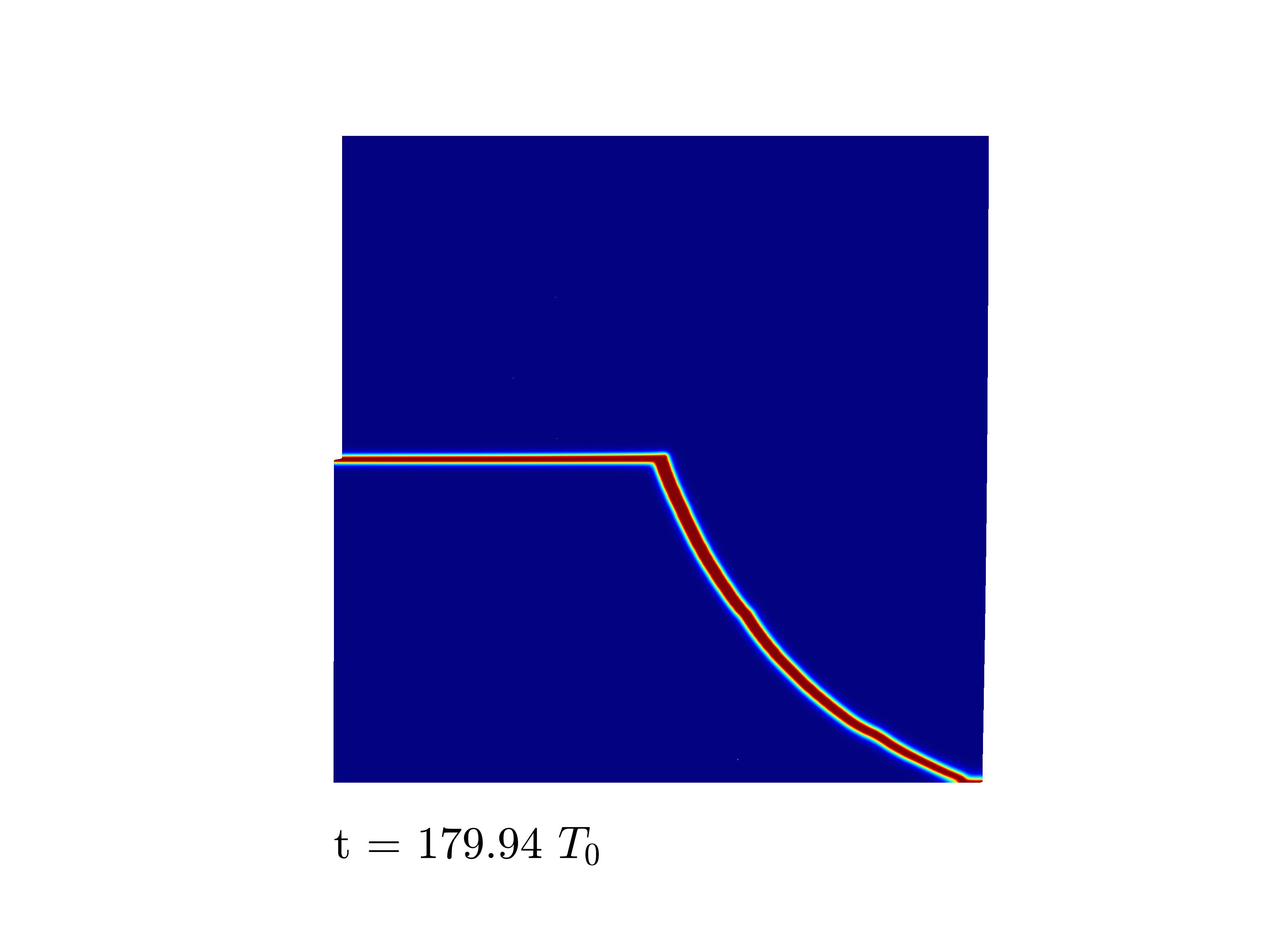}}
\caption{2D shear test: Crack propagation at various time steps. The energy split for the membrane part of the elastic energy density leads to the qualitatively correct crack path.} \label{f:shearprogr}
\end{figure}

\begin{figure}[!ht]
\captionsetup[subfigure]{labelformat=empty}
\centering
	\subfloat[$\bar{u}=0L_0$]{\includegraphics[trim= 780 400 700 320, clip,height=0.3\textwidth]{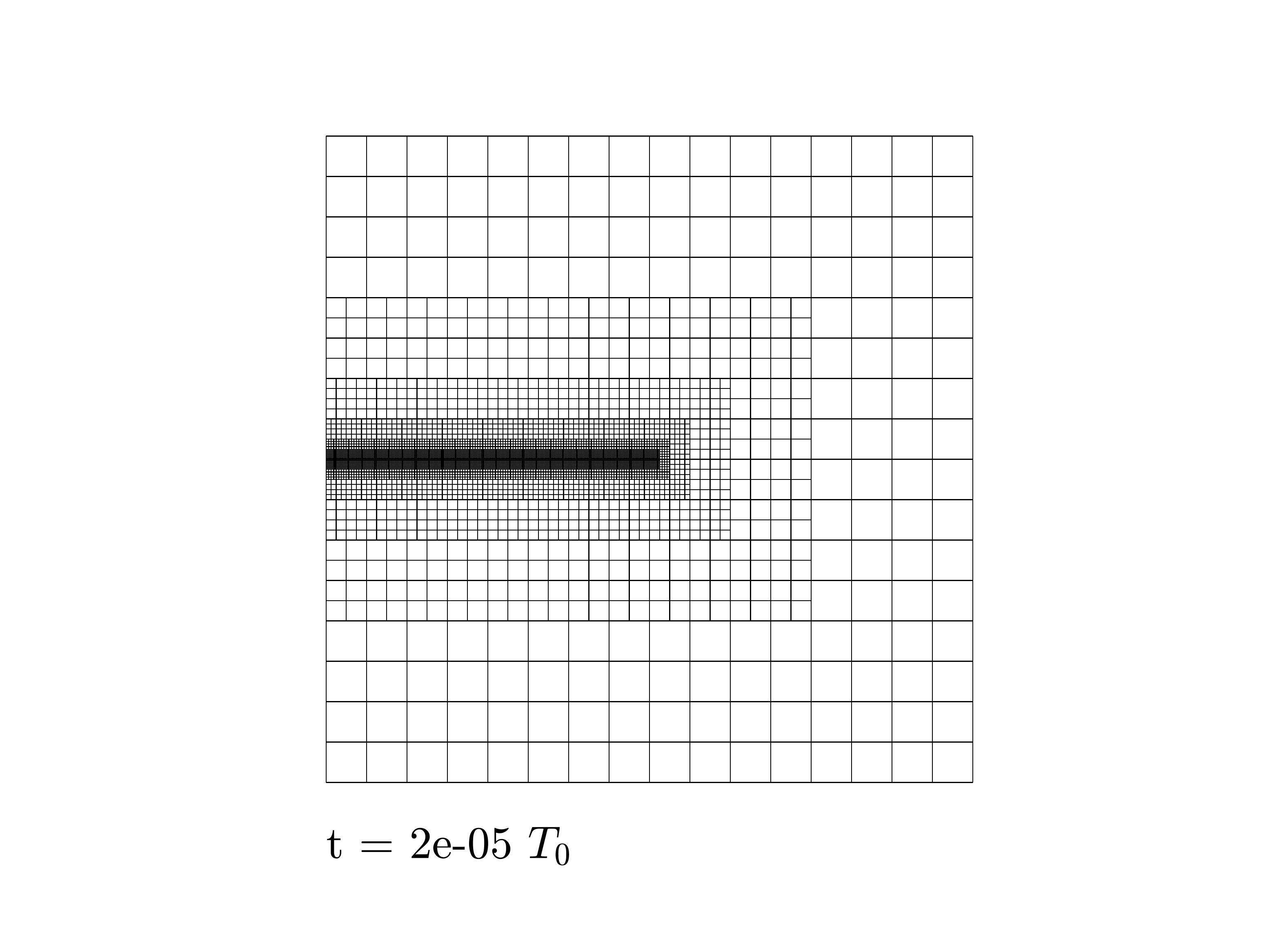}}
	\quad
	\subfloat[$\bar{u}=0.0094L_0$]{\includegraphics[trim= 780 400 700 320, clip,height=0.3\textwidth]{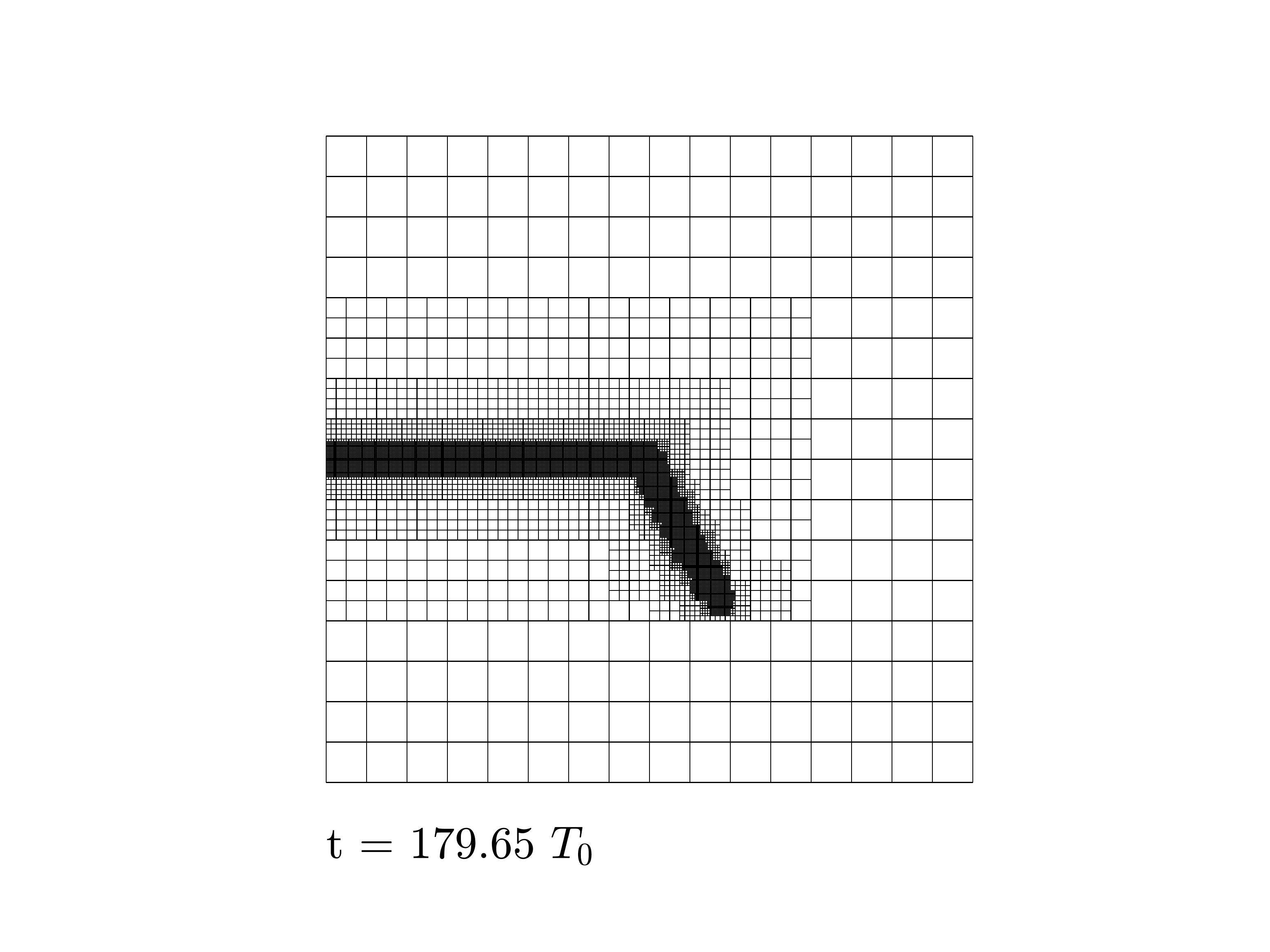}}
	\quad
	\subfloat[$\bar{u}=0.0128L_0$]{\includegraphics[trim= 780 400 700 320, clip,height=0.3\textwidth]{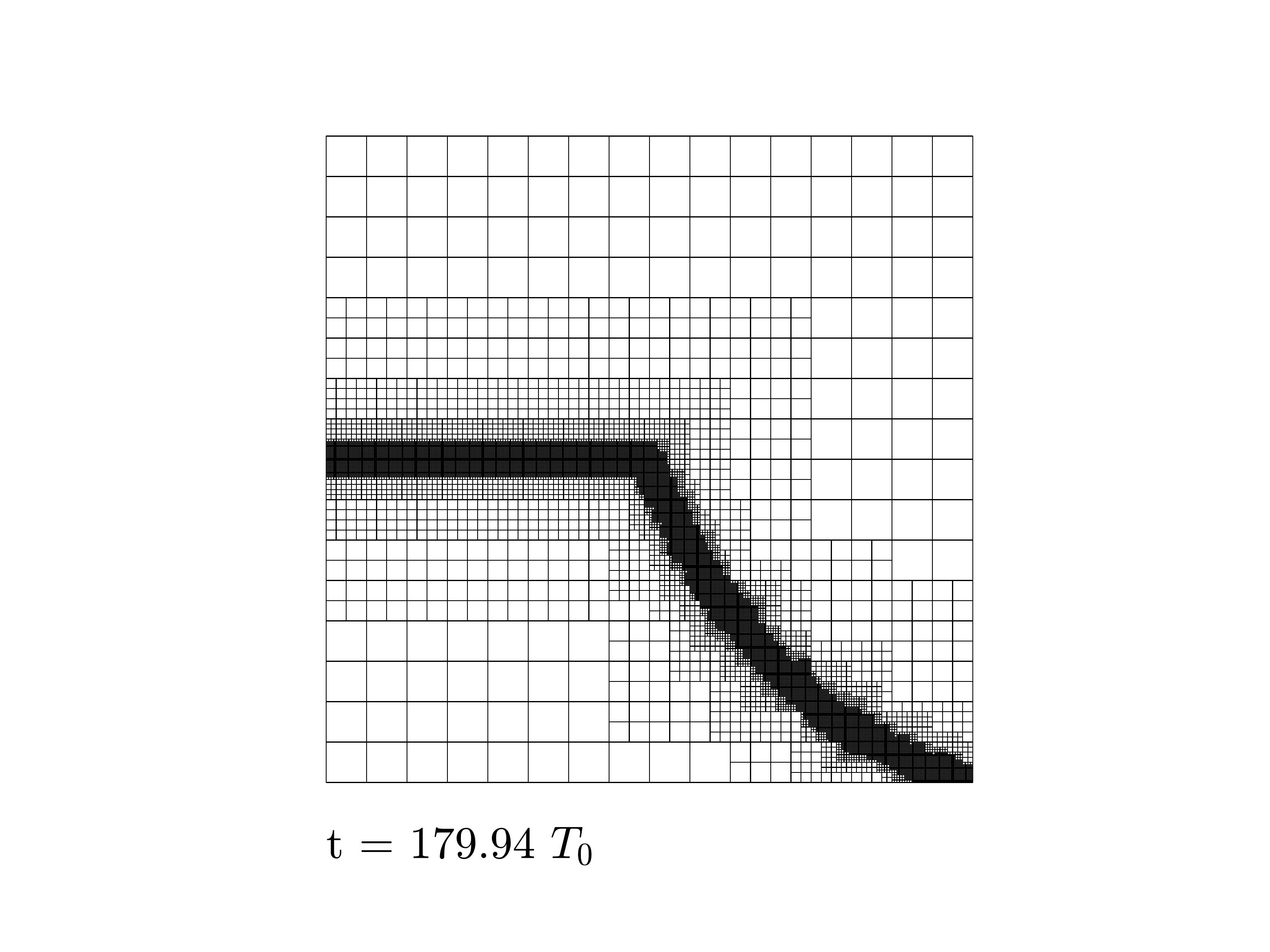}}
\caption{2D shear test: Parametric domains of LR meshes at various time steps. Only damage regions are adaptively refined and a coarse mesh is kept in regions of no damage.} \label{f:shearmesh}
\end{figure}

\begin{figure}[!ht]
\captionsetup[subfigure]{labelformat=empty}
\centering
\setlength{\abovecaptionskip}{-10pt} 
	\subfloat[]{%
		\begin{tikzpicture}
			\def\cdot{\times}
			\begin{axis}[ymode=log,grid=both,xlabel={$\bar{u}\,[L_0]$},ylabel={$\Delta t\,[T_0]$},width=0.46\textwidth]	
				\addplot[blue,line width=0.75] table [x index = {0}, y index = {1},col sep=comma,]{ifigs/shr/shr_u_dt.csv};
			\end{axis}
		\end{tikzpicture}
	}
	\quad
	\subfloat[]{%
		\begin{tikzpicture}
			\def\cdot{\times}
			\begin{axis}[grid=both,xlabel={$\bar{u}\,[L_0]$},ylabel={Energy $[E_0\,L_0^2]$},legend cell align={left},legend pos=north west,width=0.46\textwidth]	
				\addplot[blue,line width=1] table [x index = {0}, y index = {1},col sep=comma]{ifigs/shr/shr_u_nrg.csv};
				\addlegendentry{$\Pi_\mathrm{el}$};
				\addplot[red,line width=1,dashed] table [x index = {0}, y index = {2},col sep=comma]{ifigs/shr/shr_u_nrg.csv};
				\addlegendentry{$\Pi_\mathrm{frac}$};
			\end{axis}
		\end{tikzpicture}
	}
\caption{2D shear test: Computed time step sizes on the left and elastic and fracture energy over the prescribed deformation $\bar{u}$ on the right. As the crack evolves, the fracture energy increases whereas the elastic energy decreases due to the degradation of the contribution $\Psielp$ (cf. Eq.~\eqref{e:PiHF}). The fracture energy is non-zero at $\bar{u}=0\,L_0$ since the initial crack is modeled by means of an initial phase field.} \label{f:sheardtnrg}
\end{figure}
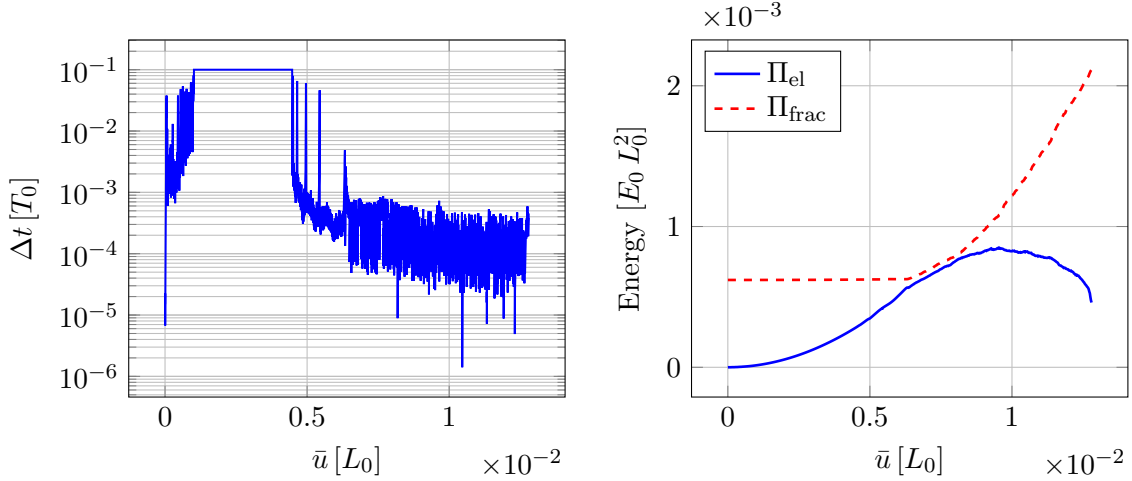

The first example investigates crack evolution in a square two-dimensional membrane that is exposed to a shear load. The geometry including boundary and loading conditions is illustrated in Fig.~\ref{fig:sheargeom}.
The mesh is initially constructed from $16\times16$ LR NURBS elements and the region next to the initial crack is refined by LR NURBS elements up to a refinement depth of $d=5$, see Fig.~\ref{f:shearmesh}. The material parameters are given in Tab.~\ref{tab:shearpar}.
\begin{table}[!ht]
	\centering
	\setlength{\tabcolsep}{8pt}
	\renewcommand{\arraystretch}{1.25}
  	\begin{tabular}{c c c c c c c }
  		$E$ $[E_0]$ & $\nu$ $[-]$  & $\Delta\bar{u}$  $[L_0]$ & $\sG_c$ $[E_0\,L_0]$  & $\ell_0$ $[L_0]$ & $T$ $[L_0]$
  		\\ \hline
  		$100$ & $0.2$ & $2\cdot10^{-6}$ & $0.001$ & $0.0025$ & $0.0125$
  \end{tabular}
  \caption{2D shear test: Material parameters and imposed load increment $\Delta\bar{u}$ per time step.}
  \label{tab:shearpar}
\end{table}
The initial phase field distribution, which is induced by an initial history field, and the crack evolution are shown in Fig. \ref{f:shearprogr}. The crack evolves towards the bottom right corner on a curved path. The qualitative behavior resembles the results shown in the literature. For instance, in \cite{borden2012} a quasi-static two-dimensional shear test has been investigated where the crack path has been locally refined \textit{a priori} based on analysis-suitable T-splines.\\
Our results show that the split of the membrane energy from Sec.~\ref{sec:energysplit} works correctly since no branch is forming towards the specimen's top edge.
Based on the adaptive spatial refinement strategy from Sec.~\ref{sec:lrsurf}, the LR mesh is refined as the crack evolves. The parametric domains of the LR meshes are illustrated in Fig.~\ref{f:shearmesh}. Only the regions of damage are refined up to the prescribed refinement depth $d=5$, while the periphery is kept coarse.
Fig.~\ref{f:sheardtnrg} shows the time step sizes employed and the contributions to the total energy in the system. The latter have been computed from
\eqb{l}
	\Pi_\mathrm{el}=\ds\int_\sS\Bigl(g(\phi)\Psielp+\Psielm\Bigr)\,\dif a\,,\quad\mathrm{and}\quad\Pi_\mathrm{frac}=\int_\sS\Psi_\mathrm{frac}\,\dif a\,.
\label{e:pielfrac}\eqe
Fig.~\ref{f:sheardtnrg} shows that at the prescribed deformation $\bar{u}\in[0.001018,0.004474]\,L_0$, the maximum time step size $\Delta t_{\max} =0.1\,T_0$ is used since the crack is not evolving. 
Thus, the fracture energy stays constant during this time. Since the initial crack is modeled by means of an initial phase field, the fracture energy is non-vanishing at $\bar{u}=0\,L_0$. The elastic energy increases steadily due to the applied deformation. As the crack evolves at $\bar{u}>0.006\,L_0$, the fracture energy increases, whereas the reduction of material stiffness leads to a decrease in elastic energy. Crack evolution takes place for $\bar{u}\in[0.006,0.0128]\,L_0$. The qualitative trend is similar to other examples shown in the literature, e.g. in \cite{borden2012} and \cite{schlueter2014}. In quasi-static simulations, there is a sudden drop in the reaction forces and energies as fracture occurs. As outlined in \cite{schlueter2014}, a bounded crack velocity prohibits such discontinuities. Due to the presence of kinetic energy in our formulation, the elastic energy does not vanish in the fully fractured state.\footnote{Also see the remark on stress waves at the beginning of this section.}

\subsection{Dynamic crack branching} \label{s:numex_dynbrnch}
\begin{figure}[!ht]
	\centering
		\subfloat[Problem setup\label{f:brnchsetup}]{\includegraphics[width=0.45\textwidth,valign=c]{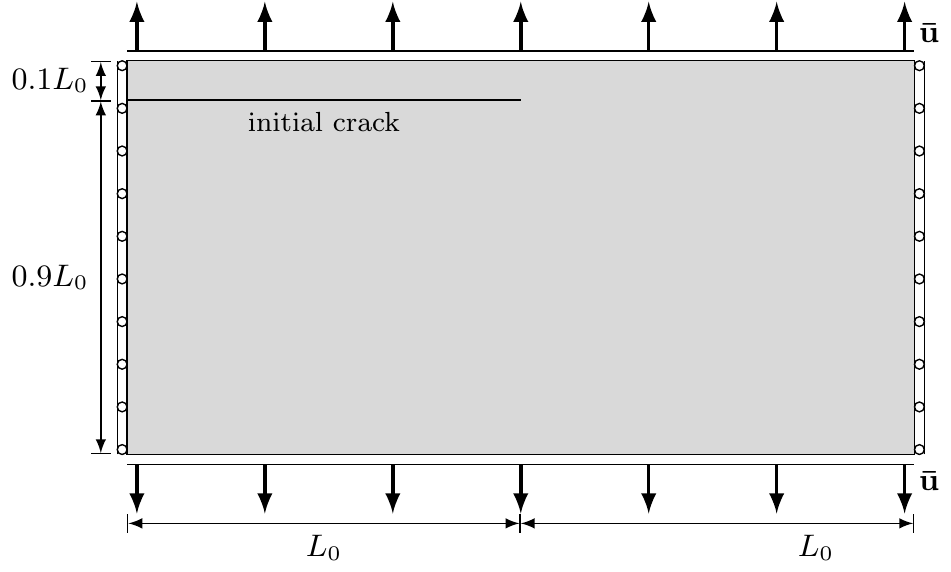}}
	\qquad
		\subfloat[Initial LR Mesh\label{f:brnchmeshinit}]{\includegraphics[width=0.45\textwidth,valign=c]{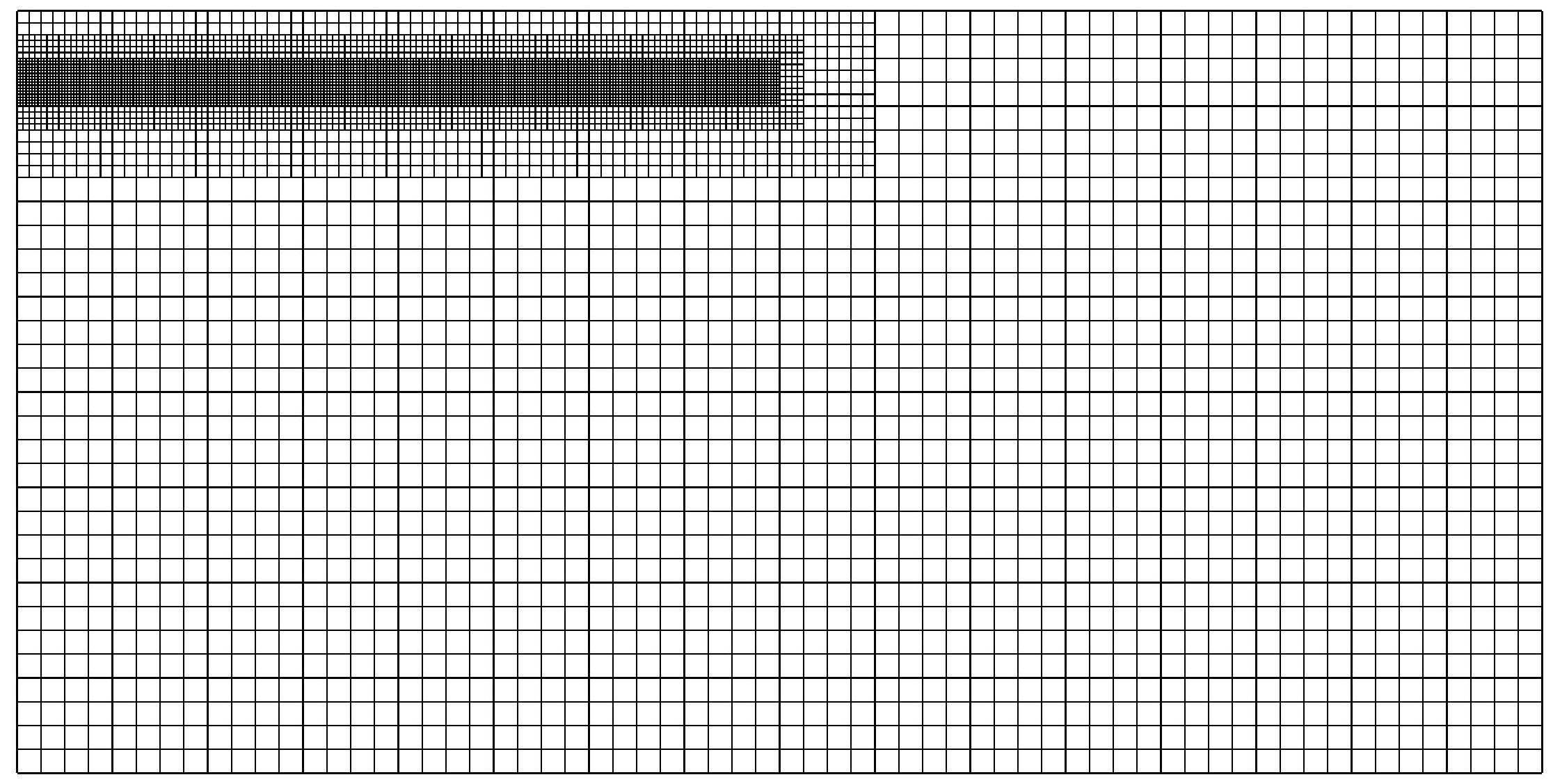}
			\vphantom{\includegraphics[width=0.45\textwidth,valign=c]{ifigs/deflcrack/deflectedcracksetup}}}
	 \caption{Dynamic crack branching: (a) Specimen geometry, boundary and loading conditions and (b) initial LR mesh in which the region around the initial crack is refined up to a refinement depth of $d=3$.}
\end{figure}

We next consider a rectangular 2D membrane with an initial crack at the top. The problem setup is shown in Fig.~\ref{f:brnchsetup}.
\begin{table}[!ht]
	\centering
	\setlength{\tabcolsep}{8pt}
	\renewcommand{\arraystretch}{1.25}
  	\begin{tabular}{c c c c c c c }
  		$E$ $[E_0]$ & $\nu$ $[-]$  & $\sG_c$ $[E_0\,L_0]$  & $\ell_0$ $[L_0]$ & $T$ $[L_0]$
  		\\ \hline
  		$100$ & $0.3$ &  $0.001$ & $0.0025$ & $0.0125$
  \end{tabular}
  \caption{Dynamic crack branching: Material parameters.}
  \label{tab:brnchpar}
\end{table}
A displacement of constant velocity is applied on the top edge upwards and on the bottom edge downwards. At each time step we impose the deformation increment $\Delta\bar{u}=\bar{v}\,\Delta t$ where the maximum time step size is set to $\Delta t_\mathrm{max}=10^{-3}\,T_0$.\footnote{We can compute the shear wave speed based on $c_\mathrm{s}=\sqrt{G/\rho}\approx6.2\,L_0/T_0$. An approximate value for the Rayleigh wave speed is then obtained as $c_\mathrm{R}\approx0.9162\cdot c_\mathrm{s}\approx5.7\,L_0/T_0$. Based on the experiments by \cite{ravichandar1984}, the crack tip velocity stays below $60\%$ of the Rayleigh wave speed. We can thus formulate a condition for the minimum time step, i.e. $\Delta t\leq\Delta t_\mathrm{max}<\Delta x_\mathrm{min}/(0.6\cdot c_\mathrm{R})\approx1.1\cdot10^{-3}\,T_0$, where the minimum element size is $\Delta x_\mathrm{min}=1/256\,L_0$.} The loading velocity is denoted $\bar{v}$. The material parameters are depicted in Tab.~\ref{tab:brnchpar}. The initial mesh is constructed from $64\times32$ LR NURBS elements and refined around the prescribed initial damage up to a refinement depth $d=3$, see Fig.~\ref{f:brnchmeshinit}.
\begin{figure}[!ht]
	\centering
		\subfloat[$\bar{v}=1.25\cdot10^{-3}\,L_0\,T_0^{-1}$\label{f:brnchphi1}]{\includegraphics[trim=950pt 1220pt 750pt 1000pt,clip,width=0.43\textwidth]{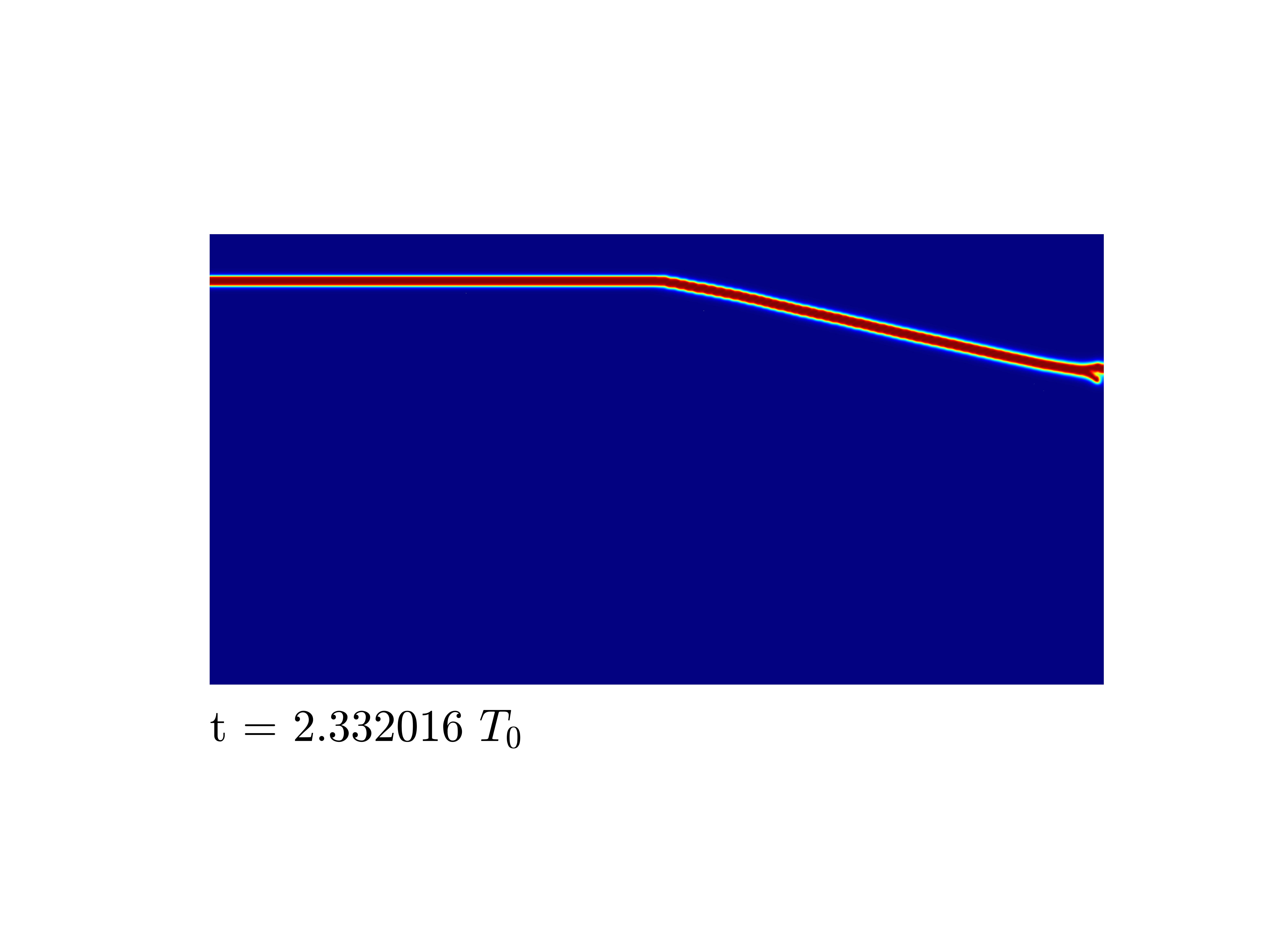}}
		\quad
		\subfloat[$\bar{v}=5\cdot10^{-3}\,L_0\,T_0^{-1}$\label{f:brnchphi2}]{\includegraphics[trim=950pt 1220pt 750pt 1000pt,clip,width=0.43\textwidth]{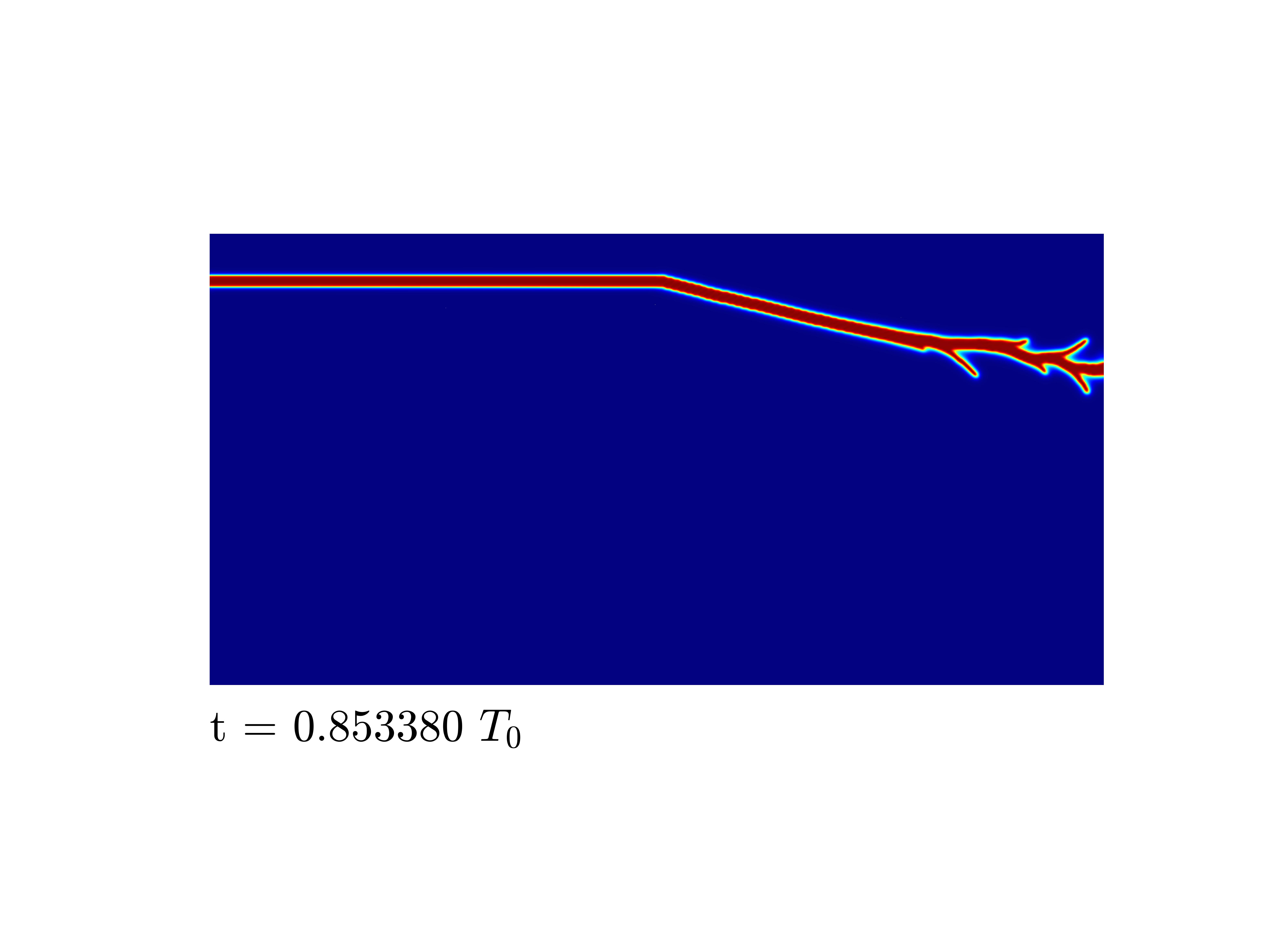}}
	\\ \vspace{-3mm}
		\subfloat[$\bar{v}=1\cdot10^{-2}\,L_0\,T_0^{-1}$\label{f:brnchphi3}]{\includegraphics[trim=950pt 1220pt 750pt 1000pt,clip,width=0.43\textwidth]{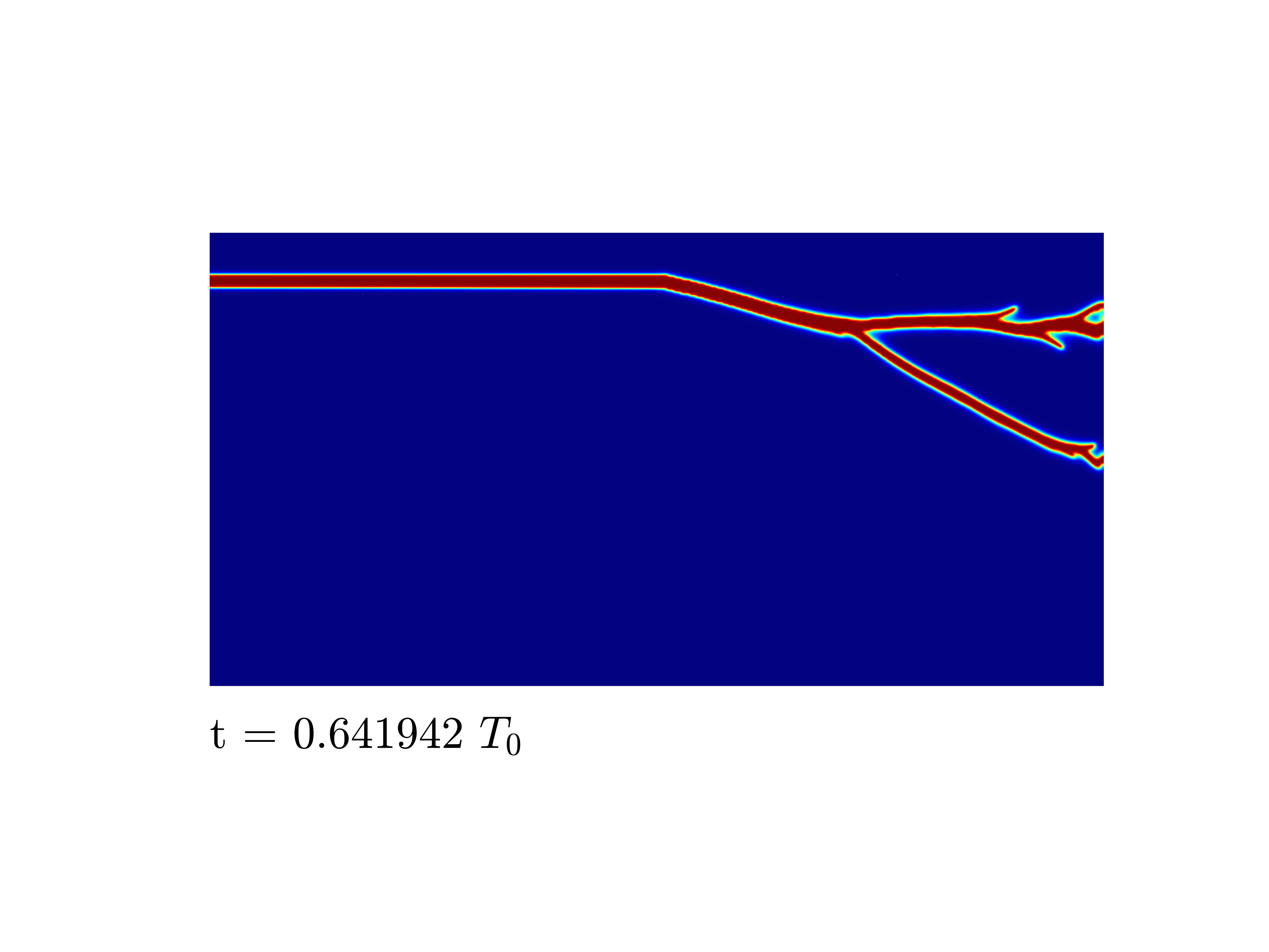}}
		\quad
		\subfloat[$\bar{v}=2\cdot10^{-2}\,L_0\,T_0^{-1}$\label{f:brnchphi4}]{\includegraphics[trim=950pt 1220pt 750pt 1000pt,clip,width=0.43\textwidth]{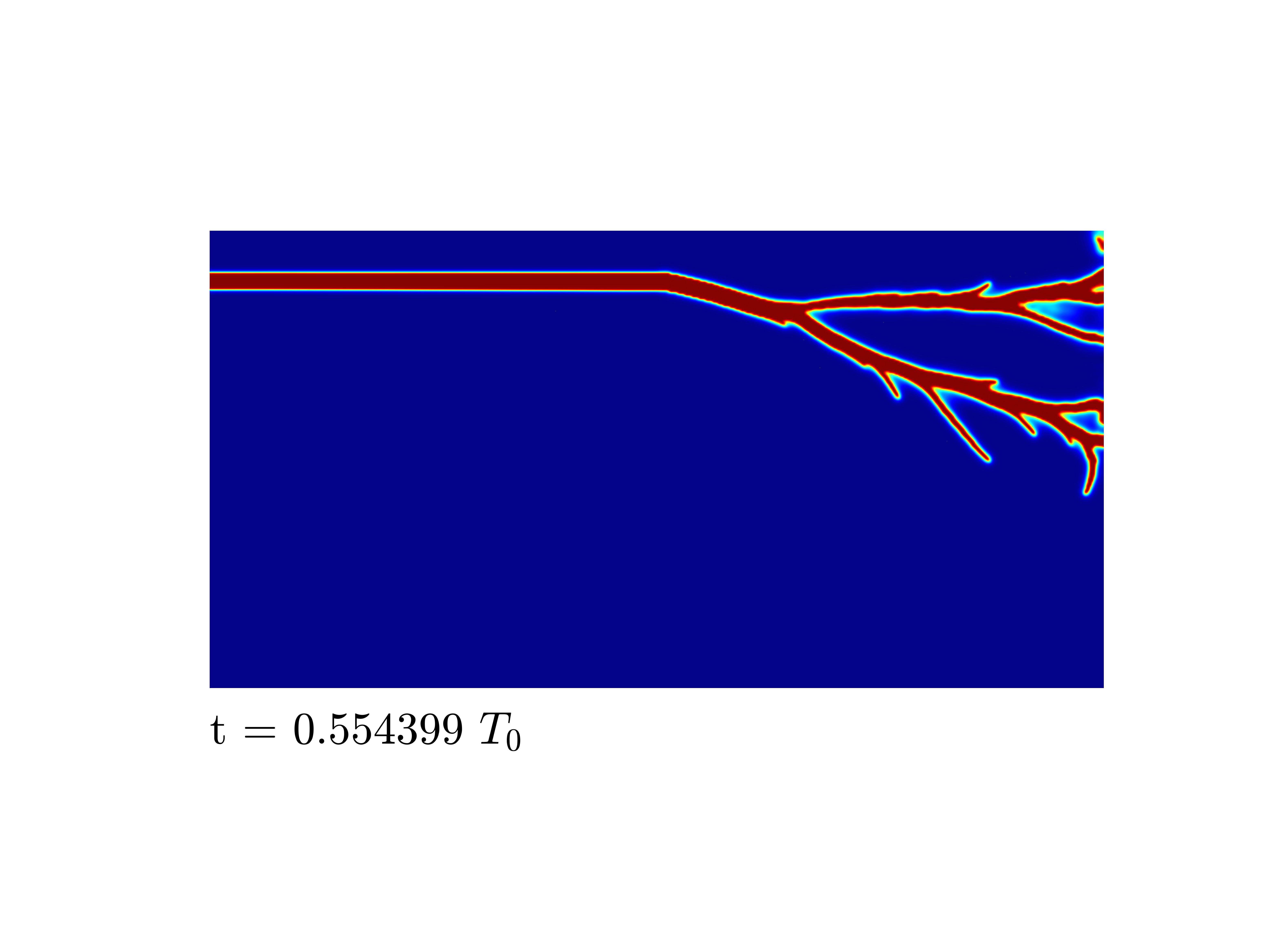}}
\caption{Dynamic crack branching: Crack evolution at the final state for different loading velocities $\bar{v}$. As the loading intensity is increased, crack branching occurs at an earlier time and closer to the left side of the membrane.}
\label{f:brnchphi}
\end{figure}
\noindent
\begin{figure}[!ht]
	\centering
		\subfloat[$\bar{v}=1.25\cdot10^{-3}\,L_0\,T_0^{-1}$\label{f:brnchmsh1}]{\includegraphics[trim=480pt 620pt 377pt 550pt,clip,width=0.43\textwidth]{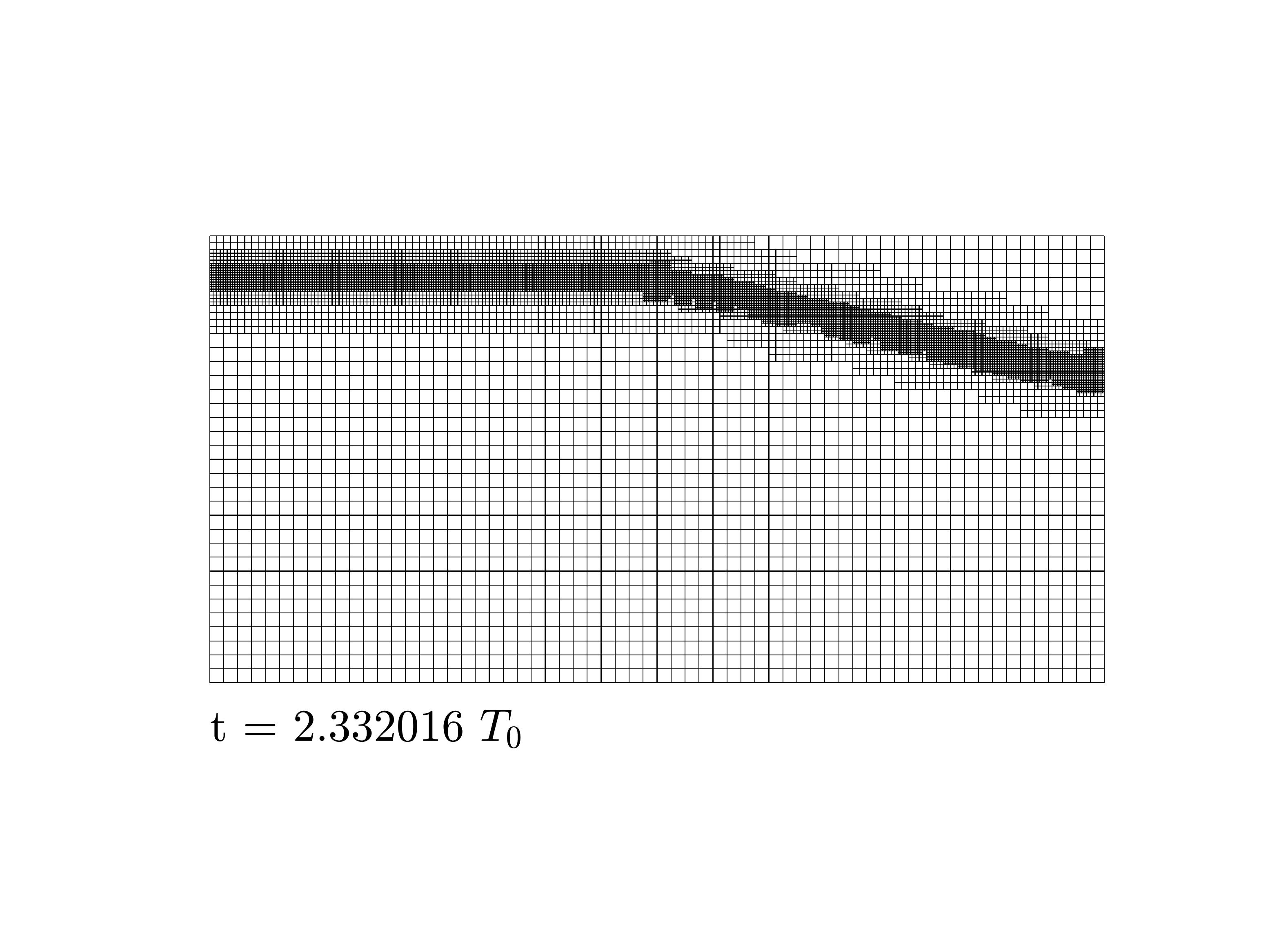}}
		\quad
		\subfloat[$\bar{v}=5\cdot10^{-3}\,L_0\,T_0^{-1}$\label{f:brnchmsh2}]{\includegraphics[trim=480pt 620pt 377pt 550pt,clip,width=0.43\textwidth]{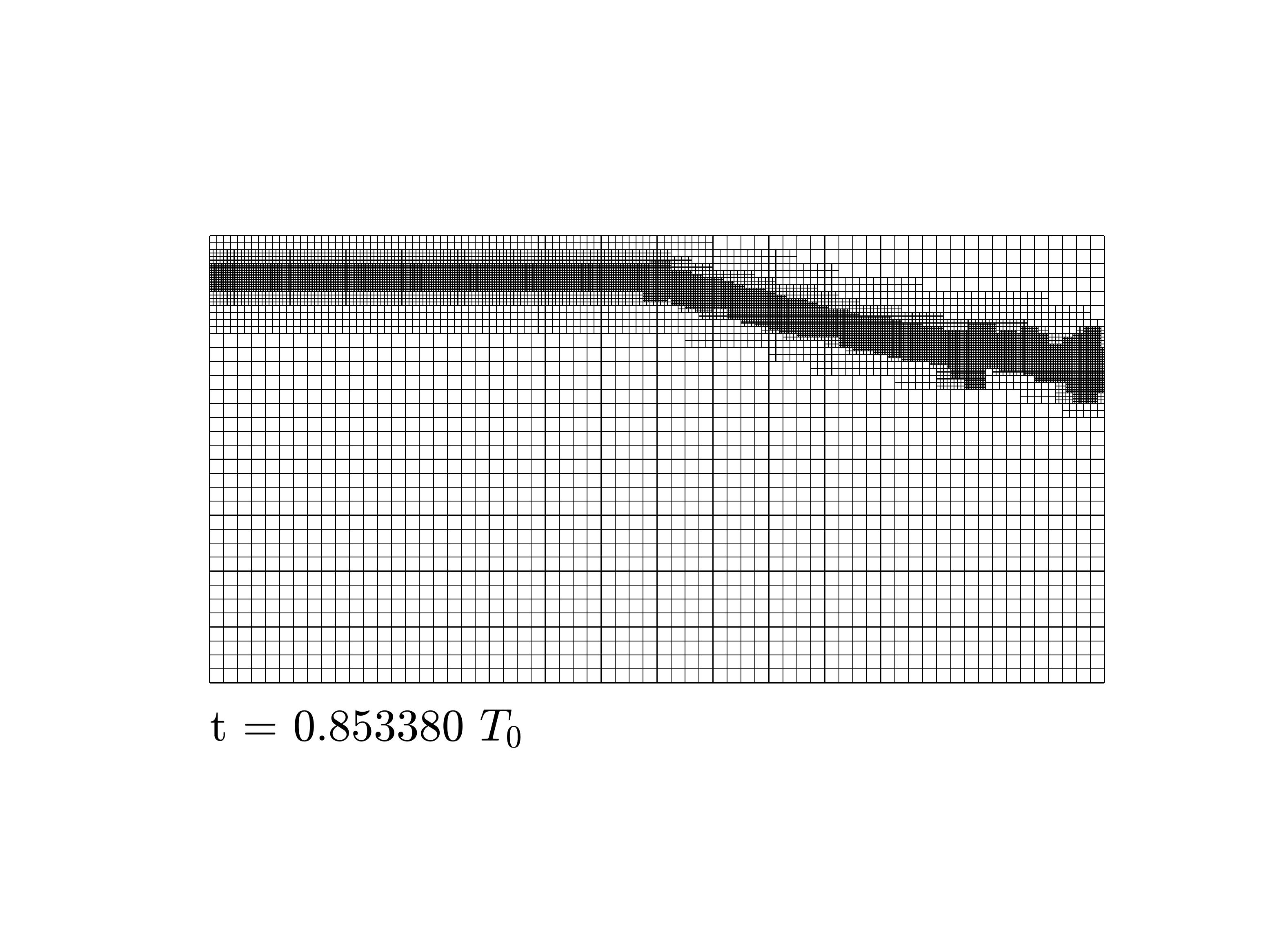}}
	\\ \vspace{-3mm}
		\subfloat[$\bar{v}=1\cdot10^{-2}\,L_0\,T_0^{-1}$\label{f:brnchmsh3}]{\includegraphics[trim=480pt 620pt 377pt 550pt,clip,width=0.43\textwidth]{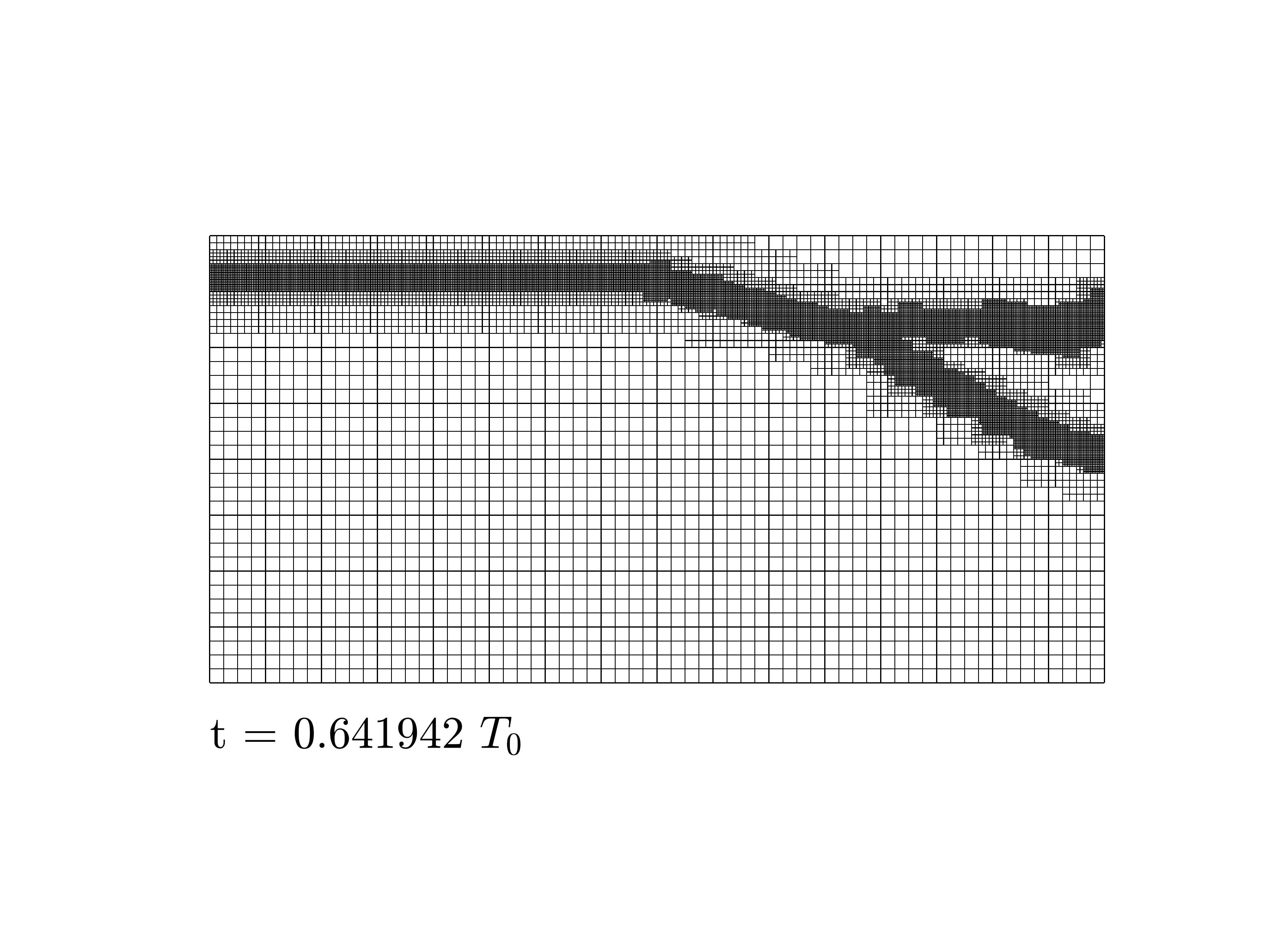}}
		\quad
		\subfloat[$\bar{v}=2\cdot10^{-2}\,L_0\,T_0^{-1}$\label{f:brnchmsh4}]{\includegraphics[trim=480pt 620pt 377pt 550pt,clip,width=0.43\textwidth]{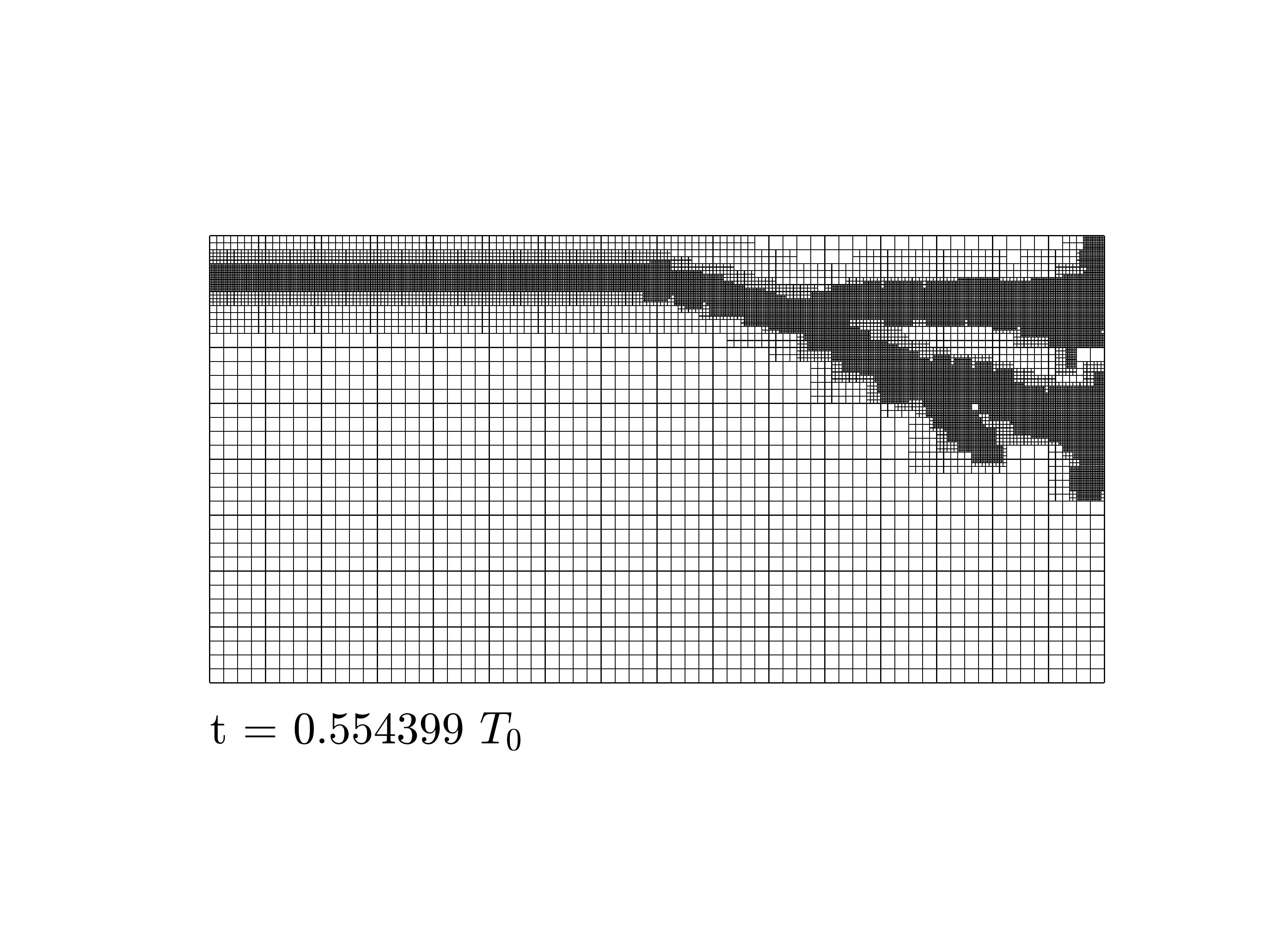}}
\caption{Dynamic crack branching: Final LR meshes as a function of the the loading velocity $\bar{v}$. The corresponding crack patterns are illustrated in Fig.~\ref{f:brnchphi}.}
\label{f:brnchmsh}
\end{figure}
The initial crack is not located on the mid-line so that the resulting asymmetric stress distribution leads to a deflection of the crack towards the bottom edge, see Fig.~\ref{f:brnchphi}. As the figure also shows, a higher loading velocity $\bar{v}$ leads to more complex fracture patterns with branching occurring sooner and more often.
This makes their prediction \textit{a priori} to the simulation very difficult. Fig.~\ref{f:brnchmsh} shows the final LR meshes in the undeformed configuration for the different crack patterns. There are large elements in regions of no fracture, whereas a highly resolved mesh is only obtained in the domain of fracture. Fig.~\ref{f:evhighint} shows three snapshots of the crack evolution and the corresponding LR meshes for the loading intensity $\bar{v}=2\cdot10^{-2}\,L_0\,T_0^{-1}$. The final states for these are shown in Figs.~\ref{f:brnchphi4} and \ref{f:brnchmsh4}. Only the periphery around the crack tip is refined, whereas no refinement is performed ahead of the crack tip. This adaptivity in space leads to an efficient prediction of fracture patterns.
\begin{figure}[!ht]
	\captionsetup[subfigure]{labelformat=empty}
	\centering
		\subfloat[]{\raisebox{-0.097\textwidth}{\rotatebox[origin=t]{90}{$t=0.250098\,T_0$}}}
		\quad
		\subfloat[]{\includegraphics[trim=480pt 620pt 377pt 550pt,clip,width=0.4\textwidth,valign=t]{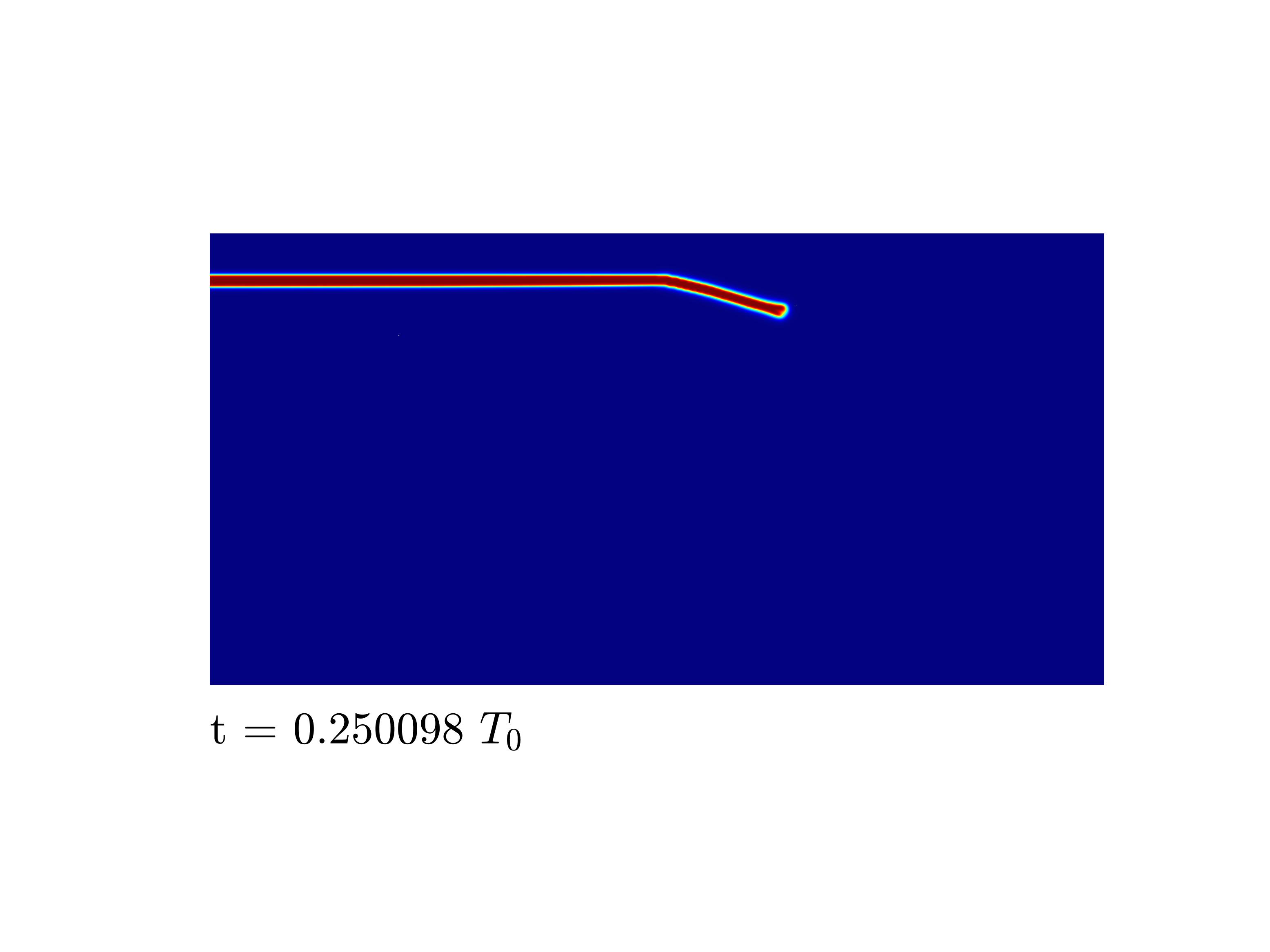}}
		\quad
		\subfloat[]{\includegraphics[trim=480pt 620pt 377pt 550pt,clip,width=0.4\textwidth,valign=t]{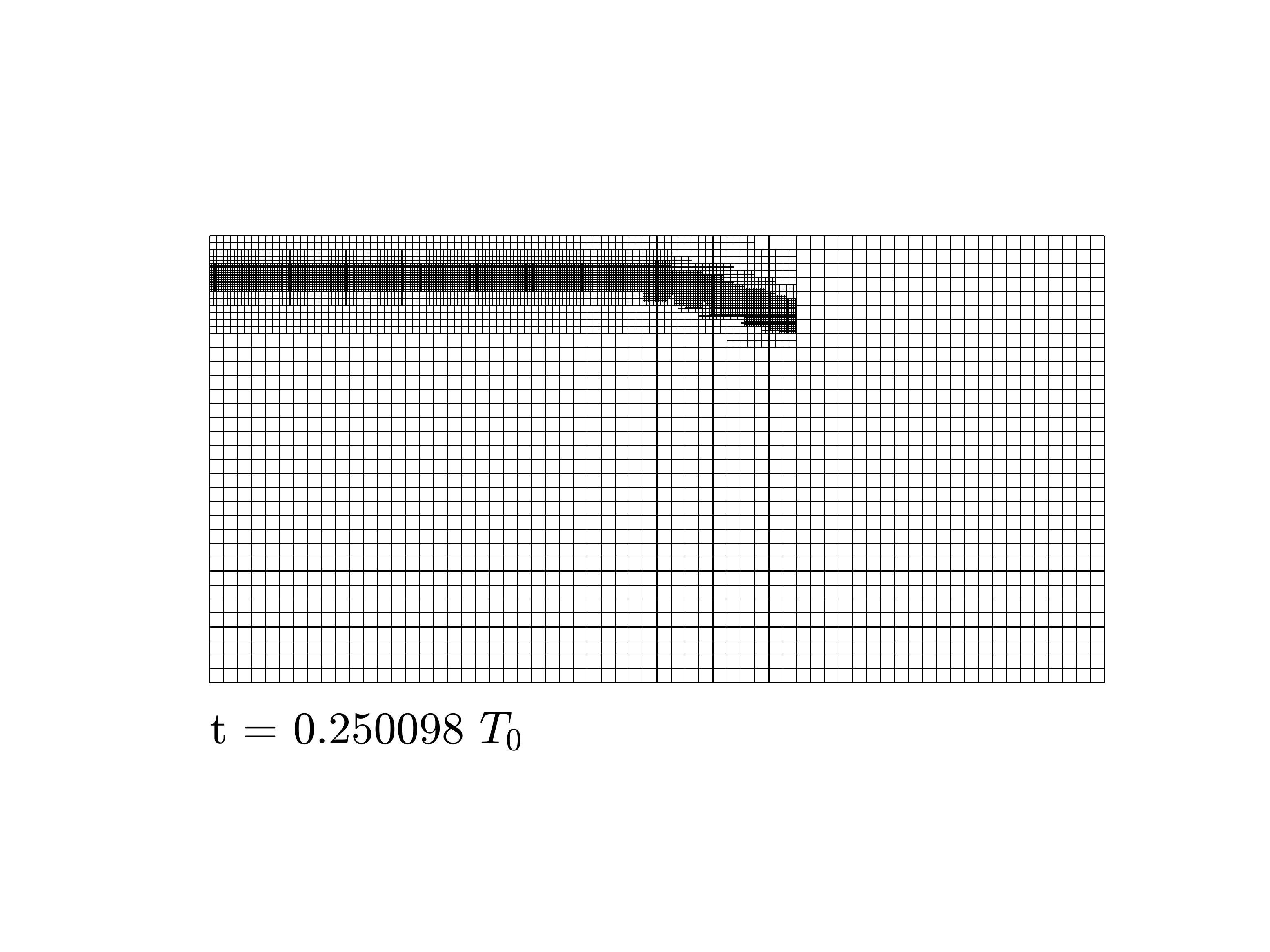}}
	\\ \vspace{-9mm}
		\subfloat[]{\raisebox{-0.097\textwidth}{\rotatebox[origin=t]{90}{$t=0.352116\,T_0$}}}
		\quad
		\subfloat[]{\includegraphics[trim=480pt 620pt 377pt 550pt,clip,width=0.4\textwidth,valign=t]{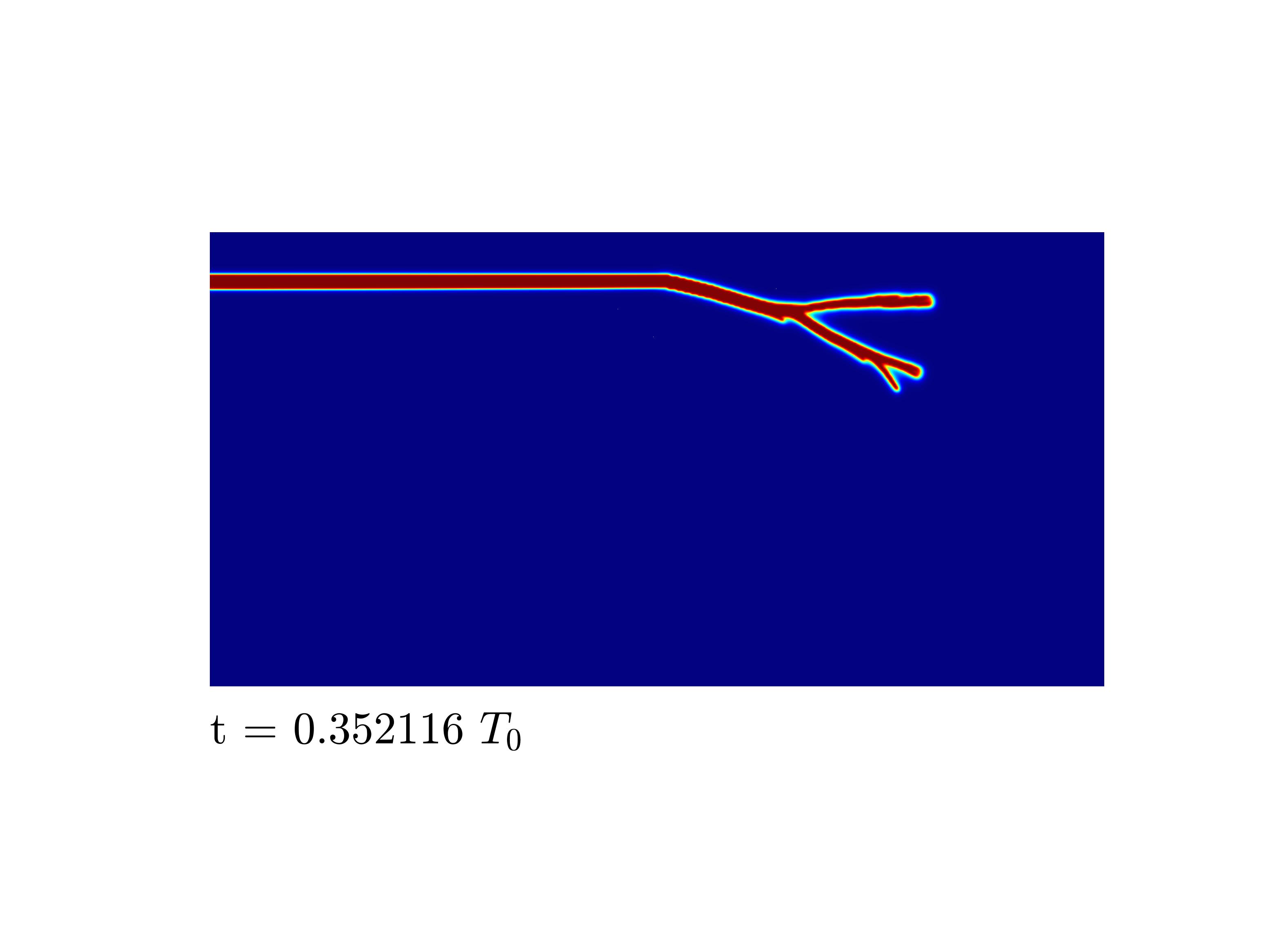}}
		\quad
		\subfloat[]{\includegraphics[trim=480pt 620pt 377pt 550pt,clip,width=0.4\textwidth,valign=t]{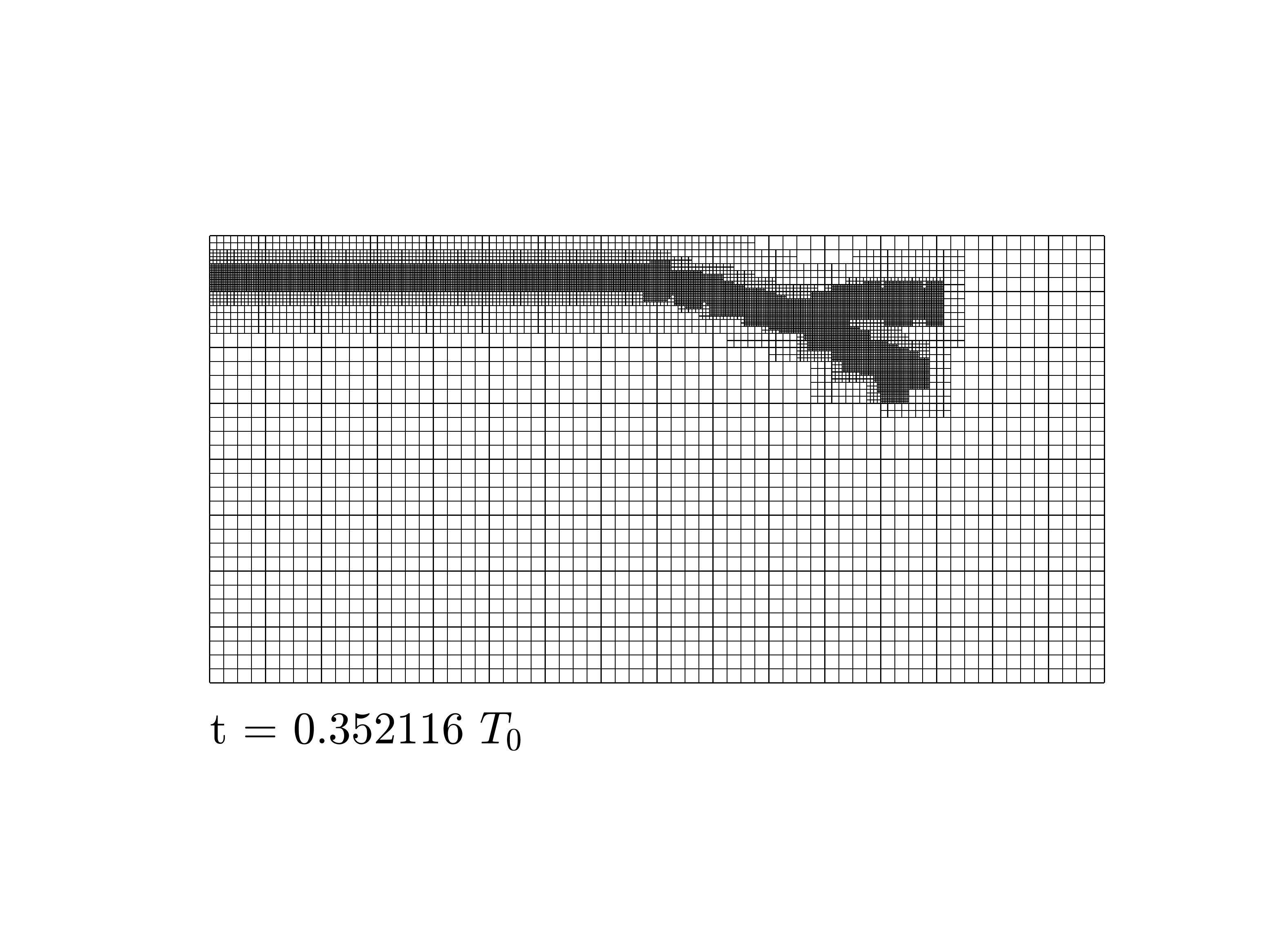}}
	\\	\vspace{-9mm}
		\subfloat[]{\raisebox{-0.097\textwidth}{\rotatebox[origin=t]{90}{$t=0.427593\,T_0$}}}
		\quad
		\subfloat[]{\includegraphics[trim=480pt 620pt 377pt 550pt,clip,width=0.4\textwidth,valign=t]{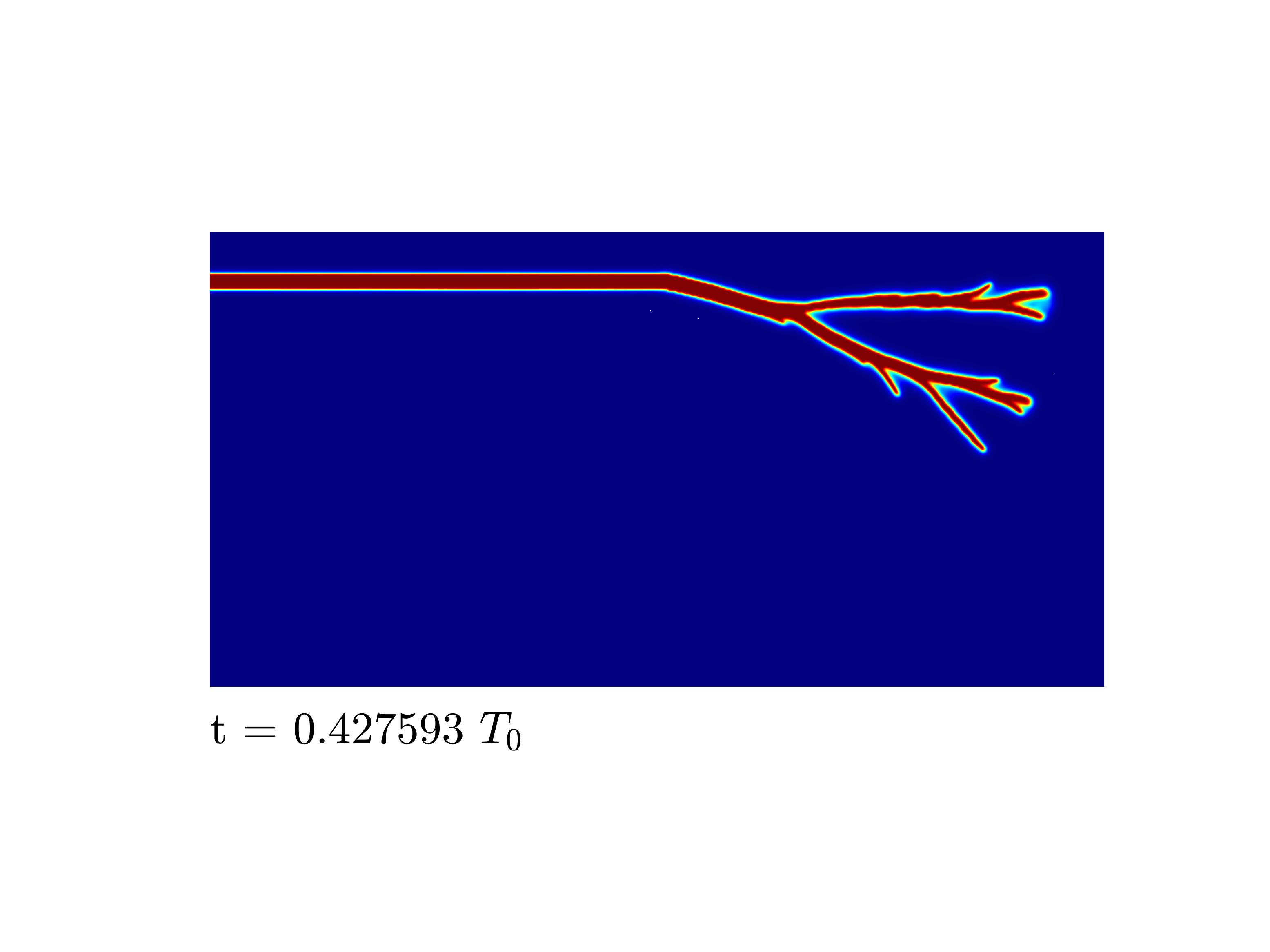}}
		\quad
		\subfloat[]{\includegraphics[trim=480pt 620pt 377pt 550pt,clip,width=0.4\textwidth,valign=t]{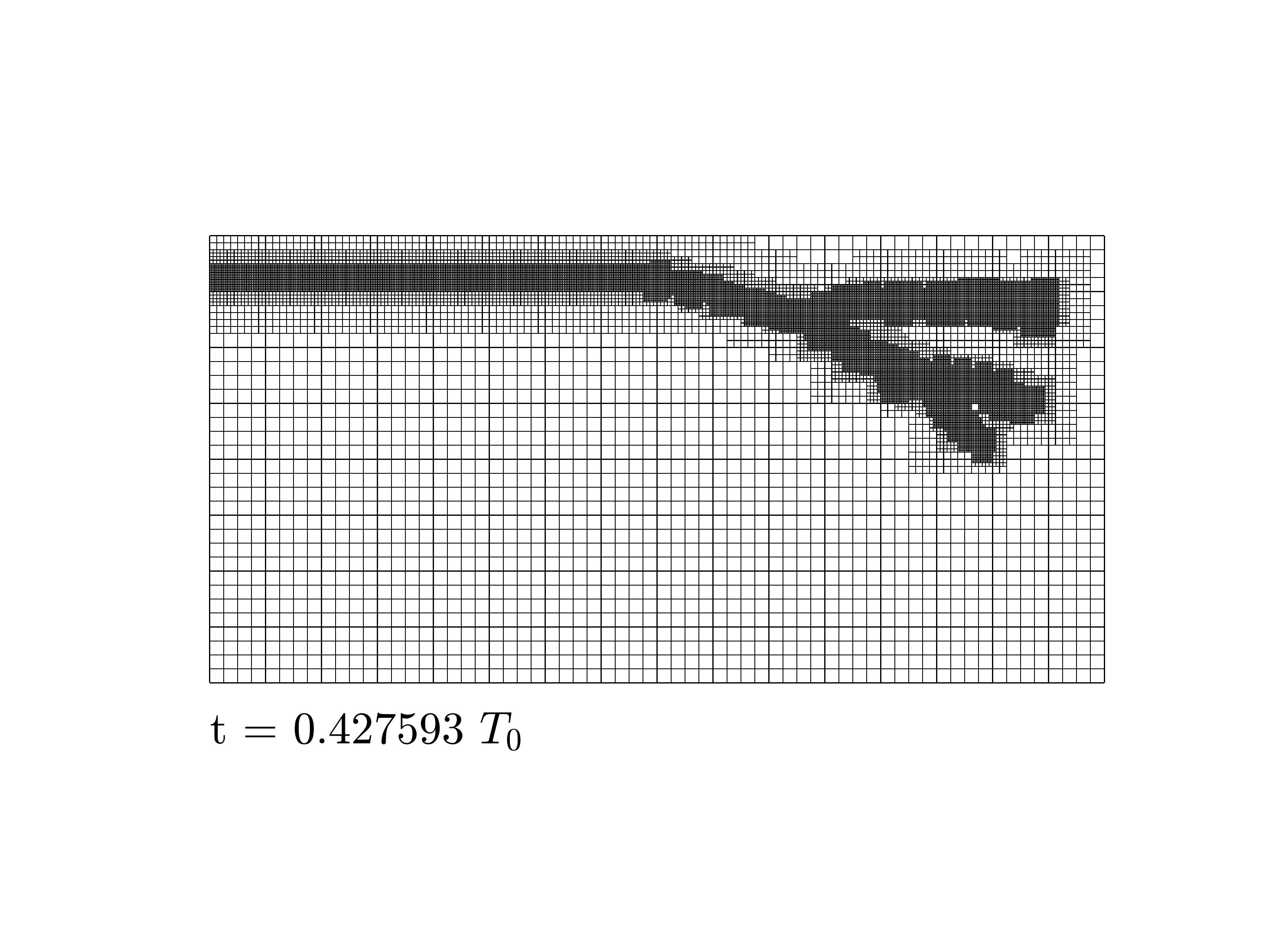}}
\caption{Dynamic crack branching: Evolution of the phase field on the left and corresponding LR meshes on the right. The loading intensity is $\bar{v}=2\cdot10^{-2}\,L_0\,T_0^{-1}$. The final phase field and LR mesh are shown in Figs.~\ref{f:brnchphi4} and \ref{f:brnchmsh4}. See also the supplementary movie at \href{https://doi.org/10.5446/42540}{https://doi.org/10.5446/42540}.}
\label{f:evhighint}
\end{figure}

\subsection{Pressurized cylinder}
\begin{figure}[!ht]
	\centering
		\subfloat[\label{fig:cyl1geom1}Top view]{\vspace{4cm}\includegraphics[scale=1]{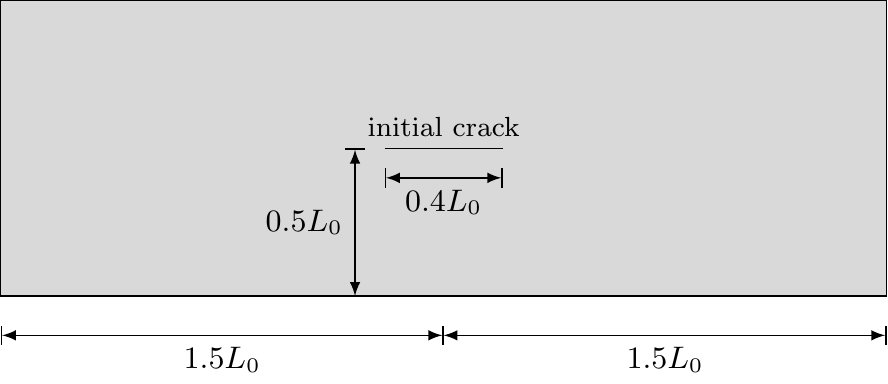}}
	\qquad
		\subfloat[\label{fig:cyl1geom2}Side view]{\raisebox{8.7mm}{\includegraphics[scale=0.91]{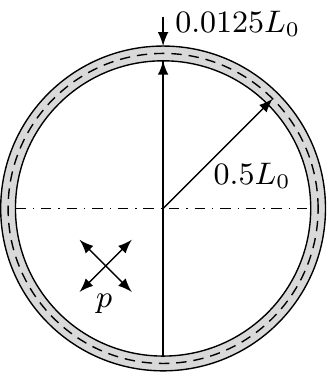}}}
\caption{Pressurized cylinder: Specimen geometry and loading conditions shown in (a) top view and (b) side view. The dashed line indicates the shell's mid plane. The shell is symmetric across the solid line in (b) which is used to reduce comutational effort. The movement of the two ends is only allowed in the axial direction and not in the radial direction.} \label{f:cyl1geom}
\end{figure}

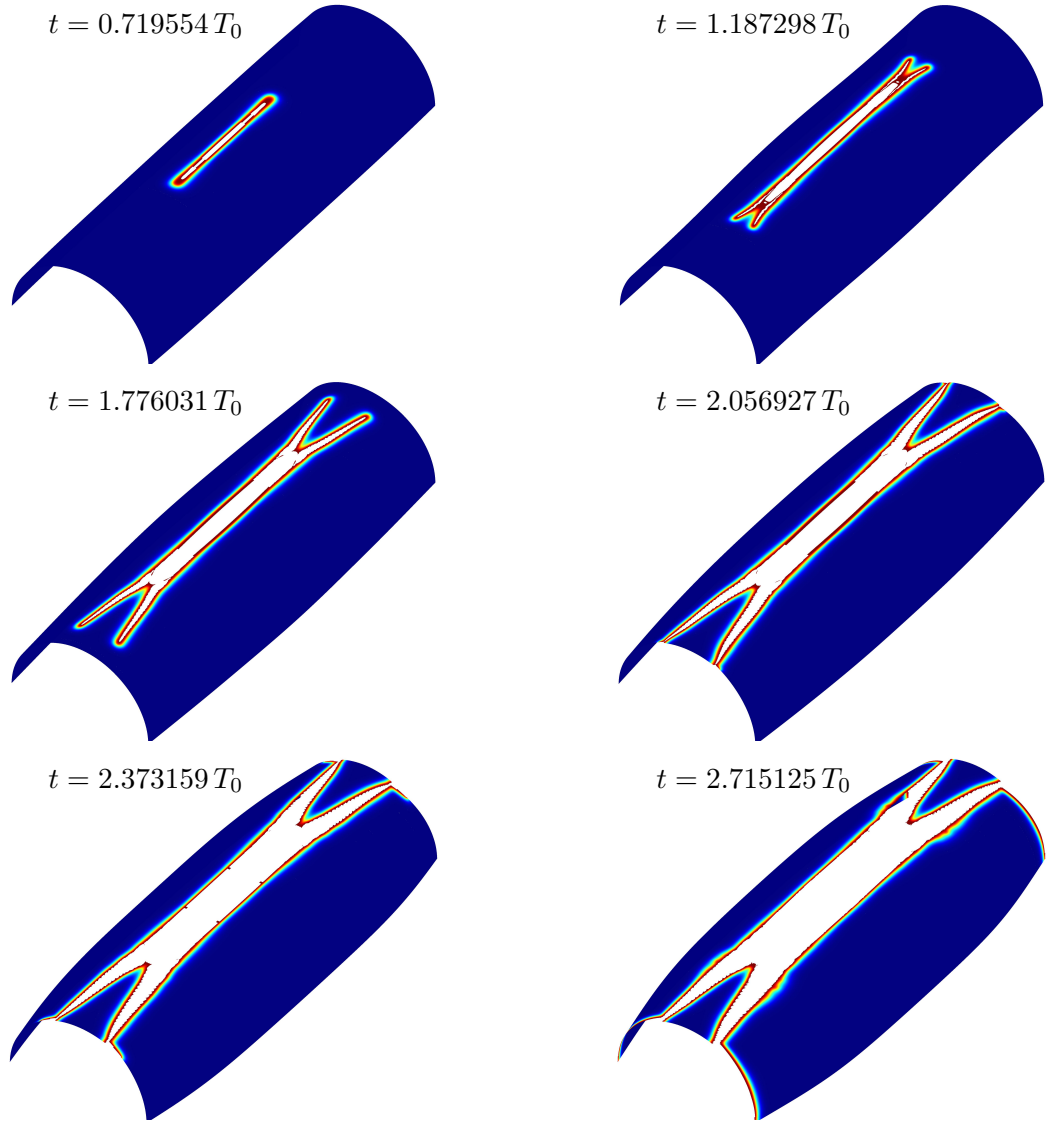
\begin{figure}[!ht]
	\centering
		\begin{tikzpicture}
			\node at (0,0) {\includegraphics[trim=950pt 900pt 750pt 800pt,clip,width=0.48\textwidth]{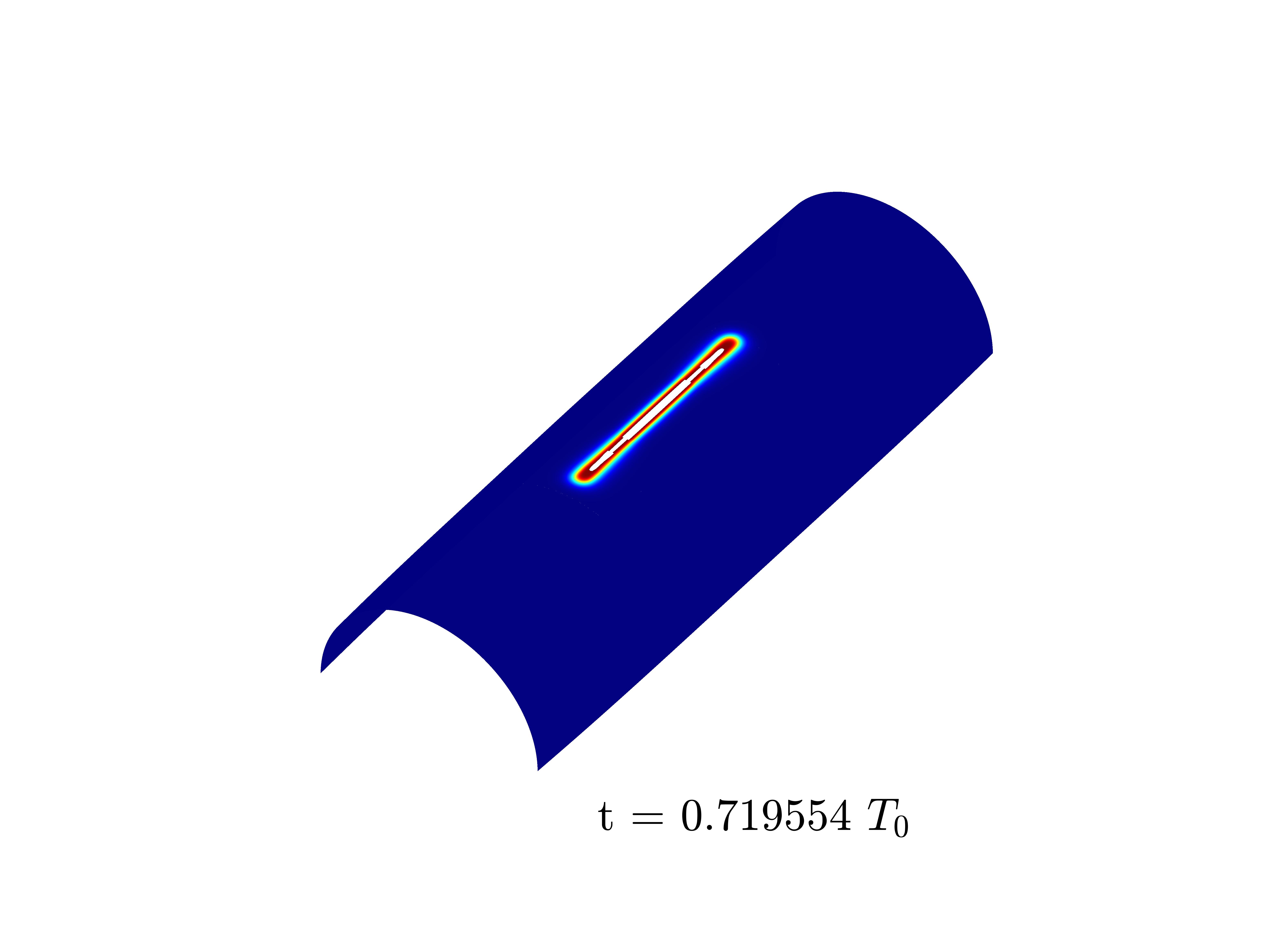}};
			\node at (-1,2) {$t=0.719554\,T_0$};
			\node at (8,0) {\includegraphics[trim=950pt 900pt 750pt 800pt,clip,width=0.48\textwidth]{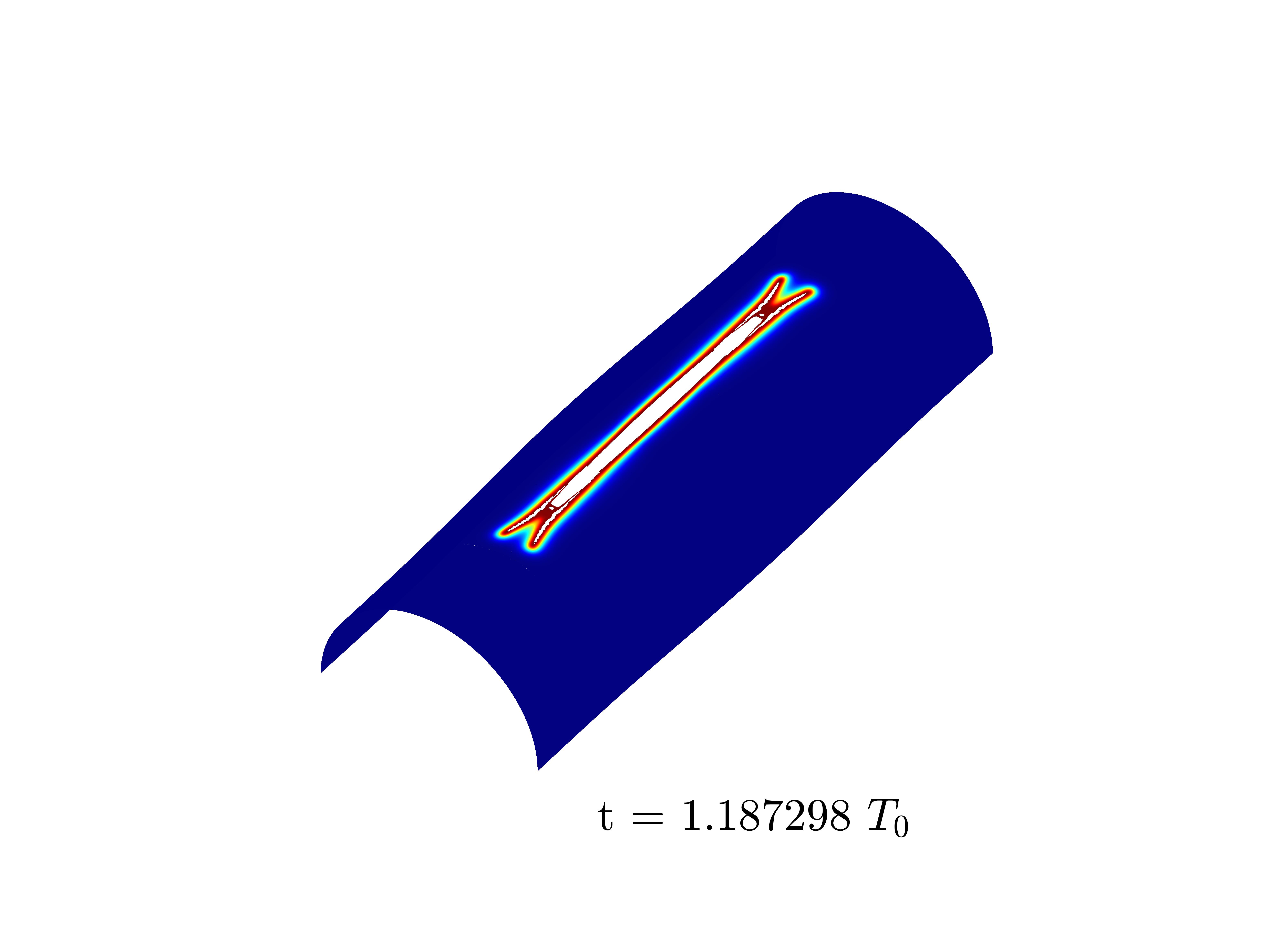}};
			\node at (7,2) {$t=1.187298\,T_0$};
			\node at (0,-5) {\includegraphics[trim=950pt 900pt 750pt 800pt,clip,width=0.48\textwidth]{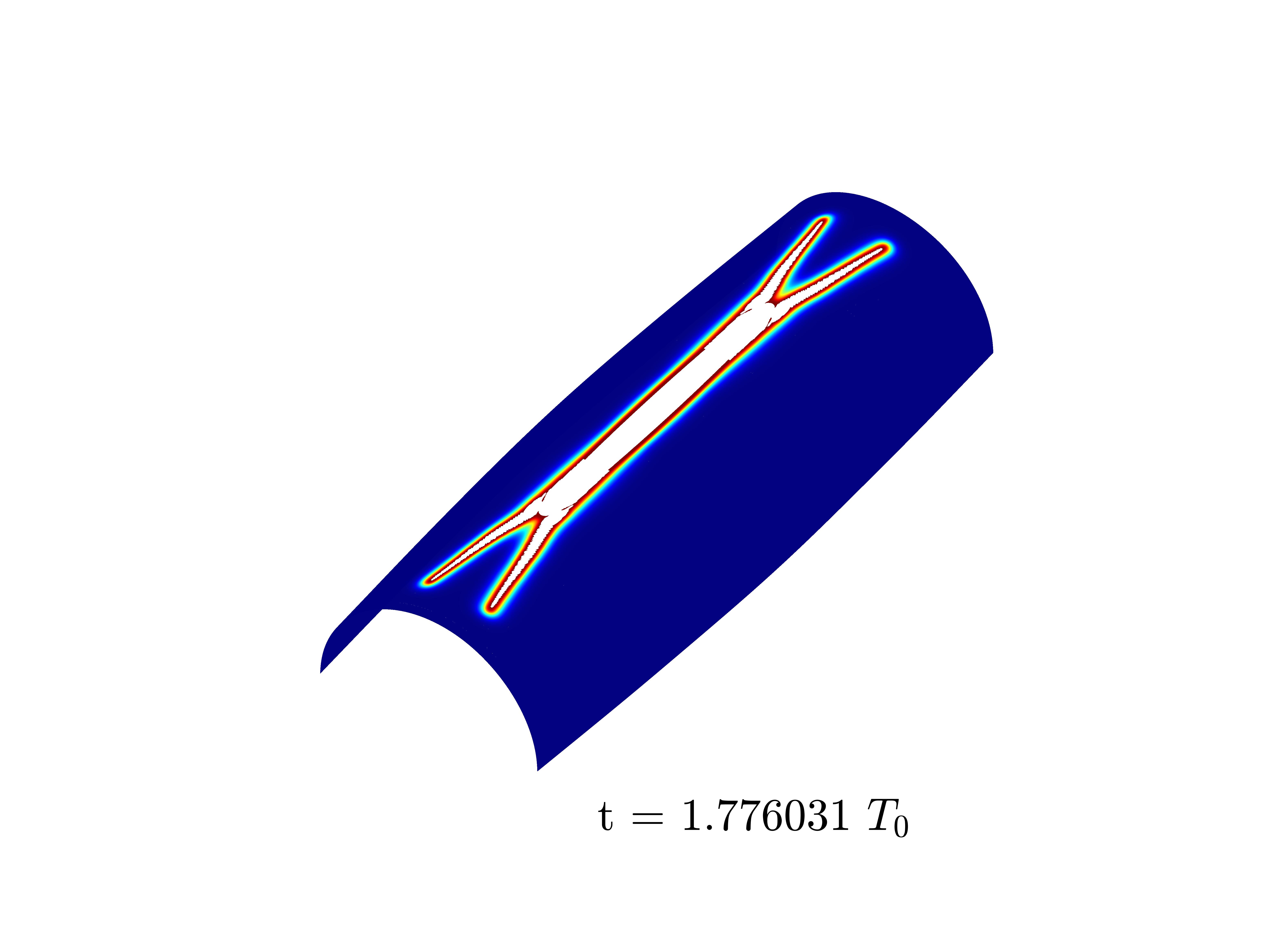}};
			\node at (-1,-3) {$t=1.776031\,T_0$};
			\node at (8,-5) {\includegraphics[trim=950pt 900pt 750pt 800pt,clip,width=0.48\textwidth]{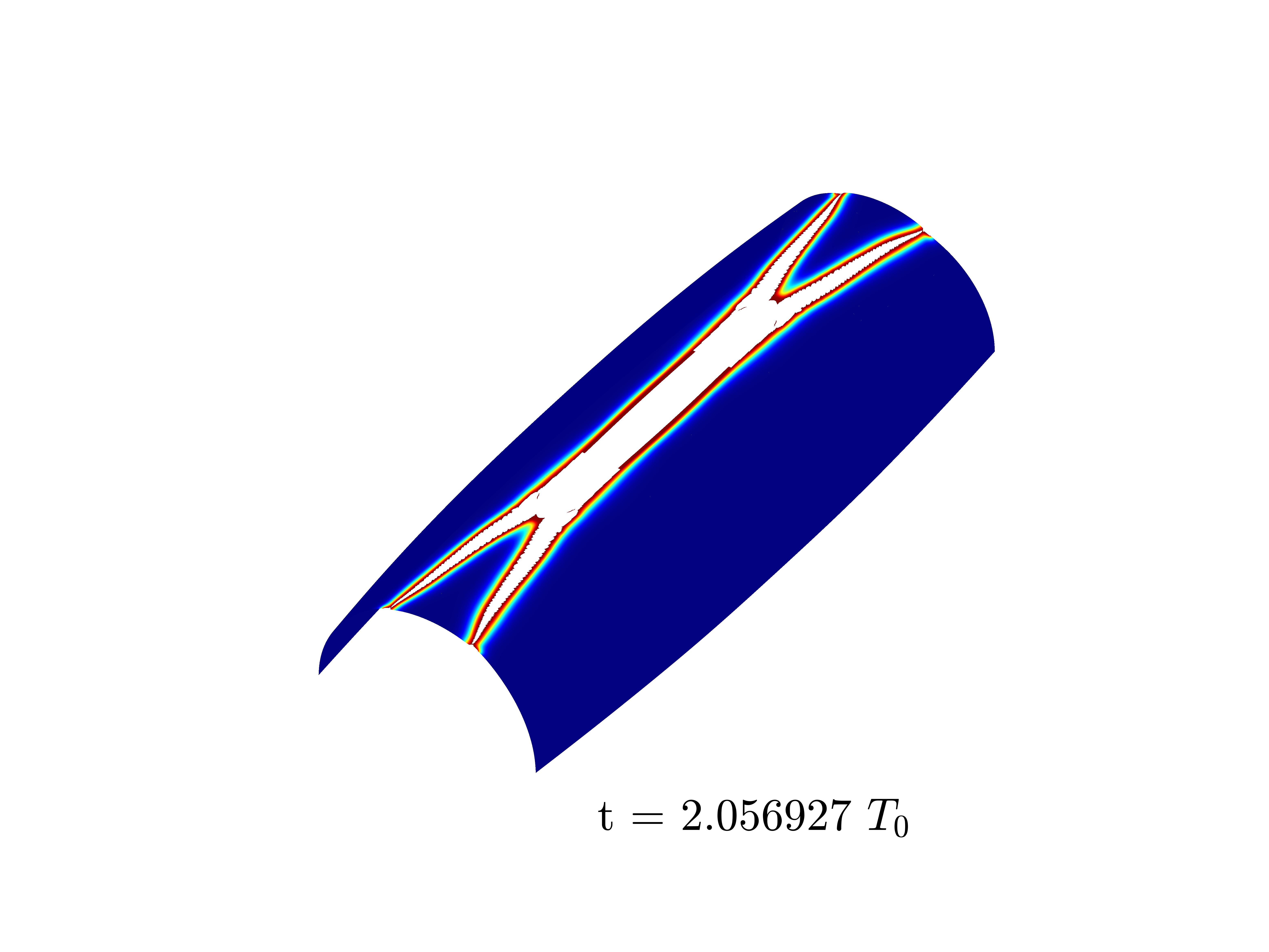}};
			\node at (7,-3) {$t=2.056927\,T_0$};
			\node at (0,-10) {\includegraphics[trim=950pt 900pt 750pt 800pt,clip,width=0.48\textwidth]{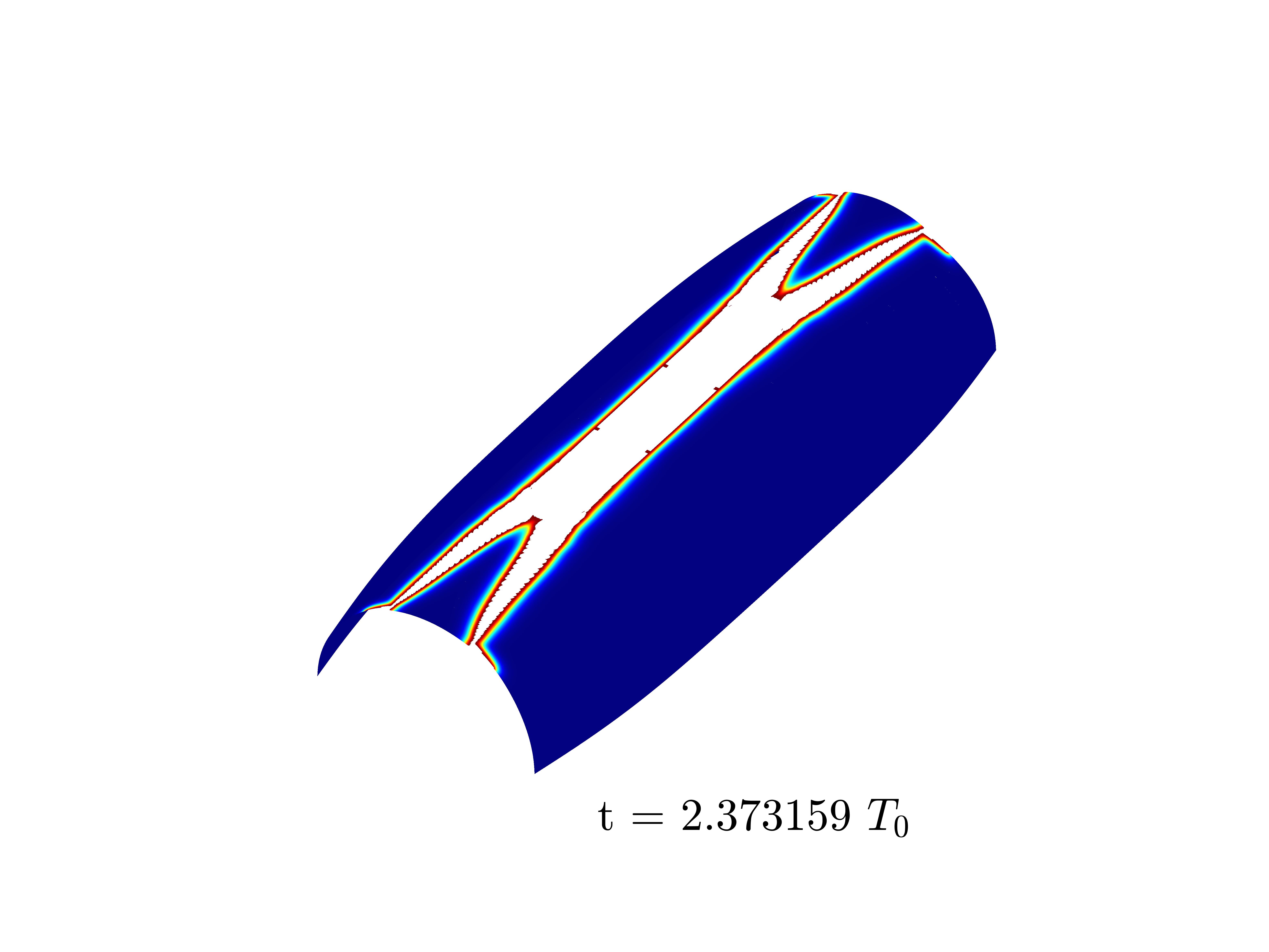}};
			\node at (-1,-8) {$t=2.373159\,T_0$};
			\node at (8,-10) {\includegraphics[trim=950pt 900pt 750pt 800pt,clip,width=0.48\textwidth]{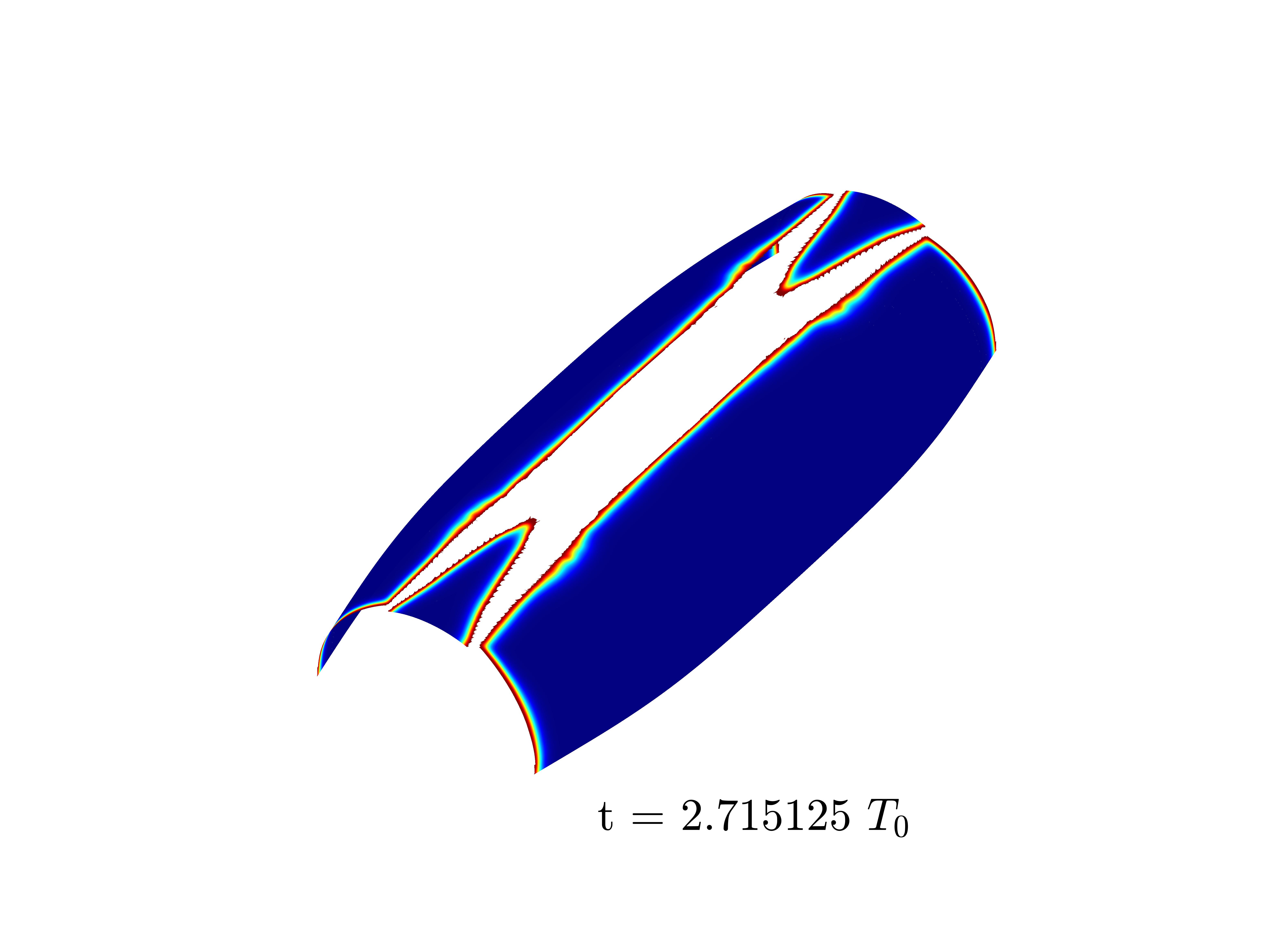}};
			\node at (7,-8) {$t=2.715125\,T_0$};
		\end{tikzpicture}
\caption{Pressurized cylinder: Crack pattern over time.
The stretch reaches up to $\approx130.49\%$ showing that the proposed formulation can model large deformations. Elements with $\phi<0.001$ have been removed for visualization. See also the supplementary movie at \href{https://doi.org/10.5446/42541}{https://doi.org/10.5446/42541}.}
\label{f:cylphi}
\end{figure}

\begin{figure}[!ht]
	\begin{tikzpicture}
		\node[inner sep=0pt] at (0,0) {\includegraphics[trim=0 350 0 350,clip,width=0.99\textwidth]{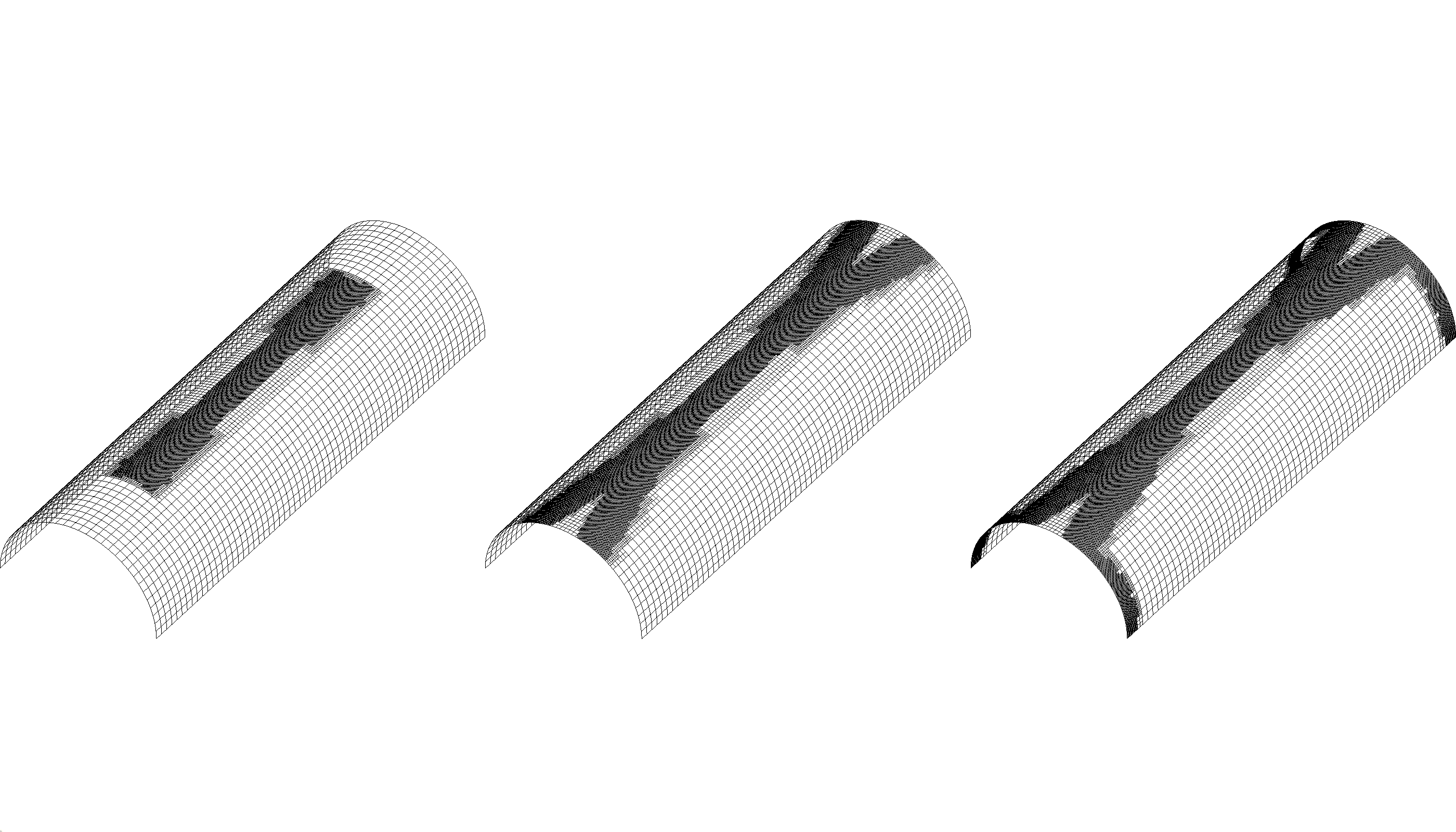}};
		\node at (-5.3,-3) {$t=1.187298\,T_0$};
		\node at (0,-3) {$t=2.056927\,T_0$};
		\node at (5.3,-3) {$t=2.715125\,T_0$};
	\end{tikzpicture}
\caption{Pressurized cylinder: LR meshes in the undeformed configuration during crack branching, before deflection and at the final state. See also the supplementary movie at \href{https://doi.org/10.5446/42541}{https://doi.org/10.5446/42541}.}
\label{f:cylmsh}
\end{figure}

\begin{figure}
	\centering
	\begin{tikzpicture}
		\def\cdot{\times}
		\begin{axis}[xmin=0,xmax=2.8,ymin=0,ymax=35000,grid=both,xlabel={$t\,[T_0]$},ylabel={$\#\,\mathrm{control\:points}\,[-]$},width=0.65\textwidth,height=0.45\textwidth]	
			\addplot[restrict expr to domain={\coordindex}{3:5600},blue,line width=1] table [x index = {0}, y index = {1},col sep=comma,]{ifigs/cylp/numnodes.csv};
		\end{axis}
	\end{tikzpicture}
\caption{Pressurized cylinder: Number of control points over time.}
\label{f:cylncp}
\end{figure}
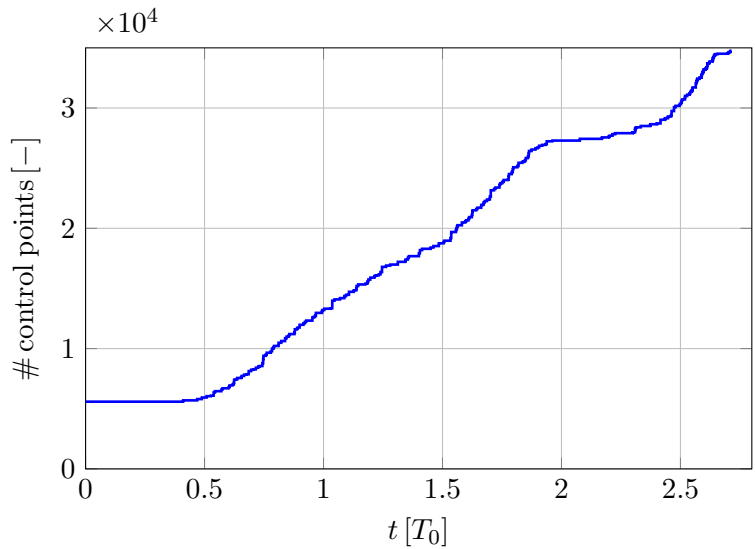

\begin{figure}[!ht]
	\centering
		\begin{tikzpicture}
			\node at (0,0) {\includegraphics[trim=950pt 900pt 750pt 800pt,clip,width=0.48\textwidth]{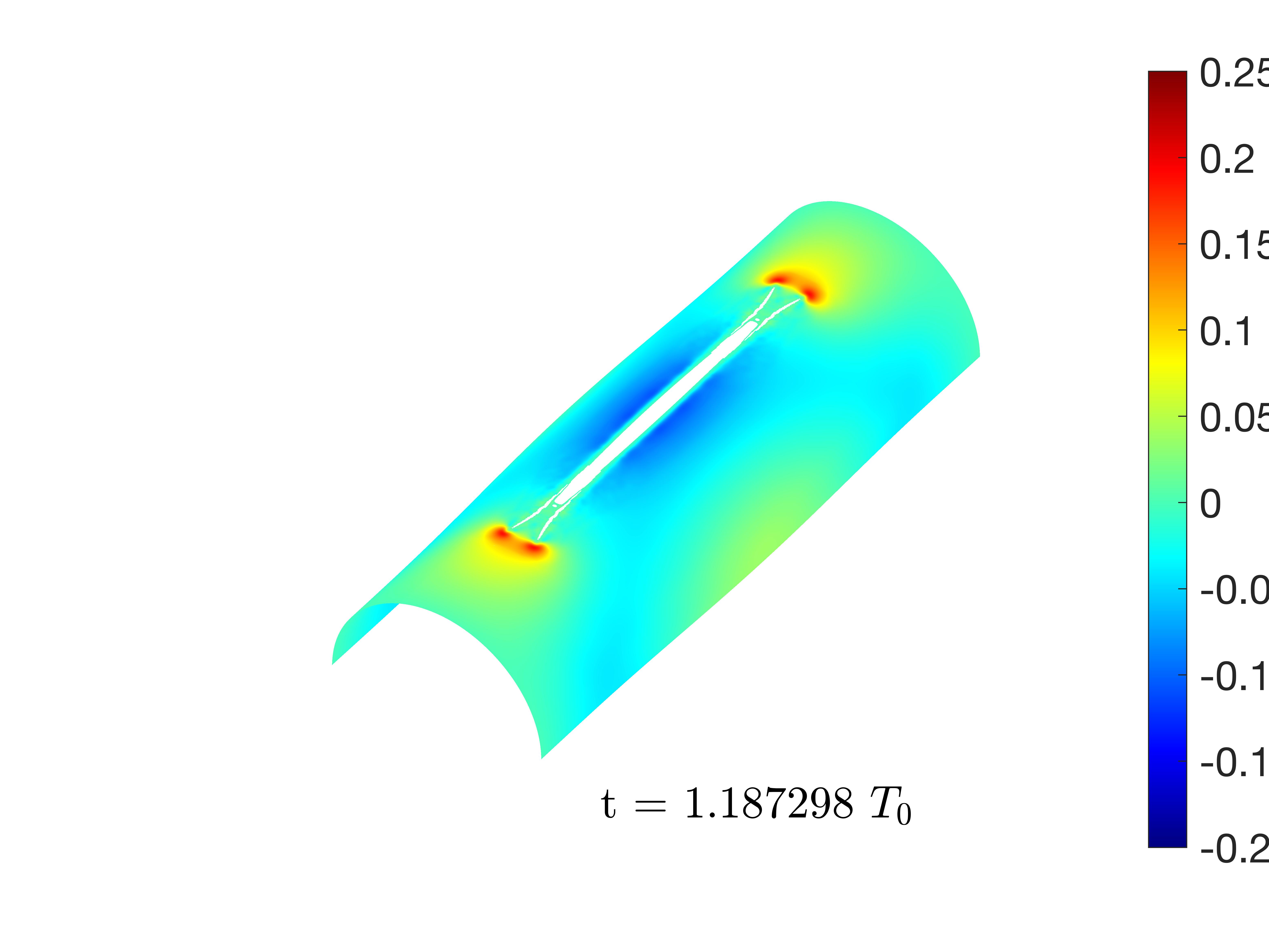}};
			\node at (-1,2) {$t=1.187298\,T_0$};
			\node at (8,0) {\includegraphics[trim=950pt 900pt 750pt 800pt,clip,width=0.48\textwidth]{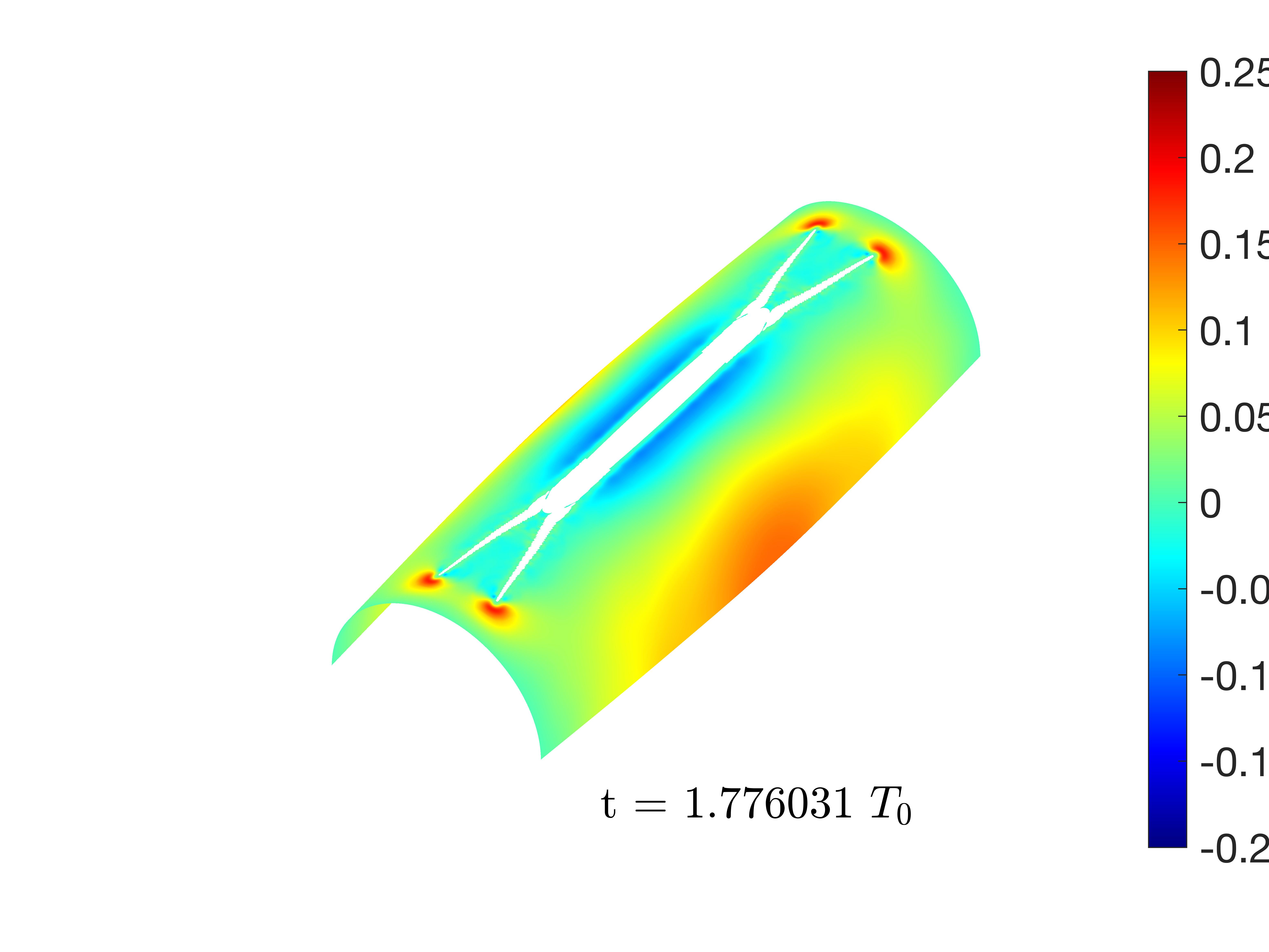}};
			\node at (7,2) {$t=1.776031\,T_0$};
			\node at (0,-5) {\includegraphics[trim=950pt 900pt 750pt 800pt,clip,width=0.48\textwidth]{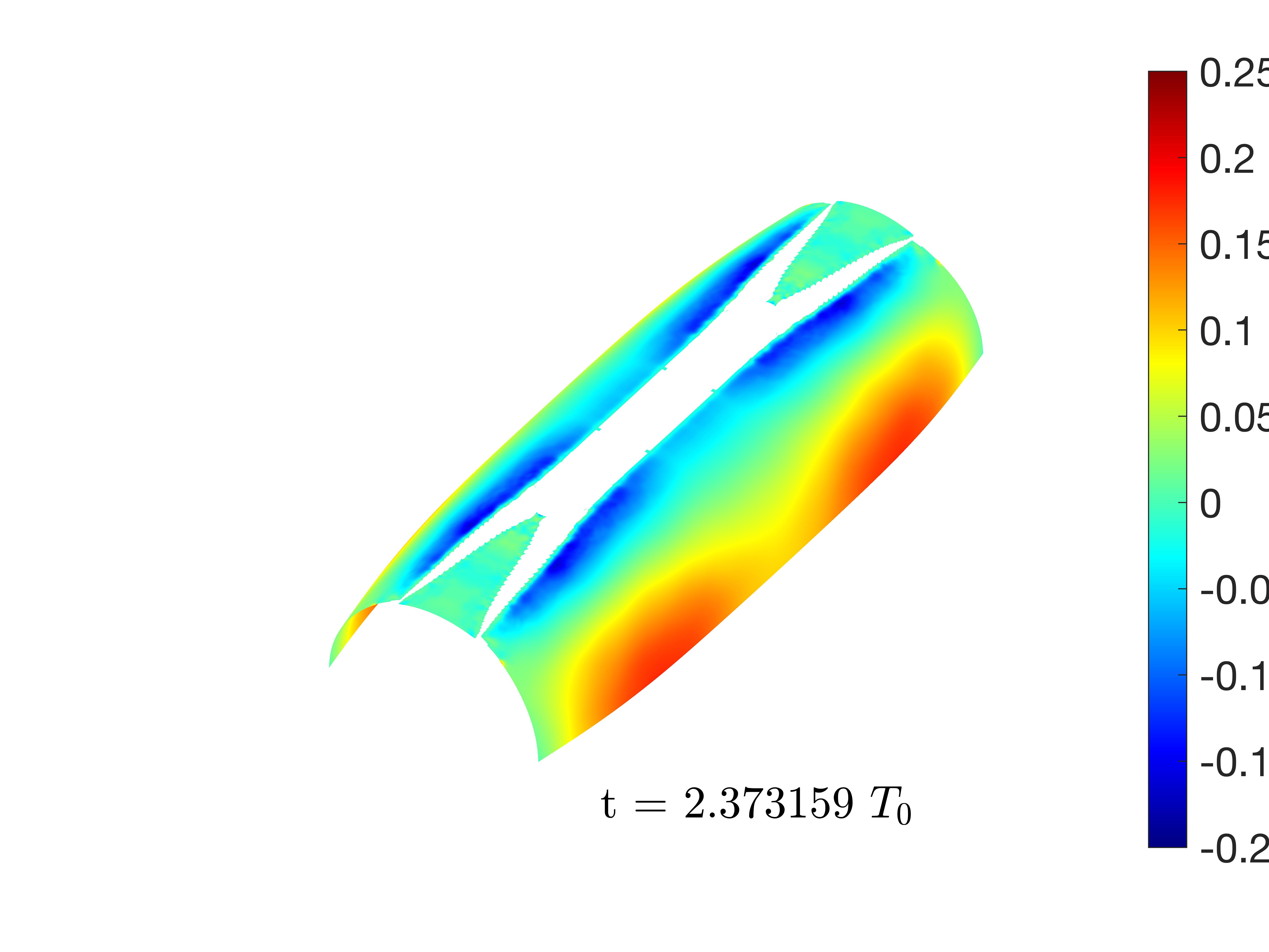}};
			\node at (-1,-3) {$t=2.373159\,T_0$};
			\node at (8,-5) {\includegraphics[trim=950pt 900pt 750pt 800pt,clip,width=0.48\textwidth]{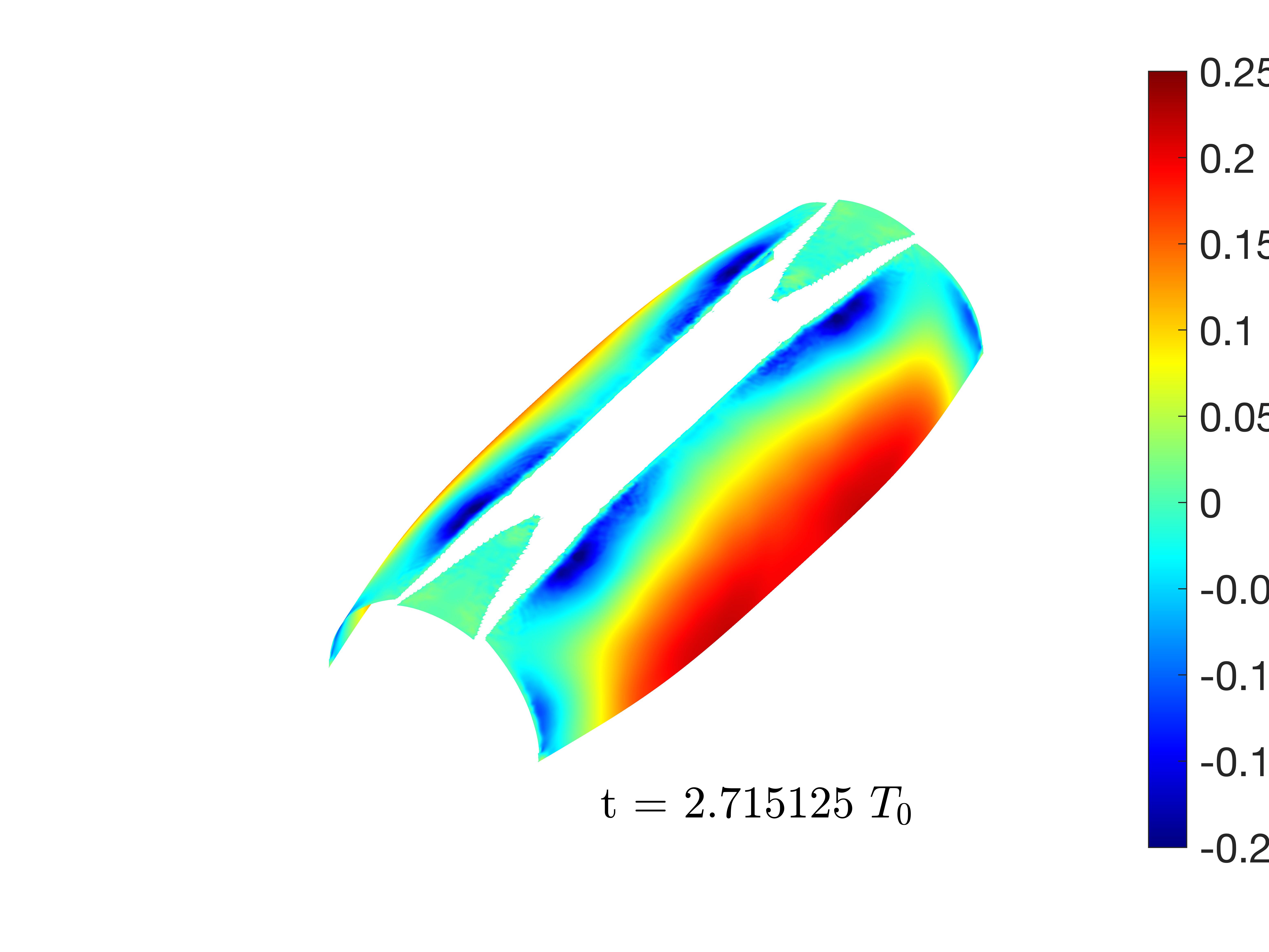}};
			\node at (7,-3) {$t=2.715125\,T_0$};
			\node at(4,-8.2) {\includegraphics[scale=0.2]{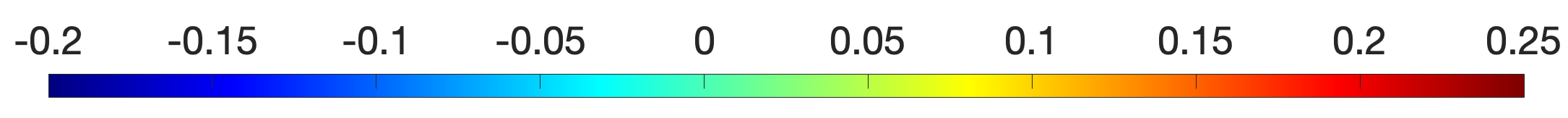}};
			\node at (4,-9) {$\gamma(\bx,t)$ $[E_0]$};
		\end{tikzpicture}
\caption{Pressurized cylinder: Surface tension $\gamma$ \eqref{e:gam} over time. Before the two branches reach the cylinder ends, the maximum values occur at the crack tip. Finally, the maximum values occur at the symmetry plane since the largest deformations occur there. Elements with $\phi<0.001$ have been removed for visualization. See also the supplementary movie at \href{https://doi.org/10.5446/42563}{https://doi.org/10.5446/42563}.}
\label{f:cylgam}
\end{figure}

In this example we study crack propagation on a curved surface. In the previous sections plane membranes without bending energy have been studied. The new problem setup is illustrated in Fig.~\ref{f:cyl1geom}. The corresponding parameters, including the imposed pressure $\bar p$ (cf. Eq.~\eqref{e:pres}), are listed in Tab.~\ref{tab:cylpar}. We note that the pressure is not ramped up over time but imposed as an initial pressure shock in the interior of the cylinder. The maximum pressure is then kept constant over time.
\begin{table}[!ht]
	\centering
	\setlength{\tabcolsep}{8pt}
	\renewcommand{\arraystretch}{1.25}
  	\begin{tabular}{c c c c c c c }
  		$E$ $[E_0]$ & $\nu$ $[-]$  & $\bar{p}$  $[E_0\,L_0^{-1}]$ & $\sG_c$ $[E_0\,L_0]$  & $\ell_0$ $[L_0]$  & $T$ $[L_0]$
  		\\ \hline
  		$10$ & $0.3$ & $-0.2$ & $0.00075$ & $0.01$ & $0.0125$
  \end{tabular}
  \caption{Pressurized cylinder: Material parameters and imposed pressure $\bar{p}$ (cf. Eq.~\eqref{e:pres}).}
  \label{tab:cylpar}
\end{table}
Fig.~\ref{f:cylphi} illustrates the phase field evolution over time. Elements with $\phi<0.001$ have been removed for visualization. The crack propagates in axial direction until it branches into two cracks at each end. These branches propagate towards the cylinder ends. The radius at these ends is fixed, which serves as a stiffener of the structure in these regions. Thus, the cracks are deflected and continue propagating in circumferential direction. This shows the ability of our model to capture crack evolution, branching and deflection on curved surfaces. Additionally, it proves that it is able to handle large deformations: The last state shown in Fig.~\ref{f:cylphi} at $t=2.715125\,T_0$ includes maximum stretches of approximately $130.49\%$.
In Fig.~\ref{f:cylmsh} the LR meshes for three different time steps are shown. In between the branches it is not refined as much as in the areas of fracture. The regions of no damage are kept coarse completely. As the crack is deflected in circumferential direction, the cylinder ends are refined up to the prescribed refinement level $d=3$. 
The initial mesh consists of $4,640$ elements and $4,572$ control points and the final mesh consists of $35,672$ elements and $34,756$ control points. A uniformly refined mesh would have $131,072$ elements and $128,777$ control points, which is almost four times higher. Fig.~\ref{f:cylncp} shows the number of control points over time.
The surface tension $\gamma(\bx,t)$ (cf. Eq.~\eqref{e:gam}) is visualized in Fig.~\ref{f:cylgam}. Elements with $\phi<0.001$ have been removed for visualization. Before the crack reaches the cylinder ends the maximum values are obtained at the crack tips. Small values are obtained behind the crack tip due to the emitted stress waves. The magnitude of the surface tension at the remaining areas is fluctuating due to reflection of stress waves and their following interaction. At the final state in Fig.~\ref{f:cylgam}, the largest stresses are obtained at the symmetry plane because the largest deformations occur there.

\section{Conclusion}\label{sec:concl}
We have coupled a higher order phase field model for brittle fracture with a nonlinear thin shell formulation based on a curvilinear surface description. Given a split of the constitutive law into membrane and bending contributions, a split of the elastic energy density has been derived for these terms separately. No spectral decomposition of the strain tensor is required in our formulation. Instead, the surface stretch indicates if there is a contribution to crack evolution or not. We have adopted a thickness integration to capture the asymmetric distribution of volumetric compression and expansion around the mid-plane that occur due to bending. As a consequence, the phase field is constant throughout the thickness and is solely defined on the deforming two-dimensional manifold. A discretization over the thickness or multiple phase fields have thus been avoided by this formulation. The interface between fractured and intact material has been adaptively refined based on the current phase field value. Quadratic LR NURBS have been used for this in the numerical examples. Time discretization is based on a fully implicit generalized-$\alpha$ scheme with adaptive time-stepping, and a monolithic Newton-Raphson procedure is used to solve the discretized coupled system.

The examples presented in Sec.~\ref{sec:num_ex} include flat membranes and curved shells. For the flat cases, the results qualitatively resemble those presented in the literature. Studying crack propagation on a cylinder indicates the ability of our formulation to capture non-trivial fracture patterns on curved surfaces. It has been observed that the phase field value serves as a suitable indicator for  local refinement since only areas along the crack paths are refined. The time step sizes are large if there is no crack evolution and are decreased as soon as the phase field starts evolving. Due to the adaptivity in space and time, the $C^1$-continuous solution is achieved within a computationally efficient framework.

Looking at the examples in Sec. \ref{sec:num_ex}, it does not seem to be necessary to keep a highly resolved mesh in regions of full damage ($\phi=0$). An adaptive coarsening strategy could be employed, which coarsens the mesh at fully damaged regions. Thus, small elements would only be retained close to the crack tip. A coarsening method for LR NURBS is given in \cite{zimmermann17}.
Additionally, stress wave propagation and reflection should be further investigated. Stress wave decay could be modeled by introducing physical viscosity into the system. The corresponding viscous energy and stresses then need to be appropriately split, similar to the energy split outlined in Sec.~\ref{sec:energysplit}. The reflection of stress waves at the boundaries could be damped by employing energy absorbing boundary layers. The same could be employed at the interfaces, where different element sizes meet to prevent reflection of stress waves at these LR mesh boundaries.

\section*{Acknowledgments}
Thomas J.R. Hughes and  Chad M. Landis were partially supported by the Office of Naval Research (Grant Nos. N00014-17-1-2119, N00014-13-1-0500, and N00014-17-1-2039).  Kranthi K. Mandadapu acknowledges support from University of California Berkeley and from the National Institutes of Health Grant R01-GM110066. Roger A. Sauer acknowledges the support from a J. Tinsley Oden fellowship in 2016 and funding from the German Research Foundation (DFG) through project GSC 111. Christopher Zimmermann and Karsten Paul were funded by the German Research Foundation (DFG) through projects GSC 111 and 33849990/GRK2379 (IRTG Modern Inverse Problems). 
Simulations were performed with computing resources granted by RWTH Aachen University under projects rwth0401 and rwth0433.

\bigskip
\section*{Appendix}
\appendix

\section{Time integration scheme}\label{sec:gena}
The system in Eq.~\eqref{e:gen_a1} with  intermediate quantities and the quantities at time step $n+1$ 
\eqb{lll}
\mx_{n+1} \is \mx_n+\Delta t\, \dot{\mx}_n+\big(\big(0.5 -\beta \big)\Delta t^2 \big)\ddot{\mx}_n+\beta\Delta t^2 \ddot{\mx}_{n+1}\,, \\[2mm]
\dot\mx_{n+1} \is \dot\mx_n+\big(\big(1 -\gamma \big)\Delta t \big)\ddot{\mx}_n+\gamma\Delta t\ddot{\mx}_{n+1}\,, \\[2mm]
\mx_{n+\alpha_\mrf} \is \big(1-\alpha_\mrf \big)\mx_n+\alpha_\mrf\mx_{n+1}\,, \\[2mm]
\dot\mx_{n+\alpha_\mrf} \is \big(1-\alpha_\mrf \big)\dot\mx_n+\alpha_\mrf\dot\mx_{n+1}\,, \\[2mm]
\ddot\mx_{n+\alpha_\mrm} \is \big(1-\alpha_\mrm \big)\ddot\mx_n+\alpha_\mrm\ddot\mx_{n+1}\,,
\label{e:gen_a2}
\eqe
has to be solved. Here, $\Delta t = t_{n+1}-t_n$ refers to the time step. Numerical dissipation is controlled by the parameters $\gamma$, $\beta$, $\alpha_\mrf$ and $\alpha_\mrm$. They are expressed in terms of $\rho_{\infty}\in [0,1]$, which resembles an algorithmic parameter that corresponds to the spectral radius of the amplification matrix as $\Delta t \rightarrow \infty$ (see \citet{chung93} for further details), i.e.
\eqb{lll}
\alpha_\mrf = \ds\frac{1}{1+\rho_{\infty}}\,,\quad
\alpha_\mrm = \ds\frac{2-\rho_{\infty}}{1+\rho_{\infty}}\,,\\[5mm]
\gamma = \ds\frac{1}{2}+\alpha_\mrm-\alpha_\mrf\,,\quad
\beta = \ds\frac{1}{4}\,(1+\alpha_\mrm-\alpha_\mrf)^2\,.
\eqe
We have found $\rho_\infty = 0.5$ to be a good choice and have used this in all computations.
To solve the nonlinear system of equations in Eq.~\eqref{e:gen_a1} using the Newton-Raphson procedure, it has to be linearized, i.e.
\eqb{lll}
\ds
\begin{bmatrix}
\mK_\mrx & \mK_\phi \\[2mm]
\bar\mK_\mrx & \bar\mK_\phi
\end{bmatrix}
\begin{bmatrix}
\Delta\mx_{n+1} \\[2mm]
\Delta\bphi_{n+1}
\end{bmatrix}
\is \ds -
\begin{bmatrix}
\mf\left(\mx_{n+\alpha_\mrf},\ddot{\mx}_{n+\alpha_\mrm},\bphi_{n+1} \right)\\[2mm]
\bar\mf\left(\mx_{n+\alpha_\mrf},\bphi_{n+1} \right)
\end{bmatrix},
\label{e:genasys}
\eqe
where the tangent matrix blocks are computed from
\eqb{lll}
\mK_\mrx \is \ds \frac{\partial \mf}{\partial \mx_{n+1}} = \ds \alpha_\mrf \frac{\partial \mf}{\partial \mx_{n+\alpha_\mrf}} + \frac{\alpha_\mrm}{\beta\Delta t^2} \frac{\partial \mf}{\partial \ddot\mx_{n+\alpha_\mrf}}\,,\\[4mm]
\mK_\phi \is \ds \frac{\partial \mf}{\partial \bphi_{n+1}}\,,\\[4mm]
\bar\mK_\mrx \is \ds \frac{\partial \bar\mf}{\partial \mx_{n+1}} = \alpha_\mrf \frac{\partial \bar\mf}{\partial \mx_{n+\alpha_\mrf}}\,, \\[4mm]
\bar\mK_\phi \is \ds \frac{\partial \bar\mf}{\partial \bphi_{n+1}}\,.
\label{e:tan_m}
\eqe
The required linearizations of the force vectors are shown in Appendix~\ref{s:lin}.
The initial guess for the Newton-Raphson iteration is set to
\eqb{lll}
\mx_{n+1}^0 = \mx_n+\Delta t\, \dot{\mx}_n+\big(\big(0.5 -\beta \big)\Delta t^2 \big)\ddot{\mx}_n+\big(\beta\Delta t^2 \big)\ddot{\mx}^0_{n+1}\,, \\[2mm]
\dot\mx_{n+1}^0 = \dot\mx_n\,,\\[2mm]
\ddot\mx_{n+1}^0 = \ddot\mx_n\ds\frac{\gamma -1}{\gamma}\,, \\[2mm]
\bphi_{n+1}^0 = \bphi_n\,,
\label{e:init_g}
\eqe
and then updated from iteration step $i\rightarrow i+1$ by
\eqb{lll}
\mx_{n+1}^{i+1} = \mx_{n+1}^{i} + \Delta\mx_{n+1}^{i+1}~,\\[2mm]
\dot\mx_{n+1}^{i+1} = \dot\mx_{n+1}^{i} + \Delta\mx_{n+1}^{i+1} \ds\frac{1}{\gamma\,\Delta t}\,, \\[2mm]
\ddot\mx_{n+1}^{i+1} = \ddot\mx_{n+1}^{i} + \Delta\mx_{n+1}^{i+1} \ds\frac{1}{\beta\,\Delta t^2}\,, \\[2mm]
\bphi_{n+1}^{i+1} = \bphi_{n+1}^{i}+\Delta \bphi_{n+1}^{i+1}\,,
\label{e:upd_g}
\eqe
until convergence is achieved. At iteration $i$ we check for the two convergence criteria
\eqb{l}
\max\ds \left\lbrace \frac{\norm{\mf^i_{n+1}}}{\norm{\mf^0_{n+1}}},\frac{\norm{\bar\mf^i_{n+1}}}{\norm{\bar\mf^0_{n+1}}} \right\rbrace \leq \text{tol}^\mathrm{dyn}\,,
\eqe
with $\norm{...}$ denoting the Euclidean norm and  $\text{tol}^\mathrm{dyn}=10^{-4}$ and
\eqb{l}
\begin{bmatrix}\mf \\[2mm]\bar\mf \end{bmatrix} \cdot \begin{bmatrix}\Delta\mx \\[2mm]\Delta\bphi\end{bmatrix} \leq \text{tol}^\mathrm{nrg}\,,
\eqe
with $\text{tol}^\mathrm{nrg}=10^{-25}$.

\section{Linearization} \label{s:lin}
This section presents the respective elemental contributions for the tangent blocks in Eq.~\eqref{e:tan_m}.
The linearization of the mechanical force vector $\mf^e:=\mf^e_\mathrm{kin}+\mf^e_\mathrm{int}-\mf^e_\mathrm{ext}$ of finite element $\Omega^e$ with respect to the respective nodal positions $\mx_e$ can be found in the work of \citet{duong2017}. Since we model the pressure as a function of the phase field variable, we need to linearize the external force vector with respect to $\phi$. This linearization of the pressure part $\mf_{\mathrm{ext}p}^e$ of the external elemental force vector reads
\eqb{l}
	\Delta_\phi\,\mf_{\mathrm{ext}p}^e := \ds\int_{\Omega^e}\mN^\mrT\,\bar p\,\bn^h\,\bar\mN\,\dif a\,\Delta\bphi_e\,.
\eqe
For the linearization of the internal force vector the four material tangents
\begin{equation}
\begin{alignedat}{4}
	c^{\alpha\beta\gamma\delta} &:= 2\ds\pa{\tau^\ab}{a_{\gamma\delta}}\,,\quad& &	d^{\alpha\beta\gamma\delta} &&:= \ds\pa{\tau^\ab}{b_{\gamma\delta}}\,,\\[4mm]
	e^{\alpha\beta\gamma\delta} &:= 2\ds\pa{M_0^\ab}{a_{\gamma\delta}}\,,\quad& &	f^{\alpha\beta\gamma\delta} &&:= \ds\pa{M_0^\ab}{b_{\gamma\delta}}\,,\\
\end{alignedat}
\end{equation}
have to be defined. Since we assume the constitutive in-plane response to be fully decoupled from the out-of-plane response, it follows that $d^{\alpha\beta\gamma\delta}=e^{\alpha\beta\gamma\delta}=0$. According to Eqs.~\eqref{e:tauabsplt} and \eqref{e:frctaumem}, the first tangent matrix can be computed based on the contributions
\eqb{lll}
	\ds\pa{\tau_\mathrm{dil}^\ab}{a_{\gamma\delta}} \is \dfrac{K}{2}\Bigr(J^2a^\ab a^\gd+\bigl(J^2-1\bigr)\,a^\abgd\Bigr)\,,\\[4mm]
	\ds\pa{\tau_\mathrm{dev}^\ab}{a_{\gamma\delta}} \is \dfrac{G}{2J}\left(\dfrac{I_1}{2}a^\ab a^\gd-I_1 a^{\ab\gd}-a^\ab A^\gd - A^\ab a^\gd\right)\,.
\eqe
Based on Eqs.~\eqref{e:M0absplt} and \eqref{e:frcM0pm3d}, the tangent matrix $f^{\ab\gd}$ can be computed with the contribution
\eqb{l}
	\ds\paqq{\tilde{\Psi}_\mathrm{bend}(\xi)}{b_\ab}{b_\gd}=\xi^2\dfrac{12}{T^3}\,c\,A^{\alpha\gamma}A^{ \beta\delta}\,.
\eqe
Since we consider the fully linearized system in Eq.~\eqref{e:genasys}, we also need to linearize the mechanical force vector with respect to the phase field, i.e.
\eqb{l}
\Delta_\phi \mf^e = \big[{\mk}^e_{\sig\phi}+{\mk}^e_{M\phi}\big]\,\Delta\bphi_e\,,
\eqe
with
\eqb{lll}
{\mk}^e_{\sig\phi} \dis \ds\int_{\Omega_0^e}g'(\phi)\,\tau^{\alpha\beta}_+\,\mN^\mrT_{\!,\alpha}\,\ba^h_\beta\,\bar\mN\,\dif A\,,  \\[4mm]
{\mk}^e_{M\phi} \dis \ds \int_{\Omega_0^e}g'(\phi)\,M_{0,+}^{\alpha\beta}\,\mN^\mrT_{\!;\alpha\beta}\,\bn^h\,\bar\mN\,\dif A\,,
\eqe
where $\tau^{\alpha\beta}:=J\sig^{\alpha\beta}$ and $M^{\alpha\beta}_0:=JM^{\alpha\beta}$ has been used to map the integrals to the element domain in the reference configuration.
According to Eq.~\eqref{e:ODEphi}, the linearization of $\bar\mf^e$  with respect to the respective nodal positions $\mx_e$ yields
\eqb{lll}
\Delta_\mrx\bar\mf^e_\mathrm{el} := \ds\int_{\Omega_0^e}\bar\mN^\mrT\ds\frac{2\ell_0}{\sG_\mrc}g'(\phi)\,\Delta_\mrx\sH\,\dif A\,\Delta\mx_\mathrm{e}\,,
\eqe
with
\eqb{l}
	\Delta_\mrx\sH := \Delta_\mrx\max\limits_{\tau\in[0,t]}\Psielp(\bx,\tau)\,,
\eqe
and
\eqb{l}
\Delta_\mrx\Psielp := {\tau^{\alpha\beta}_{\mathrm{el},+}}\,\ba_\alpha\cdot\mN_{\!,\beta} + {M^{\alpha\beta}_{0,+}}\,\bn\cdot\mN_{\!;\alpha\beta} \,.
\eqe

The linearization of  $\bar{\mf}_\mathrm{int}^e$ with respect to the phase field variables of $\Omega^e$ reads
\eqb{l}
\Delta_\phi\bar\mf^e_\mathrm{int} := \Bigl[\bar\mk^e_0 + \bar\mk^e_\mathrm{el}\Bigr]\,\Delta\bphi_\mathrm{e}\,,
\eqe
with
\eqb{l}
	\bar\mk^e_\mathrm{el} := \ds\int_{\Omega_0^e}\bar\mN^\mrT\bigg(\ds\frac{2\ell_0}{\sG_\mrc}g''(\phi)\sH\bigg)\bar\mN\,\,\dif A\,.
\eqe
The matrices $\bar\mk^e_0$ and $\bar\mk^e_\mathrm{el}$ both contribute to the tangent block $\bar\mK_\phi$ in Eq.~\eqref{e:genasys}.

\bigskip
\bibliographystyle{apalike}
\bibliography{bibliography}

\begin{thebibliography}{}

\bibitem[Ambati and De~Lorenzis, 2016]{ambati2016b}
Ambati, M. and De~Lorenzis, L. (2016).
\newblock Phase-field modeling of brittle and ductile fracture in shells with
  isogeometric {NURBS}-based solid-shell elements.
\newblock {\em Computer Methods in Applied Mechanics and Engineering}, {\bf
  312}:351---373.

\bibitem[Ambati et~al., 2015]{ambati2015}
Ambati, M., Gerasimov, T., and De~Lorenzis, L. (2015).
\newblock A review on phase-field models of brittle fracture and a new fast
  hybrid formulation.
\newblock {\em Computational Mechanics}, {\bf 55}(2):383--405.

\bibitem[Ambati et~al., 2016]{ambati2016a}
Ambati, M., Kruse, R., and De~Lorenzis, L. (2016).
\newblock A phase-field model for ductile fracture at finite strains and its
  experimental verification.
\newblock {\em Computational Mechanics}, {\bf 57}(1):149--167.

\bibitem[Amiri et~al., 2014]{amiri2014}
Amiri, F., Millán, D., Shen, Y., Rabczuk, T., and Arroyo, M. (2014).
\newblock Phase-field modeling of fracture in linear thin shells.
\newblock {\em Theoretical and Applied Fracture Mechanics}, {\bf 69}:102--109.
\newblock Introducing the new features of Theoretical and Applied Fracture
  Mechanics through the scientific expertise of the Editorial Board.

\bibitem[Amor et~al., 2009]{amor2009}
Amor, H., Marigo, J.-J., and Maurini, C. (2009).
\newblock Regularized formulation of the variational brittle fracture with
  unilateral contact: Numerical experiments.
\newblock {\em Journal of the Mechanics and Physics of Solids}, {{\bf
  57}}(8):1209--1229.

\bibitem[Areias et~al., 2016]{areias2016}
Areias, P., Rabczuk, T., and Msekh, M. (2016).
\newblock Phase-field analysis of finite-strain plates and shells including
  element subdivision.
\newblock {\em Computer Methods in Applied Mechanics and Engineering}, {\bf
  312}:322--350.

\bibitem[Badnava et~al., 2018]{badnava2018}
Badnava, H., Msekh, M.~A., Etemadi, E., and Rabczuk, T. (2018).
\newblock An h-adaptive thermo-mechanical phase field model for fracture.
\newblock {\em Finite Elements in Analysis and Design}, {\bf 138}:31--47.

\bibitem[Benson et~al., 2013]{benson2013}
Benson, D.~J., Hartmann, S., Bazilevs, Y., Hsu, M.-C., and Hughes, T. J.~R.
  (2013).
\newblock Blended isogeometric shells.
\newblock {\em Computer Methods in Applied Mechanics and Engineering}, {\bf
  255}:133--146.

\bibitem[Borden et~al., 2016]{borden2016}
Borden, M.~J., Hughes, T. J.~R., Landis, C.~M., Anvari, A., and Lee, I.~J.
  (2016).
\newblock A phase-field formulation for fracture in ductile materials: {F}inite
  deformation balance law derivation, plastic degradation, and stress
  triaxiality effects.
\newblock {\em Computer Methods in Applied Mechanics and Engineering}, {\bf
  312}:130--166.

\bibitem[Borden et~al., 2014]{borden2014}
Borden, M.~J., Hughes, T. J.~R., Landis, C.~M., and Verhoosel, C.~V. (2014).
\newblock A higher-order phase-field model for brittle fracture: Formulation
  and analysis within the isogeometric analysis framework.
\newblock {\em Computer Methods in Applied Mechanics and Engineering}, {\bf
  273}:100--118.

\bibitem[Borden et~al., 2012]{borden2012}
Borden, M.~J., Verhoosel, C.~V., Scott, M.~A., Hughes, T. J.~R., and Landis,
  C.~M. (2012).
\newblock A phase-field description of dynamic brittle fracture.
\newblock {\em Computer Methods in Applied Mechanics and Engineering}, {\bf
  217--220}:77--95.

\bibitem[Bourdin et~al., 2000]{bourdin2000}
Bourdin, B., Francfort, G., and Marigo, J.-J. (2000).
\newblock Numerical experiments in revisited brittle fracture.
\newblock {\em Journal of the Mechanics and Physics of Solids}, {\bf
  48}(4):797--826.

\bibitem[Bourdin et~al., 2011]{bourdin2011}
Bourdin, B., Larsen, C.~J., and Richardson, C.~L. (2011).
\newblock A time-discrete model for dynamic fracture based on crack
  regularization.
\newblock {\em International Journal of Fracture}, {\bf 168}(2):133--143.

\bibitem[Chen and de~Borst, 2018]{chen2018a}
Chen, L. and de~Borst, R. (2018).
\newblock {Locally Refined T-splines}.
\newblock {\em International Journal for Numerical Methods in Engineering},
  {\bf 114}(6):637--659.

\bibitem[Chen et~al., 2018]{chen2018b}
Chen, L., Verhoosel, C.~V., and de~Borst, R. (2018).
\newblock Discrete fracture analysis using locally refined {T}-splines.
\newblock {\em International Journal for Numerical Methods in Engineering},
  {\bf 116}(2):117--140.

\bibitem[Chung and Hulbert, 1993]{chung93}
Chung, J. and Hulbert, G.~M. (1993).
\newblock A time integration algorithm for structural dynamics with improved
  numerical dissipation: {T}he generalized-alpha method.
\newblock {\em Journal of Applied Mechanics}, {\bf 60}(2):371--375.

\bibitem[Ciarlet, 1993]{ciarlet1993}
Ciarlet, P.~G. (1993).
\newblock {\em Mathematical Elasticity: Three Dimensional Elasticity}.
\newblock North-Holland.

\bibitem[Dokken et~al., 2013]{dokken13}
Dokken, T., Lyche, T., and Pettersen, K.~F. (2013).
\newblock Polynomial splines over locally refined box-partitions.
\newblock {\em Computer Aided Geometric Design}, {\bf 30}(3):331--356.

\bibitem[Duong et~al., 2017]{duong2017}
Duong, T.~X., Roohbakhshan, F., and Sauer, R.~A. (2017).
\newblock A new rotation-free isogeometric thin shell formulation and a
  corresponding continuity constraint for patch boundaries.
\newblock {\em Computer Methods in Applied Mechanics and Engineering}, {\bf
  316}:43--83.

\bibitem[Echter et~al., 2013]{echter2013}
Echter, R., Oesterle, B., and Bischoff, M. (2013).
\newblock A hierarchic family of isogeometric shell finite elements.
\newblock {\em Computer Methods in Applied Mechanics and Engineering}, {\bf
  254}:170--180.

\bibitem[Forsey and Bartels, 1988]{forsey1988}
Forsey, D.~R. and Bartels, R.~H. (1988).
\newblock Hierarchical {B}-spline refinement.
\newblock {\em SIGGRAPH Comput. Graph.}, {\bf 22}(4):205--212.

\bibitem[Francfort and Marigo, 1998]{francfort1998}
Francfort, G. and Marigo, J.-J. (1998).
\newblock Revisiting brittle fracture as an energy minimization problem.
\newblock {\em Journal of the Mechanics and Physics of Solids}, {\bf
  46}(8):1319--1342.

\bibitem[Geelen et~al., 2019]{geelen2019}
Geelen, R.~J., Liu, Y., Hu, T., Tupek, M.~R., and Dolbow, J.~E. (2019).
\newblock A phase-field formulation for dynamic cohesive fracture.
\newblock {\em Computer Methods in Applied Mechanics and Engineering}, {\bf
  348}:680--711.

\bibitem[Geelen et~al., 2018]{geelen2018}
Geelen, R. J.~M., Liu, Y., Dolbow, J.~E., and Rodr\'{i}guez-Ferran, A. (2018).
\newblock An optimization-based phase-field method for continuous-discontinuous
  crack propagation.
\newblock {\em International Journal for Numerical Methods in Engineering},
  {\bf 116}(1):1--20.

\bibitem[Gerasimov and Lorenzis, 2016]{gerasimov2016}
Gerasimov, T. and Lorenzis, L.~D. (2016).
\newblock A line search assisted monolithic approach for phase-field computing
  of brittle fracture.
\newblock {\em Computer Methods in Applied Mechanics and Engineering}, {\bf
  312}:276--303.

\bibitem[Gerasimov and Lorenzis, 2019]{gerasimov2019}
Gerasimov, T. and Lorenzis, L.~D. (2019).
\newblock On penalization in variational phase-field models of brittle
  fracture.
\newblock {\em Computer Methods in Applied Mechanics and Engineering}, {\bf
  354}:990--1026.

\bibitem[Gerasimov et~al., 2018]{gerasimov2018}
Gerasimov, T., Noii, N., Allix, O., and De~Lorenzis, L. (2018).
\newblock A non-intrusive global/local approach applied to phase-field modeling
  of brittle fracture.
\newblock {\em Advanced Modeling and Simulation in Engineering Sciences}, {\bf
  5}.

\bibitem[Gomez et~al., 2014]{gomez2014}
Gomez, H., Reali, A., and Sangalli, G. (2014).
\newblock Accurate, efficient, and (iso)geometrically flexible collocation
  methods for phase-field models.
\newblock {\em Journal of Computational Physics}, {\bf 262}:153--171.

\bibitem[{Griffith}, 1921]{griffith1921}
{Griffith}, A.~A. (1921).
\newblock {{VI}. {T}he Phenomena of Rupture and Flow in Solids}.
\newblock {\em Philosophical Transactions of the Royal Society of London Series
  A}, {\bf 221}:163--198.

\bibitem[Heister et~al., 2015]{heister2015}
Heister, T., Wheeler, M.~F., and Wick, T. (2015).
\newblock A primal-dual active set method and predictor-corrector mesh
  adaptivity for computing fracture propagation using a phase-field approach.
\newblock {\em Computer Methods in Applied Mechanics and Engineering}, {\bf
  290}:466--495.

\bibitem[Hesch et~al., 2016a]{hesch2016a}
Hesch, C., Franke, M., Dittmann, M., and Temizer, {\.I}. (2016a).
\newblock Hierarchical {NURBS} and a higher-order phase-field approach to
  fracture for finite-deformation contact problems.
\newblock {\em Computer Methods in Applied Mechanics and Engineering}, {\bf
  301}:242 --58.

\bibitem[Hesch et~al., 2016b]{hesch2016b}
Hesch, C., Schuß, S., Dittmann, M., Franke, M., and Weinberg, K. (2016b).
\newblock Isogeometric analysis and hierarchical refinement for higher-order
  phase-field models.
\newblock {\em Computer Methods in Applied Mechanics and Engineering}, {\bf
  303}:185--207.

\bibitem[Hirmand and Papoulia, 2018]{hirmand2018}
Hirmand, M.~R. and Papoulia, K.~D. (2018).
\newblock A continuation method for rigid-cohesive fracture in a discontinuous
  {G}alerkin finite element setting.
\newblock {\em International Journal for Numerical Methods in Engineering},
  {\bf 115}(5):627--650.

\bibitem[Hirmand and Papoulia, 2019]{hirmand2019}
Hirmand, M.~R. and Papoulia, K.~D. (2019).
\newblock Block coordinate descent energy minimization for dynamic cohesive
  fracture.
\newblock {\em Computer Methods in Applied Mechanics and Engineering}, {\bf
  354}:663--688.

\bibitem[Hofacker and Miehe, 2013]{hofacker2012}
Hofacker, M. and Miehe, C. (2013).
\newblock A phase field model of dynamic fracture: {R}obust field updates for
  the analysis of complex crack patterns.
\newblock {\em International Journal for Numerical Methods in Engineering},
  {\bf 93}(3):276--301.

\bibitem[Hughes et~al., 2005]{hughes2005}
Hughes, T. J.~R., Cottrell, J.~A., and Bazilevs, Y. (2005).
\newblock Isogeometric analysis: {CAD}, finite elements, {NURBS}, exact
  geometry and mesh refinement.
\newblock {\em Computer Methods in Applied Mechanics and Engineering}, {\bf
  194}(39--41):4135--4195.

\bibitem[Johannessen et~al., 2014]{johannessen2014}
Johannessen, K.~A., Kvamsdal, T., and Dokken, T. ({2014}).
\newblock {Isogeometric analysis using LR {B}-splines}.
\newblock {\em Computer Methods in Applied Mechanics and Engineering}, {\bf
  269}:{471--514}.

\bibitem[Karma et~al., 2001]{karma2001}
Karma, A., Kessler, D., and Levine, H. (2001).
\newblock Phase-field model of mode {III} dynamic fracture.
\newblock {\em Physical Review Letters}, {\bf 75}.

\bibitem[K{\"a}stner et~al., 2016]{kaestner2016}
K{\"a}stner, M., Hennig, P., Linse, T., and Ulbricht, V. (2016).
\newblock Phase-field modelling of damage and fracture---{C}onvergence and
  local mesh refinement.
\newblock In Naumenko, K. and A{\ss}mus, M., editors, {\em Advanced Methods of
  Continuum Mechanics for Materials and Structures}, pages 307--324. Springer
  Singapore, Singapore.

\bibitem[Kiendl et~al., 2016]{kiendl2016}
Kiendl, J., Ambati, M., {De Lorenzis}, L., Gomez, H., and Reali, A. (2016).
\newblock Phase-field description of brittle fracture in plates and shells.
\newblock {\em Computer Methods in Applied Mechanics and Engineering}, {\bf
  312}:374--394.

\bibitem[Kiendl et~al., 2015]{kiendl2015}
Kiendl, J., Hsu, M.-C., Wu, M.~C., and Reali, A. (2015).
\newblock Isogeometric {K}irchhoff–{L}ove shell formulations for general
  hyperelastic materials.
\newblock {\em Computer Methods in Applied Mechanics and Engineering}, {\bf
  291}:280--303.

\bibitem[Krueger, 2004]{krueger2004}
Krueger, R. (2004).
\newblock {Virtual crack closure technique: History, approach, and applications
  }.
\newblock {\em Applied Mechanics Reviews}, {\bf 57}(2):109--143.

\bibitem[Kuhn and M{\"u}ller, 2010]{kuhn2010}
Kuhn, C. and M{\"u}ller, R. (2010).
\newblock A continuum phase field model for fracture.
\newblock {\em Engineering Fracture Mechanics}, {\bf 77}(18):3625--3634.
\newblock Computational Mechanics in Fracture and Damage: A Special Issue in
  Honor of Prof. Gross.

\bibitem[Kuhn et~al., 2015]{kuhn2015}
Kuhn, C., Schl{\"u}ter, A., and M{\"u}ller, R. (2015).
\newblock On degradation functions in phase field fracture models.
\newblock {\em Computational Materials Science}, {{\bf 108}}:374--384.
\newblock Selected Articles from Phase-field Method 2014 International Seminar.

\bibitem[Larsen et~al., 2010]{larsen2010a}
Larsen, C., Ortner, C., and S{\"u}li, E. (2010).
\newblock Existence of solutions to a regularized model of dynamic fracture.
\newblock {\em Math. Models Methods Appl. Sci.}, {\bf 20}:1021--1048.

\bibitem[Larsen, 2010]{larsen2010b}
Larsen, C.~J. (2010).
\newblock Models for dynamic fracture based on {G}riffith's criterion.
\newblock In Hackl, K., editor, {\em IUTAM Symposium on Variational Concepts
  with Applications to the Mechanics of Materials}, pages 131--140, Dordrecht.
  Springer Netherlands.

\bibitem[Linse et~al., 2017]{linse2017}
Linse, T., Hennig, P., K{\"a}stner, M., and de~Borst, R. (2017).
\newblock A convergence study of phase-field models for brittle fracture.
\newblock {\em Engineering Fracture Mechanics}, {\bf 184}:307--318.

\bibitem[Miehe et~al., 2010a]{miehe2010a}
Miehe, C., Hofacker, M., and Welschinger, F. (2010a).
\newblock A phase field model for rate-independent crack propagation: {R}obust
  algorithmic implementation based on operator splits.
\newblock {\em Computer Methods in Applied Mechanics and Engineering}, {\bf
  199}(45):2765--2778.

\bibitem[Miehe et~al., 2010b]{miehe2010b}
Miehe, C., Welschinger, F., and Hofacker, M. (2010b).
\newblock Thermodynamically consistent phase-field models of fracture:
  {V}ariational principles and multi-field {FE} implementations.
\newblock {\em International Journal for Numerical Methods in Engineering},
  {\bf 83}(10):1273--1311.

\bibitem[Mo{\"e}s et~al., 1999]{moes1999}
Mo{\"e}s, N., Dolbow, J., and Belytschko, T. (1999).
\newblock A finite element method for crack growth without remeshing.
\newblock {\em International Journal for Numerical Methods in Engineering},
  {\bf 46}:131--150.

\bibitem[Molinari et~al., 2007]{molinari2007}
Molinari, J.~F., Gazonas, G., Raghupathy, R., Rusinek, A., and Zhou, F. (2007).
\newblock The cohesive element approach to dynamic fragmentation: {T}he
  question of energy convergence.
\newblock {\em International Journal for Numerical Methods in Engineering},
  {\bf 69}(3):484--503.

\bibitem[Nagaraja et~al., 2018]{nagaraja2018}
Nagaraja, S., Elhaddad, M., Ambati, M., Kollmannsberger, S., De~Lorenzis, L.,
  and Rank, E. (2018).
\newblock Phase-field modeling of brittle fracture with multi-level hp-{FEM}
  and the finite cell method.
\newblock {\em Computational Mechanics}.

\bibitem[Naghdi, 1973]{naghdi1971}
Naghdi, P.~M. (1973).
\newblock The theory of shells and plates.
\newblock In Truesdell, C., editor, {\em Linear Theories of Elasticity and
  Thermoelasticity: Linear and Nonlinear Theories of Rods, Plates, and Shells},
  pages 425--640, Berlin, Heidelberg. Springer.

\bibitem[Ortiz and Pandolfi, 1999]{ortiz1999}
Ortiz, M. and Pandolfi, A. (1999).
\newblock Finite-deformation irreversible cohesive elements for
  three-dimensional crack-propagation analysis.
\newblock {\em International Journal for Numerical Methods in Engineering},
  {\bf 44}(9):1267--1282.

\bibitem[Papoulia, 2017]{papoulia2017}
Papoulia, K.~D. (2017).
\newblock Non-differentiable energy minimization for cohesive fracture.
\newblock {\em International Journal of Fracture}, {\bf 204}(2):143--158.

\bibitem[Parvizian et~al., 2007]{parvizian2007}
Parvizian, J., D{\"u}ster, A., and Rank, E. (2007).
\newblock Finite cell method.
\newblock {\em Computational Mechanics}, {\bf 41}(1):121--133.

\bibitem[Radovitzky et~al., 2011]{radovitzky2011}
Radovitzky, R., Seagraves, A., Tupek, M., and Noels, L. (2011).
\newblock A scalable 3d fracture and fragmentation algorithm based on a hybrid,
  discontinuous {G}alerkin, cohesive element method.
\newblock {\em Computer Methods in Applied Mechanics and Engineering}, {\bf
  200}(1):326--344.

\bibitem[Ravi-Chandar and Knauss, 1984]{ravichandar1984}
Ravi-Chandar, K. and Knauss, W.~G. (1984).
\newblock An experimental investigation into dynamic fracture: {III. On}
  steady-state crack propagation and crack branching.
\newblock {\em International Journal of Fracture}, {\bf 26}(2):141--154.

\bibitem[Reali and Hughes, 2015]{reali2015}
Reali, A. and Hughes, T. J.~R. (2015).
\newblock {\em An Introduction to Isogeometric Collocation Methods}, pages
  173--204.
\newblock Springer Vienna.

\bibitem[Reinoso et~al., 2017]{reinoso2017}
Reinoso, J., Paggi, M., and Linder, C. (2017).
\newblock Phase field modeling of brittle fracture for enhanced assumed strain
  shells at large deformations: formulation and finite element implementation.
\newblock {\em Computational Mechanics}, {\bf 59}(6):981--1001.

\bibitem[Remmers et~al., 2003]{remmers2003}
Remmers, J. J.~C., {de Borst}, R., and Needleman, A. (2003).
\newblock A cohesive segments method for the simulation of crack growth.
\newblock {\em Computational Mechanics}, {{\bf 31}}(1):69--77.

\bibitem[Sahu et~al., 2017]{sahu2017}
Sahu, A., Sauer, R.~A., and Mandadapu, K.~K. (2017).
\newblock Irreversible thermodynamics of curved lipid membranes.
\newblock {\em Physical Review E}, {\bf 96}:042409.

\bibitem[Sargado et~al., 2018]{sargado2018}
Sargado, J.~M., Keilegavlen, E., Berre, I., and Nordbotten, J.~M. (2018).
\newblock High-accuracy phase-field models for brittle fracture based on a new
  family of degradation functions.
\newblock {\em Journal of the Mechanics and Physics of Solids}, {\bf
  111}:458--489.

\bibitem[Sauer, 2018]{sauer2018}
Sauer, R.~A. (2018).
\newblock On the computational modeling of lipid bilayers using thin-shell
  theory.
\newblock In Steigmann, D.~J., editor, {\em The Role of Mechanics in the Study
  of Lipid Bilayers}, pages 221--286. Springer International Publishing, Cham.

\bibitem[Sauer and Duong, 2017]{sauer2017a}
Sauer, R.~A. and Duong, T.~X. (2017).
\newblock On the theoretical foundations of thin solid and liquid shells.
\newblock {\em Mathematics and Mechanics of Solids}, {\bf 22}(3):343--371.

\bibitem[Sauer et~al., 2014]{sauer2014b}
Sauer, R.~A., Duong, T.~X., and Corbett, C.~J. (2014).
\newblock A computational formulation for constrained solid and liquid
  membranes considering isogeometric finite elements.
\newblock {\em Computer Methods in Applied Mechanics and Engineering}, {\bf
  271}:48--68.

\bibitem[Sauer et~al., 2017]{sauer2017b}
Sauer, R.~A., Duong, T.~X., Mandadapu, K.~K., and Steigmann, D.~J. (2017).
\newblock A stabilized finite element formulation for liquid shells and its
  application to lipid bilayers.
\newblock {\em Journal of Computational Physics}, {\bf 330}:436--466.

\bibitem[Schillinger et~al., 2015]{schillinger2015}
Schillinger, D., Borden, M.~J., and Stolarski, H.~K. (2015).
\newblock Isogeometric collocation for phase-field fracture models.
\newblock {\em Computer Methods in Applied Mechanics and Engineering}, {\bf
  284}:583--610.
\newblock Isogeometric Analysis Special Issue.

\bibitem[Schl{\"u}ter et~al., 2014]{schlueter2014}
Schl{\"u}ter, A., Willenb{\"u}cher, A., Kuhn, C., and M{\"u}ller, R. (2014).
\newblock Phase field approximation of dynamic brittle fracture.
\newblock {\em Computational Mechanics}, {{\bf 54}}(5):1141--1161.

\bibitem[Sederberg et~al., 2003]{sederberg2003}
Sederberg, T.~W., Zheng, J., Bakenov, A., and Nasri, A. (2003).
\newblock T-splines and {T}-{NURCC}s.
\newblock {\em ACM Transactions on Graphics}, {\bf 22}(3):477--484.

\bibitem[Steigmann, 1999]{steigmann1999}
Steigmann, D.~J. (1999).
\newblock Fluid films with curvature elasticity.
\newblock {\em Archive for Rational Mechanics and Analysis}, {\bf
  150}:127--152.

\bibitem[Ulmer et~al., 2012]{ulmer2012}
Ulmer, H., Hofacker, M., and Miehe, C. (2012).
\newblock Phase field modeling of fracture in plates and shells.
\newblock {\em PAMM}, {\bf 12}(1):171--172.

\bibitem[Vavasis et~al., 2020]{vavasis2020}
Vavasis, S.~A., Papoulia, K.~D., and Hirmand, M.~R. (2020).
\newblock Second-order cone interior-point method for quasistatic and moderate
  dynamic cohesive fracture.
\newblock {\em Computer Methods in Applied Mechanics and Engineering}, {\bf
  358}:112633.

\bibitem[Zhou and Zhuang, 2018]{zhou2018}
Zhou, S. and Zhuang, X. (2018).
\newblock Adaptive phase field simulation of quasi-static crack propagation in
  rocks.
\newblock {\em Underground Space}, {\bf 3}(3):190--205.
\newblock Computational Modeling of Fracture in Geotechnical Engineering Part
  I.

\bibitem[Zimmermann and Sauer, 2017]{zimmermann17}
Zimmermann, C. and Sauer, R.~A. (2017).
\newblock Adaptive local surface refinement based on {LR NURBS} and its
  application to contact.
\newblock {\em Computational Mechanics}, {\bf 60}:1011--1031.

\bibitem[Zimmermann et~al., 2019]{zimmermann2019}
Zimmermann, C., Toshniwal, D., Landis, C.~M., Hughes, T. J.~R., Mandadapu,
  K.~K., and Sauer, R.~A. (2019).
\newblock An isogeometric finite element formulation for phase transitions on
  deforming surfaces.
\newblock {\em Computer Methods in Applied Mechanics and Engineering}, {\bf
  351}:441--477.

\end{thebibliography}

\end{document}